\documentclass[12pt,a4paper]{article}
\usepackage{fullpage}
\usepackage{float}
\usepackage[centertags]{amsmath}
\allowdisplaybreaks[4]
\numberwithin{equation}{section}
\usepackage{amsbsy}
\usepackage{amsfonts}
\usepackage{amssymb}
\usepackage[dvips]{graphicx}
\usepackage{url}
\usepackage[dvips]{hyperref}
\usepackage{tocbibind}
\usepackage[dvips]{thumbpdf}
\begin{document}

\begin{flushright}
% \text{\sf \today}
% \text{\sf 20 October 2003}
% \\
hep-ph/0310238
\end{flushright}

\vspace{1cm}

\begin{center}
\huge
\textbf{Neutrino Mixing}
\normalsize
\\[0.5cm]
\large
Carlo Giunti
\normalsize
\\[0.5cm]
INFN, Sezione di Torino, and Dipartimento di Fisica Teorica,
\\
Universit\`a di Torino,
Via P. Giuria 1, I--10125 Torino, Italy
\\[0.5cm]
\large
Marco Laveder
\normalsize
\\[0.5cm]
Dipartimento di Fisica ``G. Galilei'', Universit\`a di Padova,
\\
and INFN, Sezione di Padova,
Via F. Marzolo 8, I--35131 Padova, Italy
\\[0.5cm]
\begin{minipage}[t]{0.8\textwidth}
\begin{center}
\textbf{Abstract}
\end{center}
In this review we present the main features of the current status
of neutrino physics.
After a review of the theory of neutrino mixing and oscillations,
we discuss the current status of
solar and atmospheric neutrino oscillation experiments.
We show that the current data
can be nicely accommodated in the framework of three-neutrino mixing.
We discuss also the problem of the determination of the
absolute neutrino mass scale
through Tritium $\beta$-decay experiments
and
astrophysical observations,
and the exploration of the Majorana nature of massive neutrinos
through
neutrinoless double-$\beta$ decay experiments.
Finally, future prospects are briefly discussed.
%In this review we discuss:
%the theory of neutrino masses and mixing,
%the theory of neutrino oscillations,
%the experimental status of solar and atmospheric neutrino oscillation experiments,
%the phenomenology of three-neutrino mixing
%and future prospects.
\end{minipage}
\end{center}

\begin{flushleft}
\end{flushleft}

\newpage
\tableofcontents
\newpage

\section{Introduction}
\label{Introduction}

The last five years have seen enormous progress in our
knowledge of neutrino physics.
We have now strong experimental evidences
of the existence of neutrino oscillations,
predicted by Pontecorvo in the late 50's
\cite{Pontecorvo:1957cp,Pontecorvo-58},
which occur if neutrinos are massive and mixed particles.

In 1998 the Super-Kamiokande experiment \cite{Fukuda:1998mi}
provided a model-independent proof of the oscillations of atmospheric muon neutrinos,
which were discovered in 1988 by the
Kamiokande \cite{Hirata:1988uy} and IMB \cite{Bionta:1988an} experiments.
The values of the neutrino mixing parameters
that generate atmospheric neutrino oscillations
have been
confirmed at the end of 2002 by the first results of the
K2K long-baseline accelerator experiment \cite{Ahn:2002up},
which observed a disappearance of muon neutrinos
due to oscillations.

In 2001 the combined results of the SNO \cite{Ahmad:2001an}
and Super-Kamiokande \cite{Fukuda:2001nj}
experiments gave a model-independent indication of the oscillations of solar electron neutrinos,
which were discovered in the late 60's
by the Homestake experiment \cite{Cleveland:1998nv}.
In 2002 the SNO experiment \cite{Ahmad:2002jz} measured the total flux
of active neutrinos from the sun,
providing a model-independent evidence of oscillations of electron neutrinos
into other flavors,
which has been confirmed with higher precision by recent data
\cite{nucl-ex/0309004}.
The values of the neutrino mixing parameters
indicated by solar neutrino data
have been
confirmed at the end of 2002 by the KamLAND
very-long-baseline reactor experiment \cite{hep-ex/0212021},
which have measured a disappearance of electron antineutrinos
due to oscillations.

In this paper we review the currently favored scenario of three-neutrino mixing,
which is based on the above mentioned evidences of neutrino oscillations.
In Section~\ref{Neutrino masses and mixing}
we review the theory of neutrino masses and mixing,
showing that it is likely that massive neutrinos are Majorana particles.
In Section~\ref{Theory of neutrino oscillations}
we review the theory of neutrino oscillations
in vacuum and in matter.
In Section~\ref{Neutrino oscillation experiments}
we review the main results of neutrino oscillation experiments.
In Section~\ref{Phenomenology of three-neutrino mixing}
we discuss
the main aspects of the phenomenology of three-neutrino mixing,
including neutrino oscillations,
experiments on the measurement of the absolute scale of neutrino masses
and
neutrinoless double-$\beta$ decay experiments searching for an
evidence of the Majorana nature of massive neutrinos.
In Section~\ref{Future prospects} we discuss some
future prospects
and in Section~\ref{Conclusions}
we draw our conclusions.

For further information on the physics
of massive neutrinos,
see the books in Refs.~\cite{Mohapatra:1998rq,CWKim-book,Bahcall:1989ks},
the reviews in Refs.~\cite{Bilenky:1978nj,Bilenky:1987ty,Mikheev:1987qk,Kuo:1989qe,Castellani:1997cm,BGG-review-98,Altarelli:1999gu,Kajita:2000mr,Jung:2001dh,Altmann:2001eu,Bilenkii:2001yh,Beuthe:2001rc,Bilenkii:2001yh,Gaisser:2002jj,Giacomelli-0201032,hep-ph/0202058,Elliott:2002xe,Dolgov:2002wy,Kayser:2002qs,Miramonti-Reseghetti-2002,Bilenky:2002aw,hep-ph/0307149,hep-ph/0301276,Altarelli:2003vk}
and the references in
Ref.~\cite{Neutrino-Unbound}.

\section{Neutrino masses and mixing}
\label{Neutrino masses and mixing}

The Standard Model was formulated in the 60's
\cite{Glashow-SM-61,Weinberg-SM-67,Salam-SM-68}
on the basis of the knowledge available at that time on the
existing elementary particles and their properties.
In particular,
neutrinos were though to be massless following the so-called two-component theory of
Landau \cite{Landau-57},
Lee and Yang \cite{Lee-Yang-57},
and
Salam \cite{Salam-57},
in which the massless neutrinos are described by left-handed Weyl spinors.
This description has been reproduced in the Standard Model
of Glashow \cite{Glashow-SM-61},
Weinberg \cite{Weinberg-SM-67}
and
Salam \cite{Salam-SM-68}
assuming the non existence of right-handed neutrino fields,
which are necessary in order to generate Dirac neutrino masses
with the same Higgs mechanism that generates
the Dirac masses of quarks and charged leptons.
However,
as will be discussed in Section~\ref{Neutrino oscillation experiments},
in recent years neutrino experiments have shown
convincing evidences of the existence of neutrino oscillations,
which is a consequence of neutrino masses and mixing.
Hence,
it is time to revise the Standard Model in order to take into account neutrino masses
(notice that the Standard Model has already been revised in the early 70's
with the inclusion first of the charmed quark
and after of the third generation).

\subsection{Dirac mass term}
\label{Dirac mass term}

Considering for simplicity only one neutrino field $\nu$,
the standard Higgs mechanism generates the Dirac mass term
\begin{equation}
\mathcal{L}^{\text{D}}
=
- m_{\text{D}} \, \overline\nu \, \nu
=
- m_{\text{D}} \left( \overline{\nu_R} \, \nu_L + \overline{\nu_L} \, \nu_R \right)
\,,
\label{001}
\end{equation}
with
$ m_{\text{D}} = y \, v / \sqrt{2}$
where $y$ is a dimensionless Yukawa coupling coefficient and $v / \sqrt{2}$ is the
Vacuum Expectation Value of the Higgs field.
$\nu_L$ and $\nu_R$
are, respectively,
the chiral left-handed and right-handed components of the neutrino field,
obtained by acting on $\nu$ with the corresponding projection operator:
\begin{equation}
\nu = \nu_L + \nu_R
\,,
\quad
\nu_L = P_L \, \nu
\,,
\quad
\nu_R = P_R \, \nu
\,,
\quad
P_L
\equiv
\frac{1 - \gamma^5}{2}
\,,
\quad
P_R
\equiv
\frac{1 + \gamma^5}{2}
\,,
\label{002}
\end{equation}
such that
$ P_L P_R = P_R P_L = 0 $,
$ P_L^2 = P_L $,
$ P_R^2 = P_R $,
since $ (\gamma^5)^2 = 1 $.
Therefore, we have
\begin{equation}
P_L \, \nu_L = \nu_L
\,,
\quad
P_L \, \nu_R = 0
\,,
\quad
P_R \, \nu_L = 0
\,,
\quad
P_R \, \nu_R = \nu_R
\,.
\label{003}
\end{equation}
It can be shown that the chiral spinors $\nu_L$ and $\nu_R$
have only two independent components each,
leading to the correct number of four
for the independent components of the spinor $\nu$.

Unfortunately,
the generation of Dirac neutrino masses
through the standard Higgs mechanism is not able to explain naturally
why the neutrino are more than five order of magnitude lighter than the
electron, which is the lightest of the other elementary particles
(as discussed in Section~\ref{Neutrino oscillation experiments},
the neutrino masses are experimentally constrained below about 1-2 eV).
In other words,
there is no explanation of why the neutrino Yukawa coupling coefficients
are more than five order of magnitude smaller
than the Yukawa coupling coefficients
of quarks and charged leptons.

\subsection{Majorana mass term}
\label{Majorana mass term}

In 1937 Majorana \cite{Majorana:1937vz} discovered that
a massive neutral fermion as a neutrino can be described by a spinor $\psi$
with only two independent components
imposing the so-called Majorana condition
\begin{equation}
\psi = \psi^c
\,,
\label{004}
\end{equation}
where
$ \psi^c = \mathcal{C} \overline{\psi}^{T} = \mathcal{C} {\gamma^0}^T \psi^* $
is the operation of charge conjugation,
with the charge-conjugation matrix
$\mathcal{C}$
defined by the relations
$
\mathcal{C}
{\gamma^{\mu}}^{T}
\mathcal{C}^{-1}
=
- \gamma^{\mu}
$,
$
\mathcal{C}^{\dagger} = \mathcal{C}^{-1}
$,
$
\mathcal{C}^{T} = - \mathcal{C}
$.
Since
$
\mathcal{C}
{\gamma^5}^{T}
\mathcal{C}^{-1}
=
\gamma^5
$
and
$ \gamma^5 \gamma^\mu + \gamma^\mu \gamma^5 = 0 $,
we have
\begin{equation}
P_L \, \psi_L^c = 0
\,,
\quad
P_L \, \psi_R^c = \psi_R^c
\,,
\quad
P_R \, \psi_L^c = \psi_L^c
\,,
\quad
P_R \, \psi_R^c = 0
\,.
\label{005}
\end{equation}
In other words,
$\psi_L^c$ is right-handed
and
$\psi_R^c$ is left-handed.

Decomposing the Majorana condition (\ref{004})
into left-handed and right-handed components,
$ \psi_L + \psi_R = \psi_L^c + \psi_R^c $,
and
acting on both members of the equation
with the right-handed projector operator $P_R$,
we obtain
\begin{equation}
\psi_R = \psi_L^c
\,.
\label{006}
\end{equation}
Thus,
the right-handed component $\psi_R$
of the Majorana neutrino field $\psi$ is not independent,
but obtained from the left-handed component $\psi_L$
through charge conjugation
and the Majorana field can be written as
\begin{equation}
\psi = \psi_L + \psi_L^c
\,.
\label{007}
\end{equation}
This field depends only on the two independent components of $\psi_L$.
Using the constraint (\ref{006}) in the mass term (\ref{001}),
we obtain the Majorana mass term
\begin{equation}
\mathcal{L}^{\text{M}}
=
- \frac{1}{2} \, m_{\text{M}}
\left( \overline{\psi_L^c} \, \psi_L + \overline{\psi_L} \, \psi_L^c\right)
\,,
\label{008}
\end{equation}
where we have inserted a factor $1/2$ in order to avoid double counting
in the Euler-Lagrange derivation of the equation for the Majorana neutrino field.

\subsection{Dirac-Majorana mass term}
\label{Dirac-Majorana mass term}

In general, if both the chiral left-handed and right-handed fields exist and are independent,
in addition to the Dirac mass term (\ref{001})
also the Majorana mass terms for $\nu_L$ and $\nu_R$ are allowed:
\begin{equation}
\mathcal{L}^{\text{M}}_L
=
- \frac{1}{2} \, m_L
\left( \overline{\nu_L^c} \, \nu_L + \overline{\nu_L} \, \nu_L^c \right)
\,,
\qquad
\mathcal{L}^{\text{M}}_R
=
- \frac{1}{2} \, m_R
\left( \overline{\nu_R^c} \, \nu_R + \overline{\nu_R} \, \nu_R^c \right)
\,.
\label{009}
\end{equation}
The total Dirac\-+\-Majorana mass term
\begin{equation}
\mathcal{L}^{\text{D+M}}
=
\mathcal{L}^{\text{D}} + \mathcal{L}^{\text{M}}_L + \mathcal{L}^{\text{M}}_R
\label{0091}
\end{equation}
can be written as
\begin{equation}
\mathcal{L}^{\text{D+M}}
=
- \frac{1}{2}
\begin{pmatrix}
\overline{\nu_L^c} & \overline{\nu_R}
\end{pmatrix}
\begin{pmatrix}
m_L & m_{\text{D}}
\\
m_{\text{D}} & m_R
\end{pmatrix}
\begin{pmatrix}
\nu_L
\\
\nu_R^c
\end{pmatrix}
+
\text{H.c.}
\,.
\label{010}
\end{equation}
It is clear that the chiral fields $\nu_L$ and $\nu_R$
do not have a definite mass, since they are coupled by the Dirac mass term.
In order to find the fields with definite masses it is necessary to diagonalize
the mass matrix in Eq.~(\ref{010}).
For this task,
it is convenient to write
the Dirac\-+\-Majorana mass term in the matrix form
\begin{equation}
\mathcal{L}^{\text{D+M}}
=
\frac{1}{2} \,
\overline{N_L^c} \, M \, N_L
+
\text{H.c.}
\,,
\label{011}
\end{equation}
with the matrices
\begin{equation}
M
=
\begin{pmatrix}
m_L & m_{\text{D}}
\\
m_{\text{D}} & m_R
\end{pmatrix}
\,,
\qquad
N_L
=
\begin{pmatrix}
\nu_L
\\
\nu_R^c
\end{pmatrix}
\,.
\label{012}
\end{equation}
The column matrix $N_L$ is left-handed,
because it contains left-handed fields.
Let us write it as
\begin{equation}
N_L = U  \, n_L
\,,
\qquad
\text{with}
\qquad
n_L
=
\begin{pmatrix}
\nu_{1L}
\\
\nu_{2L}
\end{pmatrix}
\,,
\label{013}
\end{equation}
where $U$ is the unitary mixing matrix
($ U^\dagger = U^{-1} $)
and $n_L$ is the column matrix of the left-handed components of the massive neutrino fields.
The Dirac\-+\-Majorana mass term is diagonalized requiring that
\begin{equation}
U^T \, M \, U =
\begin{pmatrix}
m_1 & 0
\\
0 & m_2
\end{pmatrix}
\,,
\label{014}
\end{equation}
with $m_k$ real and positive for $k=1,2$.

Let us consider the simplest case of a real mass matrix $M$.
Since the values of $m_L$ and $m_R$ can be chosen real and positive
by an appropriate choice of phase of the chiral fields $\nu_L$ and $\nu_R$,
the  mass matrix $M$ is real if $m_{\text{D}}$ is real.
In this case,
the mixing matrix $U$ can be written as
\begin{equation}
U = \mathcal{O} \, \rho
\,,
\label{015}
\end{equation}
where $\mathcal{O}$ is an orthogonal matrix
and $\rho$ is a diagonal matrix of phases:
\begin{equation}
\mathcal{O}
=
\begin{pmatrix}
\cos\vartheta & \sin\vartheta
\\
-\sin\vartheta & \cos\vartheta
\end{pmatrix}
\,,
\qquad
\rho
=
\begin{pmatrix}
\rho_1 & 0
\\
0 & \rho_2
\end{pmatrix}
\,,
\label{016}
\end{equation}
with
$|\rho_k|^2 = 1$.
The orthogonal matrix $\mathcal{O}$ is chosen in order to have
\begin{equation}
\mathcal{O}^T \, M \, \mathcal{O} =
\begin{pmatrix}
m'_1 & 0
\\
0 & m'_2
\end{pmatrix}
\,,
\label{017}
\end{equation}
leading to
\begin{equation}
\tan 2\vartheta
=
\frac{2m_{\text{D}}}{m_R-m_L}
\,,
\qquad
m'_{2,1}
=
\frac{1}{2}
\left[
m_L + m_R
\pm
\sqrt{ \left( m_L - m_R \right)^2
+ 4 \, m_{\text{D}}^2 }
\right]
\,.
\label{018}
\end{equation}
Having chosen
$m_L$ and $m_R$ positive,
$m'_2$ is always positive,
but
$m'_1$ is negative if $ m_{\text{D}}^2 > m_L m_R $.
Since
\begin{equation}
U^T \, M \, U
=
\rho^T \, \mathcal{O}^T \, M \, \mathcal{O} \, \rho
=
\begin{pmatrix}
\rho_1^2 \, m'_1 & 0
\\
0 & \rho_2^2 \, m'_2
\end{pmatrix}
\,,
\label{019}
\end{equation}
it is clear that the role of the phases $\rho_k$ is to
make the masses $m_k$ positive, as masses must be.
Hence, we have $ \rho_2^2 = 1 $ always,
and
$ \rho_1^2 = 1 $ if $m'_1 \geq 0$
or
$ \rho_1^2 = -1 $ if $m'_1 < 0$.

An important fact to be noticed is that
the diagonalized Dirac\-+\-Majorana mass term,
\begin{equation}
\mathcal{L}^{\text{D+M}}
=
\frac{1}{2}
\sum_{k=1,2} m_k \, \overline{\nu_{kL}^c} \, \nu_{kL}
+
\text{H.c.}
\,,
\label{022}
\end{equation}
is a sum of Majorana mass terms for the massive Majorana neutrino fields
\begin{equation}
\nu_k = \nu_{kL} + \nu_{kL}^c
\qquad
(k=1,2)
\,.
\label{0221}
\end{equation}
Therefore,
the two massive neutrinos are Majorana particles.

\subsection{The see-saw mechanism}
\label{The see-saw mechanism}

It is possible to show that the Dirac\-+\-Majorana mass term
leads to maximal mixing ($\theta=\pi/4$) if
$m_L=m_R$,
or to so-called pseudo-Dirac neutrinos
if $m_L$ and $m_R$ are much smaller that $|m_{\text{D}}|$
(see Ref.~\cite{BGG-review-98}).
However, the most plausible and interesting case
is the so-called see-saw mechanism
\cite{Yanagida-SeeSaw-1979,GellMann-Ramond-Slansky-SeeSaw-1979,Mohapatra:1980ia},
which is obtained considering
$m_L=0$ and $|m_{\text{D}}| \ll m_R$.
In this case
\begin{equation}
m_1 \simeq \frac{(m_{\text{D}})^2}{m_R} \ll |m_{\text{D}}|
\,,
\qquad
m_2 \simeq m_R
\,,
\qquad
\tan\vartheta \simeq \frac{m_{\text{D}}}{m_R} \ll 1
\,,
\qquad
\rho_1^2 = -1
\,.
\label{020}
\end{equation}
What is interesting in Eq.~(\ref{020})
is that $m_1$ is much smaller than $m_{\text{D}}$,
being suppressed by the small ratio
$m_{\text{D}}/m_R$.
Since $m_2$ is of order $m_R$,
a very heavy $\nu_2$ corresponds to a very light $\nu_1$,
as in a see-saw.
Since $m_{\text{D}}$
is a Dirac mass, presumably generated with the standard Higgs mechanism,
its value is expected to be of the same order as the mass of a quark or the
charged fermion in the same generation of the neutrino we are considering.
Hence,
the see-saw explains naturally the suppression of $m_1$ with respect to
$m_{\text{D}}$,
providing the most plausible explanation of the smallness of neutrino masses.

The smallness of the mixing angle $\vartheta$ in Eq.~(\ref{020})
implies that
$\nu_{1L} \simeq - \nu_L$
and
$\nu_{2L} \simeq \nu_R^c$.
This means that the neutrino participating to weak interactions
practically coincides with the
light neutrino $\nu_1$,
whereas the heavy neutrino $\nu_2$
is practically decoupled from interactions with matter.

Besides the smallness of the light neutrino mass,
another important consequence of the see-saw mechanism
is that massive neutrinos are Majorana particles,
as we have shown above
in the general case of a Dirac\-+\-Majorana mass term.
This is a very important indication
that strongly encourages the search for the Majorana
nature of neutrinos,
mainly performed through the search of
neutrinoless double-$\beta$ decay.

An important assumption necessary for the see-saw mechanism
is $m_L=0$.
Such assumption may seem arbitrary at first sight,
but in fact it is not.
Its plausibility follows from the fact that
$\nu_L$ belongs to a weak isodoublet of the Standard Model:
\begin{equation}
L_L
=
\begin{pmatrix}
\nu_L
\\
\ell_L
\end{pmatrix}
\,.
\label{023}
\end{equation}
Since $\nu_L$ has third component of the weak isospin
$I_3=1/2$,
the combination
$\overline{\nu_L^c}\nu_L=-\nu_L^T\mathcal{C}^{\dagger}\nu_L$
in the Majorana mass term in Eq.~(\ref{009})
has $I_3=1$
and belongs to a triplet.
Since in the Standard Model there is no Higgs triplet
that could couple to
$\overline{\nu_L^c}\nu_L$
in order to form a Lagrangian term invariant under a SU(2)$_L$
transformation of the Standard Model gauge group,
a Majorana mass term for $\nu_L$
is forbidden.
In other words,
the gauge symmetries of the Standard Model imply $m_L=0$,
as needed for the see-saw mechanism.
On the other hand,
$m_{\text{D}}$
is allowed in the Standard Model,
because it is generated through the standard Higgs mechanism,
and
$m_R$ is also allowed,
because $\nu_R$
and
$\overline{\nu_R^c}\nu_R$
are singlets of the Standard Model gauge symmetries.
Hence,
quite unexpectedly,
we have an extended Standard Model with massive neutrinos
that are Majorana particles
and in which the smallness of neutrino masses
can be naturally explained through the see-saw mechanism.

The only assumption which remains unexplained in this scenario
is the heaviness of $m_R$ with respect to $m_{\text{D}}$.
This assumption cannot be motivated in the framework of
the Standard Model, because
$m_R$ is only a parameter which could have any value.
However, there are rather strong arguments
that lead us to believe that the Standard Model
is a theory that describes the world only
at low energies.
In this case it is natural to expect that
the mass $m_R$ is generated at ultra-high energy
by the symmetry breaking
of the theory beyond the Standard Model.
Hence, it is plausible that the value of $m_R$
is many orders of magnitude larger than the
scale of the electroweak symmetry breaking
and of $m_{\text{D}}$, as required for the
working of the see-saw mechanism.

\subsection{Effective dimension-five operator}
\label{Effective dimension-five operator}

If we consider the possibility of a theory beyond
the Standard Model, another question
regarding the neutrino masses arises:
is it possible that a Lagrangian term exists at
the high energy of the theory beyond
the Standard Model which generates at low energy
an effective Majorana mass term for $\nu_L$?
The answer is yes
\cite{Weinberg:1979sa,Weinberg:1980bf,Weldon:1980gi}:
the operator with lowest dimension
invariant\footnote{
Since the high-energy theory
reduces to the Standard Model at low energies,
its gauge symmetries
must include the gauge symmetries of the Standard Model.
}
under
$ \text{SU}(2)_L \times \text{U}(1)_Y $
that can generate
a Majorana mass term for $\nu_L$
after electroweak symmetry breaking is the dimension-five operator\footnote{
In units where $\hbar = c = 1$
scalar fields have dimension of energy,
fermion fields have dimension of $(\text{energy})^{3/2}$
and all Lagrangian terms have dimension $(\text{energy})^{4}$.
The ``dimension-five'' character of the operator in Eq.~(\ref{024})
refers to the power of energy of the dimension of the operator
$
( L_L^T \, \sigma_2 \, \Phi )
\, \mathcal{C}^{-1} \,
( \Phi^T \, \sigma_2 \, L_L )
$,
which is divided by the mass $\mathcal{M}$
in order to obtain a Lagrangian term with correct dimension.
}
\begin{equation}
\frac{ g }{ \mathcal{M} }
( L_L^T \, \sigma_2 \, \Phi )
\, \mathcal{C}^{-1} \,
( \Phi^T \, \sigma_2 \, L_L )
+
\text{H.c.}
\,,
\label{024}
\end{equation}
where $g$ is a dimensionless coupling coefficient
and $\mathcal{M}$
is the high-energy scale at which the new theory
breaks down to the Standard Model.
The dimension-five operator in Eq.~(\ref{024})
does not belong to the Standard Model
because it is not renormalizable.
It must be considered as an effective operator
which is the low-energy manifestation of the renormalizable
new theory beyond the Standard Model,
in analogy with the old non-renormalizable Fermi theory of weak interactions,
which is the low-energy manifestation of the Standard Model.

At the electroweak symmetry breaking
\begin{equation}
\Phi
=
\begin{pmatrix}
\phi^+
\\
\phi^0
\end{pmatrix}
\ \xrightarrow{\text{Symmetry Breaking}} \
\begin{pmatrix}
0
\\
v / \sqrt{2}
\end{pmatrix}
\,,
\label{025}
\end{equation}
the operator in Eq.~(\ref{024})
generates the Majorana mass term for $\nu_L$ in Eq.~(\ref{009}),
with
\begin{equation}
m_L = \frac{ g \, v^2 }{ \mathcal{M} }
\,.
\label{026}
\end{equation}
This relation is very important,
because it shows that the Majorana mass $m_L$
is suppressed with respect to $v$ by the small ratio
$ v / \mathcal{M} $.
In other words,
since the Dirac mass term $m_{\text{D}}$
is equal to $v/\sqrt{2}$ times a Yukawa coupling coefficient,
the relation (\ref{026}) has a see-saw form.
Therefore,
the effect of the dimension-five effective operator in Eq.~(\ref{024})
does not spoil the natural suppression of the light neutrino mass
provided by the see-saw mechanism.
Indeed, considering $ m_L \sim m_{\text{D}}^2 / m_R $
and
taking into account that $m_1=|m'_1|$,
from Eq.~(\ref{018}) we obtain
\begin{equation}
m_1 \simeq \left| m_L - \frac{(m_{\text{D}})^2}{m_R} \right|
\,.
\label{027}
\end{equation}
Equations~(\ref{026}) and (\ref{027})
show that the see-saw mechanism is operating even if $m_L$ is not zero,
but it is generated by the dimension-five operator in Eq.~(\ref{024}).
On the other hand,
if the chiral right-handed neutrino field $\nu_R$
does not exist,
the standard see-saw mechanism cannot be implemented,
but a Majorana neutrino mass $m_L$ can be generated by
the dimension-five operator in Eq.~(\ref{024}),
and
Eq.~(\ref{026})
shows that the suppression of the light neutrino mass
is natural and of see-saw type.

\subsection{Three-neutrino mixing}
\label{Three-neutrino mixing}

So far we have considered for simplicity only one neutrino,
but it is well known from a large variety of experimental data
that there are three neutrinos that participate to weak interactions:
$\nu_e$,
$\nu_\mu$,
$\nu_\tau$.
These neutrinos are called ``active flavor neutrinos''.
From the precise measurement of the invisible width
of the $Z$-boson
produced by the decays
$ \displaystyle Z \to \sum_{\alpha} \nu_\alpha \bar\nu_\alpha $
we also know that the number of active flavor neutrinos
is exactly three
(see Ref.~\cite{PDG}), excluding the possibility of existence of
additional heavy active flavor neutrinos\footnote{
More precisely,
what is excluded is the existence of
additional active flavor neutrinos with mass
$ \lesssim m_Z / 2 \simeq 46 \, \text{GeV} $
\cite{Bulanov:2003ka}.
For a recent discussion of the possible existence of heavier active flavor neutrinos see
Ref.~\cite{hep-ph/0210153}.
}.
The active flavor neutrinos take part in the
charged-current (CC) and neutral current (NC)
weak interaction Lagrangians
\begin{align}
\mathcal{L}_I^{\text{CC}}
=
- \frac{g}{2\sqrt{2}}
\,
j^{\text{CC}}_{\rho} \, W^{\rho}
+
\text{H.c.}
\,,
\null & \null
\quad
\text{with}
\quad
j^{\text{CC}}_{\rho}
=
2
\sum_{\alpha=e,\mu,\tau}
\overline{\nu_{\alpha L}} \, \gamma_{\rho} \, \alpha_L
\,,
\label{0271}
\\
\mathcal{L}_I^{\text{NC}}
=
- \frac{g}{2\cos{\vartheta_{\text{W}}}}
\,
j^{\text{NC}}_{\rho} \, Z^{\rho}
\,,
\null & \null
\quad
\text{with}
\quad
j^{\text{NC}}_{\rho}
=
\sum_{\alpha=e,\mu,\tau}
\overline{\nu_{\alpha L}} \, \gamma_{\rho} \, \nu_{\alpha L}
\,,
\label{0272}
\end{align}
where
$j^{\text{CC}}_{\rho}$ and $j^{\text{NC}}_{\rho}$
are, respectively,
the charged and neutral leptonic currents,
$\vartheta_{\text{W}}$ is the weak mixing angle
($\sin^2\vartheta_{\text{W}} \simeq 0.23$)
and
$g=e/\sin\vartheta_{\text{W}}$
($e$ is the positron electric charge).
 
Let us consider
three left-handed chiral fields
$\nu_{e L}$,
$\nu_{\mu L}$,
$\nu_{\tau L}$
that describe the three active flavor neutrinos
and three corresponding right-handed chiral fields
$\nu_{s_1 R}$,
$\nu_{s_2 R}$,
$\nu_{s_3 R}$
that describe three sterile neutrinos\footnote{
Let us remark,
however,
that the number of sterile neutrinos is not constrained by experimental data,
because they cannot be detected,
and could well be different from three.
},
which do not take part in weak interactions.
The corresponding Dirac\-+\-Majorana mass term
is given by Eq.~(\ref{0091}) with
\begin{align}
\null & \null
\mathcal{L}^{\text{D}}
=
-
\sum_{s,\beta}
\overline{\nu_{s R}}
\,
M^{\text{D}}_{s\beta}
\,
\nu_{\beta L}
+
\text{H.c.}
\,,
\label{028}
\\
\null & \null
\mathcal{L}^{\text{M}}_L
=
- \frac{1}{2}
\sum_{\alpha,\beta}
\overline{\nu_{{\alpha}L}^c}
\,
M^{L}_{\alpha\beta}
\,
\nu_{{\beta}L}
+
\text{H.c.}
\,,
\label{029}
\\
\null & \null
\mathcal{L}^{\text{M}}_R
=
- \frac{1}{2}
\sum_{s,s'}
\overline{\nu_{s R}^c}
\,
M^{R}_{\alpha\beta}
\,
\nu_{s' R}
+
\text{H.c.}
\,,
\label{030}
\end{align}
where
$M^{\text{D}}$
is a complex matrix
and
$M^L$, $M^R$
are symmetric complex matrices.
The Dirac\-+\-Majorana mass term can be written as the one in Eq.~(\ref{011})
with the column matrix of left-handed fields
\begin{equation}
N_L
=
\begin{pmatrix}
\nu_L
\\ \displaystyle
\nu_R^c
\end{pmatrix}
\,,
\qquad
\text{with}
\qquad
\nu_L
=
\begin{pmatrix}
\nu_{eL}
\\ \displaystyle
\nu_{{\mu}L}
\\ \displaystyle
\nu_{{\tau}L}
\end{pmatrix}
\qquad
\text{and}
\qquad
\nu_R^c
=
\begin{pmatrix}
\nu_{s_1 R}^c
\\ \displaystyle
\nu_{s_2 R}^c
\\ \displaystyle
\nu_{s_3 R}^c
\end{pmatrix}
\,,
\label{031}
\end{equation}
and the $6\times6$ mass matrix
\begin{equation}
M
=
\begin{pmatrix}
M^L & (M^{\text{D}})^T
\\ \displaystyle
M^{\text{D}} & M^R
\end{pmatrix}
\,.
\label{032}
\end{equation}
The mass matrix is diagonalized by a unitary transformation
analogous to the one in Eq.~(\ref{013}):
\begin{equation}
N_L = \mathbb{V}  \, n_L
\,,
\qquad
\text{with}
\qquad
n_L
=
\begin{pmatrix}
\nu_{1L}
\\
\vdots
\\
\nu_{6L}
\end{pmatrix}
\,,
\label{033}
\end{equation}
where $\mathbb{V}$ is the unitary $6\times6$ mixing matrix
and $n_{kL}$ are the left-handed components of the massive neutrino fields.
The mixing matrix $\mathbb{V}$ is determined by the diagonalization relation
\begin{equation}
\mathbb{V}^T \, M \, \mathbb{V}
=
\text{diag}\!\left(m_1,\ldots,m_6\right)
\,,
\label{034}
\end{equation}
with $m_k$ real and positive for $k=1,\ldots,6$
(see Ref.~\cite{Bilenky:1987ty} for a proof that it can be done).
After diagonalization the Dirac\-+\-Majorana mass term
is written as
\begin{equation}
\mathcal{L}^{\text{D+M}}
=
- \frac{1}{2}
\sum_{k=1}^{6}
m_k
\,
\overline{\nu_{kL}^c}
\,
\nu_{kL}
+
\text{H.c.}
\,,
\label{035}
\end{equation}
which is a sum of Majorana mass terms for the massive Majorana neutrino fields
\begin{equation}
\nu_k = \nu_{kL} + \nu_{kL}^c
\qquad
(k=1,\ldots,6)
\,.
\label{036}
\end{equation}
Hence,
as we have already seen in Section~\ref{Dirac-Majorana mass term}
in the case of one neutrino generation,
a Dirac\-+\-Majorana mass term
implies that massive neutrinos are Majorana particles.
The mixing relation (\ref{033})
can be written as
\begin{equation}
\nu_{{\alpha}L}
=
\sum_{k=1}^{6}
\mathbb{V}_{{\alpha}k} \, \nu_{kL}
\qquad
(\alpha=e,\mu,\tau)
\,,
\qquad
\nu_{sR}^c
=
\sum_{k=1}^{6}
\mathbb{V}_{sk} \, \nu_{kL}
\qquad
(s=s_1,s_2,s_3)
\,,
\label{037}
\end{equation}
which shows that active and sterile neutrinos
are linear combinations of the same massive neutrino fields.
This means that in general active-sterile oscillations are possible
(see Section~\ref{Theory of neutrino oscillations}).

The most interesting possibility offered by the
Dirac\-+\-Majorana mass term
is the implementation of the see-saw mechanism
for the explanation of the smallness of the light neutrino masses,
which is however considerably more complicated than
in the case of one generation discussed in
Section~\ref{The see-saw mechanism}.
Let us assume that
$M^L=0$,
in compliance with the gauge symmetries of the Standard Model
and the absence of a Higgs triplet\footnote{
For the sake of simplicity
we do not consider here the possible existence of
effective dimension-five operators of the type discussed in
Section~\ref{Effective dimension-five operator},
which in any case do not spoil the effectiveness
see-saw mechanism.
}.
Let us further assume that 
the eigenvalues of
$M^R$
are much larger than those of
$M^{\text{D}}$,
as expected if the Majorana mass term (\ref{030})
for the sterile neutrinos is generated at a very high energy
scale characteristic of the theory beyond the Standard Model.
In this case,
we can write the mixing matrix $\mathbb{V}$ as
\begin{equation}
\mathbb{V} = \mathbb{W} \, \mathbb{U}
\,,
\label{038}
\end{equation}
where both $\mathbb{W}$ and $\mathbb{U}$ are unitary matrices,
and use $\mathbb{W}$ for an approximate block-diagonalization of the mass matrix $M$
at leading order in the expansion in powers of
$(M^R)^{-1}M^{\text{D}}$:
\begin{equation}
\mathbb{W}^T
\,
M
\,
\mathbb{W}
\simeq
\begin{pmatrix}
M_{\text{light}} & 0
\\
0 & M_{\text{heavy}}
\end{pmatrix}
\,.
\label{039}
\end{equation}
The matrix $\mathbb{W}$ is given by
\begin{equation}
\mathbb{W}
=
1
-
\frac{1}{2}
\begin{pmatrix}
(M^{\text{D}})^\dagger (M^R {(M^R)}^\dagger)^{-1} M^{\text{D}}
&
2 (M^{\text{D}})^\dagger {(M^R)^\dagger}^{-1}
\\
- 2 (M^R)^{-1} M^{\text{D}}
&
(M^R)^{-1} M^{\text{D}} (M^{\text{D}})^\dagger {(M^R)^\dagger}^{-1} 
\end{pmatrix}
\,,
\label{040}
\end{equation}
and is unitary up to corrections of order
$(M^R)^{-1}M^{\text{D}}$.
The two $3\times3$ mass matrices
$M_{\text{light}}$
and
$M_{\text{heavy}}$
are given by
\begin{equation}
M_{\text{light}}
\simeq
-(M^{\text{D}})^T \, (M^R)^{-1} \, M^{\text{D}}
\,,
\qquad
M_{\text{heavy}}
\simeq
M^R
\,.
\label{041}
\end{equation}
Therefore,
the see-saw mechanism is implemented by the suppression of
$M_{\text{light}}$
with respect to $M^{\text{D}}$
by the small ratio
$(M^{\text{D}})^T \, (M^R)^{-1}$.
The light and heavy mass sectors are
practically decoupled because of the smallness of the
off-diagonal $3\times3$ block elements in Eq.~(\ref{040}).

For the low-energy phenomenology
it is sufficient to consider only the light $3\times3$
mass matrix $M_{\text{light}}$
which is diagonalized by the $3\times3$ upper-left submatrix
of $\mathbb{U}$ that we call $U$,
such that
\begin{equation}
U^T \, M_{\text{light}} \, U
=
\text{diag}\!\left(m_1,m_2,m_3\right)
\,,
\label{043}
\end{equation}
where $m_1,m_2,m_3$ are the three light neutrino mass eigenvalues.
Neglecting the small mixing with the heavy sector,
the effective mixing of the active flavor neutrinos
relevant for the low-energy phenomenology is given by
\begin{equation}
\nu_{{\alpha}L}
=
\sum_{k=1}^{3}
U_{{\alpha}k} \, \nu_{kL}
\qquad
(\alpha=e,\mu,\tau)
\,,
\label{044}
\end{equation}
where
$\nu_{1L}$,
$\nu_{2L}$,
$\nu_{3L}$
are the left-handed components of
the three light massive Majorana neutrino fields.
This scenario, called ``three-neutrino mixing'',
can accommodate the experimental evidences of neutrino oscillations
in solar and atmospheric neutrino experiments reviewed in
Section~\ref{Neutrino oscillation experiments}.
The phenomenology of three-neutrino mixing
is discussed in Section~\ref{Phenomenology of three-neutrino mixing}.

The $3\times3$ unitary mixing matrix $U$
can be parameterized in terms of $3^2=9$ parameters
which can be divided in $3$ mixing angles and
$6$ phases.
However,
only $3$ phases are physical.
This can be seen by considering
the charged-current Lagrangian (\ref{0271})\footnote{
Unitary mixing has no effect on the neutral-current
weak interaction Lagrangian,
which is diagonal in the massive neutrino fields,
$ \displaystyle
\mathcal{L}_I^{\text{NC}}
=
- \frac{g}{2\cos{\vartheta_{\text{W}}}}
\sum_{k=1}^{3}
\overline{\nu_{kL}} \, \gamma^{\rho} \, \nu_{kL} \, Z_{\rho}
$
(GIM mechanism).
},
which can be written as
\begin{equation}
\mathcal{L}_I^{\text{CC}}
=
- \frac{g}{\sqrt{2}}
\sum_{\alpha=e,\mu,\tau}
\sum_{k=1}^{3}
\overline{\alpha_L} \, \gamma^{\rho} \, U_{\alpha k} \, \nu_{kL} \, W_{\rho}^{\dagger}
+
\text{H.c.}
\,,
\label{045}
\end{equation}
in terms of the light massive neutrino fields $\nu_k$ ($k=1,2,3$).
Three of the six phases in $U$ can be eliminated by rephasing the
charged lepton fields
$e$, $\mu$, $\tau$,
whose phases are arbitrary
because all other terms in the Lagrangian are invariant under such change of phases
(see
Refs.~\cite{Murnaghan-book-62,Schechter-Valle-COMMENT-80,Schechter-Valle-MASSES-80}
and the appendices of Refs.~\cite{GKM-atm-98,Giunti:2002pp}).
On the other hand,
the phases of the Majorana massive neutrino fields cannot be changed,
because the Majorana mass term in Eq.~(\ref{035})
are not invariant\footnote{
In Field Theory, Noether's theorem establishes that
invariance of the Lagrangian under a global change of phase
of the fields corresponds to the
conservation of a quantum number:
lepton number $L$ for leptons and baryon number $B$ for quarks.
The non-invariance of the Majorana mass term in Eq.~(\ref{035})
under rephasing of $\nu_{kL}$
implies the violation of lepton number conservation.
Indeed,
a Majorana mass term induces $|\Delta{L}|=2$ processes
as neutrinoless double-$\beta$ decay
(see Refs.~\cite{Mohapatra:1998rq,CWKim-book,Bilenky:1987ty,BGG-review-98,Elliott:2002xe}).
}
under rephasing of $\nu_{kL}$.
Therefore,
the number of physical phases in the mixing matrix $U$ is three
and it can be shown that two of these phases can be factorized
in a diagonal matrix of phases on the right of $U$.
These two phases are usually called ``Majorana phases'',
because they appear only if the massive neutrinos
are Majorana particles
(if the massive neutrinos are Dirac particles these two phases
can be eliminated by rephasing the massive neutrino fields,
since a Dirac mass term is invariant under rephasing of the fields).
The third phase is usually called ``Dirac phase'',
because it is present also if the massive neutrinos are Dirac particles,
being the analogous of the phase in the quark mixing matrix.
These complex phases in the mixing matrix generate violations
of the CP symmetry
(see Refs.~\cite{Mohapatra:1998rq,CWKim-book,Bilenky:1987ty,BGG-review-98}).

The most common parameterization of the mixing matrix is
\begin{align}
U
=
\null & \null
R_{23} \, W_{13} \, R_{12} \, D(\lambda_{21},\lambda_{31})
\nonumber
\\
=
\null & \null
\begin{pmatrix}
1 & 0 & 0
\\
0 & c_{23} & s_{23}
\\
0 & -s_{23} & c_{23}
\end{pmatrix}
\begin{pmatrix}
c_{13} & 0 & s_{13} e^{-i\varphi_{13}}
\\
0 & 1 & 0
\\
- s_{13} e^{i\varphi_{13}} & 0 & c_{13}
\end{pmatrix}
\begin{pmatrix}
c_{12} & s_{12} & 0
\\
-s_{12} & c_{12} & 0
\\
0 & 0 & 1
\end{pmatrix}
\begin{pmatrix}
1 & 0 & 0
\\
0 & e^{i\lambda_{21}} & 0
\\
0 & 0 & e^{i\lambda_{31}}
\end{pmatrix}
\nonumber
\\
=
\null & \null
\begin{pmatrix}
c_{12}
c_{13}
&
s_{12}
c_{13}
&
s_{13}
e^{-i\varphi_{13}}
\\
-
s_{12}
c_{23}
-
c_{12}
s_{23}
s_{13}
e^{i\varphi_{13}}
&
c_{12}
c_{23}
-
s_{12}
s_{23}
s_{13}
e^{i\varphi_{13}}
&
s_{23}
c_{13}
\\
s_{12}
s_{23}
-
c_{12}
c_{23}
s_{13}
e^{i\varphi_{13}}
&
-
c_{12}
s_{23}
-
s_{12}
c_{23}
s_{13}
e^{i\varphi_{13}}
&
c_{23}
c_{13}
\end{pmatrix}
\begin{pmatrix}
1 & 0 & 0
\\
0 & e^{i\lambda_{21}} & 0
\\
0 & 0 & e^{i\lambda_{31}}
\end{pmatrix}
\,,
\label{046}
\end{align}
with
$c_{ij} \equiv \cos\vartheta_{ij}$,
$s_{ij} \equiv \sin\vartheta_{ij}$,
where 
$\vartheta_{12}$,
$\vartheta_{23}$,
$\vartheta_{13}$
are the three mixing angles,
$\varphi_{13}$
is the Dirac phase,
$\lambda_{21}$
and
$\lambda_{31}$
are the Majorana phases.
In Eq.~(\ref{046})
$R_{ij}$ is a real rotation in the $i$-$j$ plane,
$W_{13}$ is a complex rotation in the $1$-$3$ plane
and
$D(\lambda_{21},\lambda_{31})$
is the diagonal matrix with the Majorana phases.

Let us finally remark that,
although in the case of Majorana neutrinos
there is no difference between neutrinos and antineutrinos
and one should only distinguish between states with positive
and negative helicity,
it is a common convention to call neutrino a particles created
together with a positive charged lepton
and having almost exactly negative helicity,
and
antineutrino a particles created
together with a negative charged lepton
and having almost exactly positive helicity.
This convention follows from the fact
that Dirac neutrinos are created
together with a positive charged lepton
and almost exactly negative helicity,
and
Dirac antineutrinos are created
together with a negative charged lepton
and almost exactly positive helicity.

\section{Theory of neutrino oscillations}
\label{Theory of neutrino oscillations}

In order to derive neutrino oscillations it is useful to
realize from the beginning that detectable neutrinos,
relevant in oscillation experiments,
are always ultrarelativistic particles.
Indeed,
as discussed in Section~\ref{Neutrino oscillation experiments},
the neutrino masses are experimentally constrained below about 1-2 eV,
whereas only neutrinos more energetic than about 200 keV can be detected in:
\renewcommand{\labelenumi}{\theenumi.}
\renewcommand{\theenumi}{\arabic{enumi}}
\begin{enumerate}
\item
Charged current weak processes
which have an energy threshold
larger than some fraction of MeV.
For example\footnote{
In a scattering process
$\nu + A \to B + C$
the Lorentz-invariant Mandelstam variable
$ s = \left( p_\nu + p_A \right)^2 = \left( p_B + p_C \right)^2 $
calculated for the initial state in the laboratory frame
in which the target particle $A$ is at rest is
$s = 2 E m_A + m_A^2$.
The value of $s$ calculated for the final state in the center-of-mass frame
is given by
$ s = \left( E_B + E_C \right)^2 \geq ( m_B + m_C )^2$.
Confronting the two expressions for $s$ we obtain the neutrino
energy threshold in the laboratory frame
$ \displaystyle
E_{\text{th}}
=
\frac{ ( m_B + m_C )^2 }{ 2 m_A } - \frac{ m_A }{ 2 }
$.
}:
\begin{itemize}
\item
$
E_{\text{th}}
=
0.233 \, \text{MeV}
$
for
$ \nu_e + {}^{71}\text{Ga} \to {}^{71}\text{Ge} + e^- $
in the GALLEX \cite{Hampel:1998xg},
SAGE \cite{Abdurashitov:2002nt}
and
GNO \cite{Altmann:2000ft}
solar neutrino experiments.
\item
$
E_{\text{th}}
=
0.81 \, \text{MeV}
$
for
$ \nu_e + {}^{37}\text{Cl} \to {}^{37}\text{Ar} + e^- $
in the Homestake \cite{Cleveland:1998nv}
solar neutrino experiment.
\item
$
E_{\text{th}}
=
1.8 \, \text{MeV}
$
for
$ \bar\nu_e + p \to n + e^+ $
in reactor neutrino experiments
(for example Bugey \cite{Bugey}, CHOOZ \cite{CHOOZ-99} and KamLAND \cite{hep-ex/0212021}).
\end{itemize}
\item
The elastic scattering process
$ \nu + e^- \to \nu + e^- $,
whose cross section is proportional to the neutrino energy
($
\sigma(E)
\sim
\sigma_0 E / m_e
$,
with
$
\sigma_0
\sim
10^{-44} \, \text{cm}^2
$).
Therefore,
an energy threshold
of some MeV's
is needed in order to have a signal above the background.
For example,
$
E_{\text{th}}
\simeq
5 \, \text{MeV}
$
in the Super-Kamiokande \cite{Fukuda:2002pe,hep-ex/0309011}
solar neutrino experiment.
\end{enumerate}
The comparison
of the experimental limit on neutrino masses
with the energy threshold in the processes of neutrino detection
implies that detectable neutrinos are extremely relativistic.

\subsection{Neutrino oscillations in vacuum}
\label{Neutrino oscillations in vacuum}

Active neutrinos are created and detected with a definite flavor
in weak charged-current interactions described by the Lagrangian (\ref{0271}).
The state that describes an active neutrino with flavor $\alpha$
created together with a charged lepton $\alpha^+$
in a decay process of type\footnote{
This is the most common neutrino creation process.
Other processes can be treated with the same method,
leading to the same result (\ref{103})
for the state describing a ultrarelativistic flavor neutrino.
}
\begin{equation}
A \to B + \alpha^+ + \nu_\alpha
\label{1021}
\end{equation}
is given by\footnote{
The flavor neutrino fields $\nu_\alpha$
are not quantizable because they do not have a definite mass
and are coupled by the mass term.
Therefore,
the state $| \nu_\alpha \rangle$ is not a quantum of the
field $\nu_\alpha$.
It is an appropriate superposition of the massive states
$| \nu_k \rangle$,
quanta of the respective fields $\nu_k$,
which describes a neutrino created in the process (\ref{1021})
\cite{Giunti:1992cb}.
}
\begin{equation}
| \nu_\alpha \rangle
\propto
\sum_{k=1}^3 | \nu_k \rangle
\,
\langle \nu_k , \alpha^+ | j^{\rho}_{\text{CC}} | 0 \rangle
\,
J^{A \to B}_{\rho}
\,,
\label{101}
\end{equation}
where
$ J^{A \to B}_{\rho} $ is the current describing the $A \to B$ transition.
Neglecting the effect of neutrino masses in the production process,
which is negligible for ultrarelativistic neutrinos,
from Eqs.~(\ref{0271}) and (\ref{044}) it follows that
\begin{equation}
\langle \nu_k , \alpha^+ | j^{\rho}_{\text{CC}} | 0 \rangle
\,
J^{A \to B}_{\rho}
\propto
U_{\alpha k}^*
\,.
\label{102}
\end{equation}
Therefore,
the normalized state describing a neutrino with flavor $\alpha$ is
\begin{equation}
| \nu_\alpha \rangle
=
\sum_{k=1}^3 U_{\alpha k}^* \, | \nu_k \rangle
\,.
\label{103}
\end{equation}
This state describes the neutrino at the production point at the production time.
The state describing the neutrino at detection,
after a time $T$ at a distance $L$ of propagation in vacuum,
is obtained by acting on $| \nu_{\alpha} \rangle$
with the space-time translation operator\footnote{
We consider for simplicity only one space dimension
along neutrino propagation.
}
$\exp\left( -i \widehat{E} T + i \widehat{P} \cdot L \right)$,
where
$\widehat{E}$ and $\widehat{P}$
are the energy and momentum operators,
respectively.
The resulting state is
\begin{equation}
| \nu_\alpha (L,T) \rangle
=
\sum_{k=1}^3 U_{\alpha k}^* \, e^{ - i E_k T + i p_k L } \, | \nu_k \rangle
\,,
\label{104}
\end{equation}
where $E_k$ and $p_k$ are, respectively,
the energy and momentum\footnote{
Since the energy and momentum of the massive neutrino $\nu_k$
satisfy the relativistic dispersion relation
$ E_k^2 = p_k^2 + m_k^2 $,
elementary dimensional considerations imply that
at first order in the contribution of the mass $m_k$ we have
$ \displaystyle
E_k
\simeq
E
+
\xi
\,
\frac{ m_{k}^2 }{ 2 E }
$
and
$ \displaystyle
p_k
\simeq
E
-
\left( 1 - \xi \right)
\frac{ m_{k}^2 }{ 2 E }
$,
where $E$ is the neutrino energy in the massless limit
and $\xi$ is a dimensionless quantity that depends on the neutrino
production process.
}
of the massive neutrino $\nu_k$, which are determined by the process
in which the neutrino has been produced.
Using the expression of $ | \nu_k \rangle $
in terms of the flavor neutrino states
obtained inverting Eq.~(\ref{044}),
$ \displaystyle
| \nu_k \rangle
=
\sum_{\beta=e,\mu,\tau} U_{\beta k} \, | \nu_\beta \rangle
$,
we obtain
\begin{equation}
| \nu_\alpha (L,T) \rangle
=
\sum_{\beta=e,\mu,\tau}
\left(
\sum_{k=1}^3
U_{\alpha k}^* \, e^{ - i E_k T + i p_k L } \, U_{\beta k}
\right)
| \nu_\beta \rangle
\,,
\label{105}
\end{equation}
which shows that at detection
the state describes a superposition of different neutrino flavors.
The coefficient of
$| \nu_\beta \rangle$
is the amplitude of
$\nu_\alpha\to\nu_\beta$ transitions,
whose probability is given by
\begin{equation}
P_{\nu_\alpha\to\nu_\beta}(L,T)
=
| \langle \nu_\beta | \nu_\alpha (L,T) \rangle |^2
=
\left|
\sum_{k=1}^3
U_{\alpha k}^* \, e^{ - i E_k T + i p_k L } \, U_{\beta k}
\right|^2
\,.
\label{106}
\end{equation}
The transition probability
(\ref{106})
depends on the space and time of neutrino propagation,
but in real experiments
the propagation time is not measured.
Therefore it is necessary to connect the propagation time to the
propagation distance,
in order to obtain an expression for the transition probability
depending only on the known distance between neutrino source and detector.
This is not a problem for ultrarelativistic neutrinos
whose propagation time $T$ is equal to the distance $L$
up to negligible corrections depending on the ratio of the neutrino mass
and energy\footnote{
A rigorous derivation of the neutrino transition probability in space
that justifies the $T=L$ approximation
requires a wave packet description
(see Refs.\cite{Giunti:2000kw,Beuthe:2001rc,Giunti:2003ax} and references therein).
},
leading to the approximation
\begin{equation}
E_k t - p_k x
\simeq
\left( E_k - p_k \right) L
=
\frac{ E_k^2 - p_k^2 }{ E_k + p_k } \, L
=
\frac{ m_k^2 }{ E_k + p_k } \, L
\simeq
\frac{ m_k^2 }{ 2 E } \, L
\,,
\label{107}
\end{equation}
where $E$ is the neutrino energy in the massless limit.
This approximation for the phase of the neutrino oscillation amplitude
is very important, because it shows that the phase of ultrarelativistic
neutrinos depends only on the ratio $ m_k^2 L / E $
and not on the specific values of $E_k$ and $p_k$,
which in general depend on the specific characteristics of the production process.
The resulting oscillation probability is, therefore,
valid in general,
regardless of the production process.

With the approximation (\ref{107}),
the transition probability in space can be written as
\begin{align}
P_{\nu_{\alpha}\to\nu_{\beta}}(L)
=
\null & \null
\left|
\sum_{k}
U_{{\alpha}k}^{*}
\,
e^{-i m_k^2 L / 2 E}
\,
U_{{\beta}k}
\right|^2
\nonumber
\\
=
\null & \null
\sum_{k}
|U_{{\alpha}k}|^2
\,
|U_{{\beta}k}|^2
+
2
\,
\text{Re}
\sum_{k>j}
U_{{\alpha}k}^{*}
\,
U_{{\beta}k}
\,
U_{{\alpha}j}
\,
U_{{\beta}j}^{*}
\,
\exp\left(
-i\frac{\Delta{m}^2_{kj} L}{2E}
\right)
\,,
\label{108}
\end{align}
where
$ \Delta{m}^2_{kj} \equiv m_k^2 - m_j^2$.
Equation (\ref{108})
shows that
the constants of nature that determine
neutrino oscillations are the elements of the mixing matrix
and the differences of the squares of the neutrino masses.
Different experiments are characterized by different neutrino energy $E$
and different source-detector distance $L$.

In Eq.~(\ref{108})
we have separated the constant term
\begin{equation}
\overline{P}_{\nu_{\alpha}\to\nu_{\beta}}
=
\sum_{k}
|U_{{\alpha}k}|^2
\,
|U_{{\beta}k}|^2
\label{109}
\end{equation}
from the oscillating term which is produced by the interference
of the contributions of the different massive neutrinos.
If the energy $E$ or the distance $L$
are not known with sufficient precision,
the oscillating term is averaged out and only
the constant flavor-changing probability
(\ref{109})
is measurable.

In the simplest case of two-neutrino mixing\footnote{
This is a limiting case of three-neutrino mixing obtained if
two mixing angles are negligible.
}
between
$\nu_\alpha$, $\nu_\beta$
and
$\nu_1$, $\nu_2$,
there is only one squared-mass difference
$ \Delta{m} \equiv \Delta{m}^2_{21} \equiv m_2^2 - m_1^2$
and
the mixing matrix can be parameterized\footnote{
Here we neglect a possible Majorana phase,
which does not have any effect on oscillations
(see the end of Section~\ref{Neutrino oscillations in matter}).
}
in terms of one mixing angle $\vartheta$,
\begin{equation}
U
=
\begin{pmatrix}
\cos\vartheta
&
\sin\vartheta
\\
-\sin\vartheta
&
\cos\vartheta
\end{pmatrix}
\,.
\label{110}
\end{equation}
The resulting transition probability between different flavors can be written as
\begin{equation}
P_{\nu_{\alpha}\to\nu_{\beta}}(L)
=
\sin^2 2\vartheta
\,
\sin^2\left(
\frac{\Delta{m}^2 L}{4E}
\right)
\,.
\label{111}
\end{equation}
This expression is historically very important,
because the data of neutrino oscillation experiments have been
always analyzed as a first approximation in the two-neutrino mixing framework
using Eq.~(\ref{111}).
The two-neutrino transition probability
can also be written as
\begin{equation}
P_{\nu_{\alpha}\to\nu_{\beta}}(L)
=
\sin^2 2\vartheta
\,
\sin^2\left(
1.27
\,
\frac
{ \left( \Delta{m}^2 / \text{eV}^2 \right) \left( L / \text{km} \right) }
{ \left( E / \text{GeV} \right) }
\right)
\,,
\label{1111}
\end{equation}
where we have used typical units of short-baseline
accelerator experiments (see below).
The same numerical factor applies
if $L$ is expressed in meters and $E$ in MeV,
which are typical units of short-baseline
reactor experiments.

The transition probability in Eq.~(\ref{1111})
is useful in order to understand the classification of different
types of neutrino experiments.
Since neutrinos interact very weakly with matter,
the event rate in neutrino experiments is low and often at the limit of the background.
Therefore,
flavor transitions are observable only if
the transition probability is not too low,
which means that it is necessary that
\begin{equation}
\frac{\Delta{m}^2 L}{4E}
\gtrsim
0.1 - 1
\,.
\label{112}
\end{equation}
Using this inequality we classify neutrino oscillation experiments
according to the ratio $L/E$ which establishes the range of
$\Delta{m}^2$ to which an experiment is sensitive:
\begin{description}
\item[Short-baseline (SBL) experiments.]
In these experiments
$L/E \lesssim 1 \, \text{eV}^{-2}$.
Since the source-detector distance in these experiment is not too large,
the event rate is relatively high and oscillations can be
detected for $\Delta{m}^2 L / 4E \gtrsim 0.1$,
leading a sensitivity to
$\Delta{m}^2 \gtrsim 0.1 \, \text{eV}^2$.
There are two types of SBL experiments:
reactor $\bar\nu_e$ disappearance experiments
with
$ L \sim 10 \, \text{m} $,
$ E \sim 1 \, \text{MeV} $
as,
for example,
Bugey \cite{Bugey};
accelerator $\nu_\mu$ experiments
with
$ L \lesssim 1 \, \text{km} $,
$ E \gtrsim 1 \, \text{GeV} $,
as,
for example,
CDHS \cite{Dydak:1984zq} ($\nu_\mu\to\nu_\mu$),
CCFR \cite{Naples:1998va} ($\nu_\mu\to\nu_\mu$, $\nu_\mu\to\nu_e$ and $\nu_e\to\nu_\tau$),
CHORUS \cite{Eskut:2000de} ($\nu_\mu\to\nu_\tau$ and $\nu_e\to\nu_\tau$),
NOMAD \cite{hep-ex/0306037} ($\nu_\mu\to\nu_\tau$ and $\nu_\mu\to\nu_e$),
LSND \cite{Aguilar:2001ty} ($\bar\nu_\mu\to\bar\nu_e$ and $\nu_\mu\to\nu_e$),
KARMEN \cite{Armbruster:2002mp} ($\bar\nu_\mu\to\bar\nu_e$).
\item[Long-baseline (LBL) and atmospheric experiments.]
In these experiments
$ L/E \lesssim 10^{4} $ $ \text{eV}^{-2} $.
Since the source-detector distance is large,
these are low-statistics experiments
in which flavor transitions can be
detected if $\Delta{m}^2 L / 4E \gtrsim 1$,
giving a sensitivity to
$\Delta{m}^2 \gtrsim 10^{-4} \, \text{eV}^2$.
There are two types of LBL experiments analogous to the two types of SBL experiments:
reactor $\bar\nu_e$ disappearance experiments
with
$ L \sim 1 \, \text{km} $,
$ E \sim 1 \, \text{MeV} $
(CHOOZ \cite{Apollonio:2003gd} and Palo Verde \cite{Boehm:2001ik});
accelerator $\nu_\mu$ experiments
with
$ L \lesssim 10^3 \, \text{km} $,
$ E \gtrsim 1 \, \text{GeV} $
(K2K \cite{Ahn:2002up} for $\nu_\mu\to\nu_\mu$ and $\nu_\mu\to\nu_e$,
MINOS \cite{Diwan:2002pu} for $\nu_\mu\to\nu_\mu$ and $\nu_\mu\to\nu_e$,
CNGS \cite{hep-ex/0209082} for $\nu_\mu\to\nu_\tau$).
Atmospheric experiments
(Kamiokande \cite{Fukuda:1994mc},
IMB \cite{Becker-Szendy:1992hq},
Super-Kamiokande \cite{Fukuda:1998mi},
Soudan-2 \cite{hep-ex/0307069},
MACRO \cite{Ambrosio:2003yz})
detect neutrinos which travel a distance from about 20 km (downward-going)
to about $ 12780 \, \text{km} $ (upward-going)
and cover a wide energy spectrum,
from about 100 MeV to about 100 GeV
(see Section~\ref{Atmospheric neutrino experiments and K2K}).
\item[Very long-baseline (VLBL) and solar experiments.]
The only existing VLBL is the
reactor $\bar\nu_e$ disappearance experiment
KamLAND \cite{hep-ex/0212021}
with
$ L \sim 180 \, \text{km} $,
$ E \sim 3 \, \text{MeV} $,
yielding
$L/E \sim 3 \times 10^{5} \, \text{eV}^{-2}$.
Since the statistics is very low,
the KamLAND experiment is sensitive to
$\Delta{m}^2 \gtrsim 3 \times 10^{-5} \, \text{eV}^2$.
A sensitivity to such low values of $\Delta{m}^2$
is very important in order to have an overlap with
the sensitivity of solar neutrino experiments
which extends over the very wide range
$
10^{-8} \, \text{eV}^{-2}
\lesssim
\Delta{m}^2
\lesssim
10^{-4} \, \text{eV}^{-2}
$
because of matter effects (discussed below).
Solar neutrino experiments
(Homestake \cite{Cleveland:1998nv},
Kamiokande \cite{Fukuda:1996sz},
GALLEX \cite{Hampel:1998xg},
GNO \cite{Altmann:2000ft},
SAGE \cite{Abdurashitov:2002nt},
Super-Kamiokande \cite{Fukuda:2002pe,hep-ex/0309011},
SNO \cite{Ahmad:2001an,Ahmad:2002jz,nucl-ex/0309004})
can also measure vacuum oscillations
over the sun--earth distance
$ L \sim 1.5 \times 10^8 \, \text{km} $,
with a neutrino energy
$ E \sim 1 \, \text{MeV} $,
yielding
$L/E \sim 10^{12} \, \text{eV}^{-2}$
and a sensitivity to
$\Delta{m}^2 \gtrsim 10^{-12} \, \text{eV}^2$.
\end{description}

\subsection{Neutrino oscillations in matter}
\label{Neutrino oscillations in matter}

So far we have considered only neutrino oscillations in vacuum.
In 1978 Wolfenstein
\cite{Wolfenstein:1978ue}
realized that when neutrinos propagate in matter
oscillations are modified by the coherent interactions with the medium
which produce effective potentials that are different for different
neutrino flavors.

\begin{figure}[t]
\begin{center}
\setlength{\tabcolsep}{0cm}
\begin{tabular}{lcr}
\includegraphics*[bb=189 540 422 770, width=0.30\textwidth]{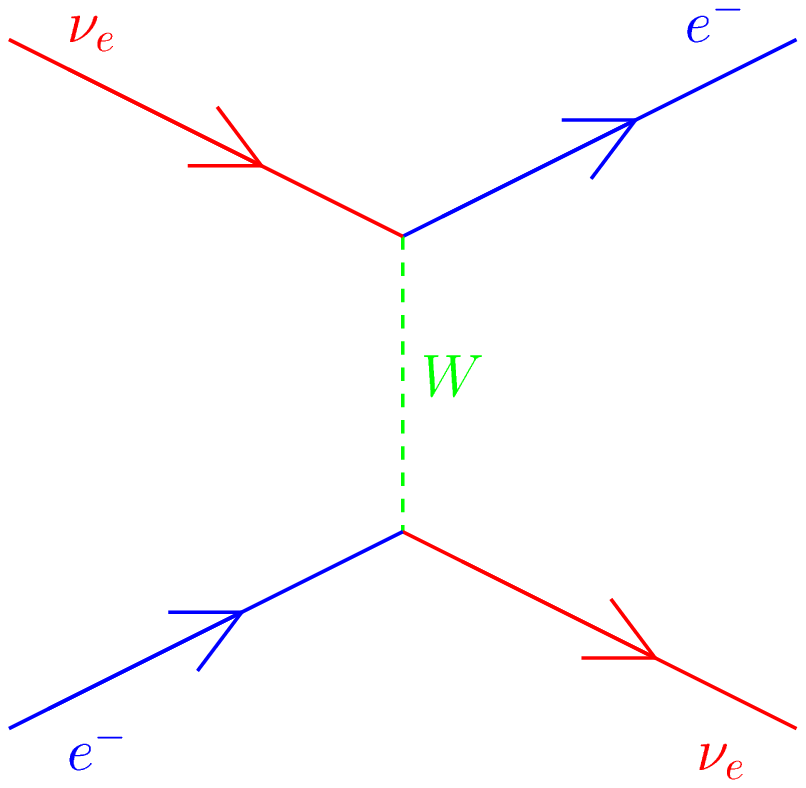}
&
\hspace{0.20\textwidth}
&
\includegraphics*[bb=189 540 422 770, width=0.30\textwidth]{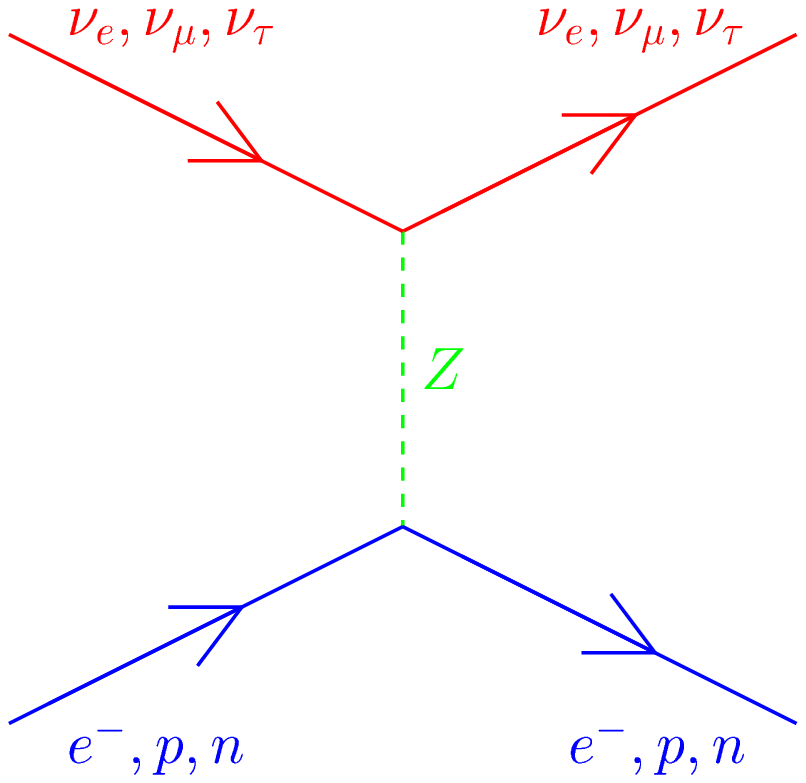}
\end{tabular}
\end{center}
\caption{ \label{V}
Feynman diagrams of the coherent forward elastic scattering
processes that generate
the CC potential $V_{\text{CC}}$ through $W$ exchange
and
the NC potential $V_{\text{NC}}$ through $Z$ exchange.
}
\end{figure}

Let us consider for simplicity\footnote{
A more complicated wave packet treatment is necessary for the
derivation of neutrino oscillations in matter
taking into account different energies and momenta of the
different massive neutrino components
\cite{Giunti:1992sx}.
}
a flavor neutrino state with definite momentum $p$,
\begin{equation}
|\nu_\alpha(p)\rangle
=
\sum_k U_{\alpha k}^{*} \, |\nu_k(p)\rangle
\,.
\label{121}
\end{equation}
The massive neutrino states $|\nu_k(p)\rangle$ with momentum $p$
are eigenstates of the vacuum Hamiltonian $\mathcal{H}_0$:
\begin{equation}
\mathcal{H}_0 \, |\nu_k(p)\rangle
=
E_k \, |\nu_k(p)\rangle
\,,
\qquad
\text{with}
\qquad
E_k = \sqrt{ p^2 + m_k^2 }
\,.
\label{122}
\end{equation}
The total Hamiltonian in matter is
\begin{equation}
\mathcal{H} = \mathcal{H}_0 + \mathcal{H}_I
\,,
\qquad
\text{with}
\qquad
\mathcal{H}_I \, |\nu_\alpha(p)\rangle
=
V_\alpha \, |\nu_\alpha(p)\rangle
\,,
\label{123}
\end{equation}
where
$V_\alpha$
is the effective potential felt by the active flavor neutrino $\nu_\alpha$
($\alpha=e,\mu,\tau$)
because of coherent interactions with the medium due to
forward elastic weak CC and NC scattering
whose Feynman diagrams are shown in Fig.~\ref{V}.
The CC and NC potential are
\cite{Langacker:1983ih}
\begin{equation}
V_{\text{CC}}
=
\sqrt{2} G_{\text{F}} N_{e}
\,,
\qquad
V_{\text{NC}}
=
-
\frac{ \sqrt{2} }{ 2 } 
G_{\text{F}} N_{n}
\,,
\label{124}
\end{equation}
where $G_{\text{F}}$ is the Fermi constant,
and
$N_{e}$ and $N_{n}$
are,
respectively,
the electron and neutron number densities.
As shown in Fig.~\ref{V},
the CC potential $V_{\text{CC}}$ is felt only by the electron neutrino,
whereas the NC potential is felt equally by the three active flavor neutrinos.
Moreover,
since the NC potential due to scattering on electrons and protons are
equal and opposite, they cancel each other (the medium is assumed to be electrically neutral)
and only the NC potential due to scattering on neutrons contributes to $V_{\text{NC}}$.
Summarizing,
we can write
\begin{equation}
V_{\alpha} = V_{\text{CC}} \, \delta_{\alpha e} + V_{\text{NC}}
\,.
\label{125}
\end{equation}
For antineutrinos the signs of all potentials are reversed.

In the Schr\"odinger picture the neutrino state with initial flavor $\alpha$
obeys the evolution equation
\begin{equation}
i \, \frac{\text{d}}{\text{d}t} \,
|\nu_\alpha(p,t)\rangle
=
\mathcal{H}
|\nu_\alpha(p,t)\rangle
\,,
\qquad
\text{with}
\qquad
|\nu_\alpha(p,0)\rangle
=
|\nu_\alpha(p)\rangle
\,.
\label{126}
\end{equation}
Let us consider the amplitudes of
$\nu_\alpha\to\nu_\beta$
flavor transitions
\begin{equation}
\psi_{\alpha\beta}(p,t)
=
\langle\nu_\beta(p)|\nu_\alpha(p,t)\rangle
\,,
\qquad
\text{with}
\qquad
\psi_{\alpha\beta}(p,0)
=
\delta_{\alpha\beta}
\,.
\label{127}
\end{equation}
In other words,
$\psi_{\alpha\beta}(p,t)$
is the probability amplitude that
a neutrino born at $t=0$ with flavor $\alpha$
is found to have flavor $\beta$ after the time $t$.

From Eqs.~(\ref{122}), (\ref{123}) and (\ref{126}),
the time evolution equation of the flavor transition amplitudes is
\begin{equation}
i \, \frac{\text{d}}{\text{d}t} \,
\psi_{\alpha\beta}(p,t)
=
\sum_\rho
\left(
\sum_{k}
U_{\beta k}
\,
E_k
\,
U_{\rho k}^*
+
\delta_{\beta\rho} \, V_{\beta}
\right)
\psi_{\alpha\rho}(p,t)
\,.
\label{128}
\end{equation}
Considering ultrarelativistic neutrinos
for which
\begin{equation}
E_k \simeq E + \frac{m_k^2}{2E}
\,,
\qquad
p = E
\,,
\qquad
t = x
\,,
\label{129}
\end{equation}
we have the evolution equation in space
\begin{equation}
i \, \frac{\text{d}}{\text{d}x} \,
\psi_{\alpha\beta}(x)
=
\left(
p + \frac{m_1^2}{2E} + V_{\text{NC}}
\right)
\psi_{\alpha\beta}(x)
+
\sum_\rho
\left(
\sum_{k}
U_{\beta k}
\,
\frac{\Delta{m}^2_{k1}}{2E}
\,
U_{\rho k}^*
+
\delta_{\beta e} \, \delta_{\rho e} \, V_{\text{CC}}
\right)
\psi_{\alpha\rho}(x)
\,,
\label{130}
\end{equation}
where we put in evidence the term
$
\left(
p + m_1^2/2E + V_{\text{NC}}
\right)
\psi_{\alpha\beta}(x)
$
which generates a phase common to all flavors.
This phase is irrelevant for the flavor transitions
and can be eliminated by the phase shift
\begin{equation}
\psi_{\alpha\beta}(x)
\to
\psi_{\alpha\beta}(x)
\,
e^{
- i \left( p + m_1^2/2E \right) x
- i \int_0^x V_{\text{NC}}(x') \, \text{d}x'
}
\,,
\label{131}
\end{equation}
which does not have any effect on the probability of
$\nu_\alpha\to\nu_\beta$
transitions,
\begin{equation}
P_{\nu_\alpha\to\nu_\beta}(x)
=
|\psi_{\alpha\beta}(x)|^2
\,.
\label{132}
\end{equation}
Therefore, the relevant evolution equation for the flavor transition amplitudes is
\begin{equation}
i \, \frac{\text{d}}{\text{d}x} \,
\psi_{\alpha\beta}(x)
=
\sum_\rho
\left(
\sum_{k}
U_{\beta k}
\,
\frac{\Delta{m}^2_{k1}}{2E}
\,
U_{\rho k}^*
+
\delta_{\beta e} \, \delta_{\rho e} \, V_{\text{CC}}
\right)
\psi_{\alpha\rho}(x)
\,,
\label{133}
\end{equation}
which shows that neutrino oscillation in matter,
as neutrino oscillation in vacuum,
depends on the differences of the squared neutrino masses,
not on the absolute value of neutrino masses.
Equation~(\ref{133}) can be written in matrix form as
\begin{equation}
i \, \frac{\text{d}}{\text{d}x} \,
\Psi_{\alpha}
=
\frac{1}{2E}
\left(
U
\,
\Delta{\mathbb{M}}^2
\,
U^{\dagger}
+
\mathbb{A}
\right)
\Psi_{\alpha}
\,,
\label{134}
\end{equation}
with, in the case of three-neutrino mixing,
\begin{equation}
\Psi_{\alpha}
=
\begin{pmatrix}
\psi_{\alpha e}
\\
\psi_{\alpha\mu}
\\
\psi_{\alpha\tau}
\end{pmatrix}
\,,
\qquad
\Delta{\mathbb{M}}^2
=
\begin{pmatrix}
0 & 0 & 0
\\
0 & \Delta{m}^2_{21} & 0
\\
0 & 0 & \Delta{m}^2_{31}
\end{pmatrix}
\,,
\qquad
\mathbb{A}
=
\begin{pmatrix}
A_{\text{CC}} & 0 & 0
\\
0 & 0 & 0
\\
0 & 0 & 0
\end{pmatrix}
\,,
\label{135}
\end{equation}
where
\begin{equation}
A_{\text{CC}}
\equiv
2 \, E \, V_{\text{CC}}
=
2 \, \sqrt{2} \, E \, G_{\text{F}} \, N_e
\,.
\label{136}
\end{equation}

Since the case of three neutrino mixing is too complicated
for an introductory discussion,
let us consider the simplest case of two neutrino mixing
between
$\nu_e$, $\nu_\mu$
and
$\nu_1$, $\nu_2$.
Neglecting an irrelevant common phase,
the evolution equation (\ref{134})
can be written as
\begin{equation}
i \frac{ \text{d} }{ \text{d}x }
\begin{pmatrix}
\psi_{e e}
\\
\psi_{e \mu}
\end{pmatrix}
=
\frac{1}{4E}
\begin{pmatrix}
- \Delta{m}^2 \cos{2\vartheta} + 2 A_{\text{CC}}
&
\Delta{m}^2 \sin{2\vartheta}
\\
\Delta{m}^2 \sin{2\vartheta}
&
\Delta{m}^2 \cos{2\vartheta}
\end{pmatrix}
\begin{pmatrix}
\psi_{e e}
\\
\psi_{e \mu}
\end{pmatrix}
\,,
\label{137}
\end{equation}
where
$ \Delta{m}^2 \equiv m_2^2 - m_1^2 $
and
$\vartheta$
is the mixing angle,
such that
\begin{equation}
\nu_e
=
\cos\vartheta \, \nu_1
+
\sin\vartheta \, \nu_2
\,,
\qquad
\nu_\mu
=
-
\sin\vartheta \, \nu_1
+
\cos\vartheta \, \nu_2
\,.
\label{138}
\end{equation}
If the initial neutrino is a $\nu_e$,
as in solar neutrino experiments,
the initial condition for the evolution equation (\ref{137}) is
\begin{equation}
\begin{pmatrix}
\psi_{e e}(0)
\\
\psi_{e \mu}(0)
\end{pmatrix}
=
\begin{pmatrix}
1
\\
0
\end{pmatrix}
\,,
\label{139}
\end{equation}
and the probabilities of $\nu_e\to\nu_\mu$ transitions
and $\nu_e$ survival are
\begin{equation}
P_{\nu_e\to\nu_\mu}(x)
=
|\psi_{e \mu}(x)|^2
\,,
\qquad
P_{\nu_e\to\nu_e}(x)
=
|\psi_{e e}(x)|^2
=
1 - P_{\nu_e\to\nu_\mu}(x)
\,.
\label{140}
\end{equation}

In practice the evolution equation of the flavor transition amplitudes
can always be solved numerically with sufficient degree of precision
given enough computational power.
Let us discuss the analytical solution of Eq.~(\ref{137})
in the case of a matter density profile which is sufficiently smooth.
This solution is useful in order to understand the qualitative aspects of the problem.

The effective Hamiltonian matrix in Eq.~(\ref{137})
can be diagonalized by the orthogonal transformation
\begin{equation}
\Psi_e
=
U_{\text{M}} \, \Psi
\,,
\quad
\text{with}
\quad
\Psi_e
=
\begin{pmatrix}
\psi_{e e}
\\
\psi_{e \mu}
\end{pmatrix}
\,,
\quad
U_{\text{M}}
=
\begin{pmatrix}
\cos\vartheta_{\text{M}} & \sin\vartheta_{\text{M}}
\\
-\sin\vartheta_{\text{M}} & \cos\vartheta_{\text{M}}
\end{pmatrix}
\,,
\quad
\Psi
=
\begin{pmatrix}
\psi_{1}
\\
\psi_{2}
\end{pmatrix}
\,,
\label{141}
\end{equation}
where
$\psi_{k}$
can be thought of as 
the amplitude of the effective massive neutrino $\nu_k$ in matter
(although such probability is not measurable,
because only flavor neutrinos can be detected).
The angle
$\vartheta_{\text{M}}$
is the effective mixing angle in matter,
given by
\begin{equation}
\tan2\vartheta_{\text{M}}
=
\dfrac{\tan2\vartheta}{1-\dfrac{A_{\text{CC}}}{\Delta{m}^2\cos2\vartheta}}
\,.
\label{142}
\end{equation}
The interesting new phenomenon,
discovered by Mikheev and Smirnov in 1985 \cite{Mikheev:1986wj}
(and beautifully explained by Bethe in 1986 \cite{Bethe:1986ej})
is that there is a resonance for
\begin{equation}
A_{\text{CC}} = \Delta{m}^2 \cos2\vartheta
\,,
\label{143}
\end{equation}
which corresponds to the electron number density
\begin{equation}
N_e^{\text{R}}
=
\frac{ \Delta{m}^2 \cos2\vartheta }{ 2 \sqrt{2} E G_{\text{F}} }
\,.
\label{1431}
\end{equation}
In the resonance the effective mixing angle is equal to $45^\circ$,
\textit{i.e.} the mixing is maximal,
leading to the possibility of total transitions between the two flavors
if the resonance region is wide enough.
This mechanism is called ``MSW effect''
in honor of Mikheev, Smirnov and Wolfenstein.

The effective squared-mass difference in matter is
\begin{equation}
\Delta{m}^2_{\text{M}}
=
\sqrt{
\left( \Delta{m}^2\cos2\vartheta - A_{\text{CC}} \right)^2
+
\left( \Delta{m}^2\sin2\vartheta \right)^2
}
\,.
\label{144}
\end{equation}

Neglecting an irrelevant common phase,
the evolution equation for the amplitudes of the effective massive neutrinos in matter
is
\begin{equation}
i \frac{ \text{d} }{ \text{d}x }
\begin{pmatrix}
\psi_{1}
\\
\psi_{2}
\end{pmatrix}
=
\left[
\frac{1}{4E}
\begin{pmatrix}
- \Delta{m}^2_{\text{M}}
&
0
\\
0
&
\Delta{m}^2_{\text{M}}
\end{pmatrix}
+
\begin{pmatrix}
0 & -i \text{d}\vartheta_{\text{M}} / \text{d}x
\\
i \text{d}\vartheta_{\text{M}} / \text{d}x & 0
\end{pmatrix}
\right]
\begin{pmatrix}
\psi_{1}
\\
\psi_{2}
\end{pmatrix}
\,,
\label{145}
\end{equation}
with the initial condition
\begin{equation}
\begin{pmatrix}
\psi_{1}(0)
\\
\psi_{2}(0)
\end{pmatrix}
=
\begin{pmatrix}
\cos\vartheta_{\text{M}}^{0} & -\sin\vartheta_{\text{M}}^{0}
\\
\sin\vartheta_{\text{M}}^{0} & \cos\vartheta_{\text{M}}^{0}
\end{pmatrix}
\begin{pmatrix}
1
\\
0
\end{pmatrix}
=
\begin{pmatrix}
\cos\vartheta_{\text{M}}^{0}
\\
\sin\vartheta_{\text{M}}^{0}
\end{pmatrix}
\,,
\label{146}
\end{equation}
where
$\vartheta_{\text{M}}^{0}$
is the effective mixing angle in matter at the point of neutrino production.

If the matter density is constant,
$ \text{d}\vartheta_{\text{M}} / \text{d}x = 0 $
and the evolutions of the amplitudes of the effective massive neutrinos in matter
are decoupled,
leading to the transition probability
\begin{equation}
P_{\nu_e\to\nu_\mu}(x)
=
\sin^2 2\vartheta_{\text{M}}
\,
\sin^2\left(
\frac{\Delta{m}^2_{\text{M}} x}{4E}
\right)
\,,
\label{147}
\end{equation}
which has the same structure as the two-neutrino
transition probability in vacuum (\ref{111}),
with the mixing angle and the squared-mass difference replaced by their effective values in matter.

\begin{figure}[t]
\begin{center}
\setlength{\tabcolsep}{0cm}
\begin{tabular*}{\textwidth}{l@{\extracolsep{\fill}}r}
\includegraphics*[bb=123 551 482 755, width=0.48\textwidth]{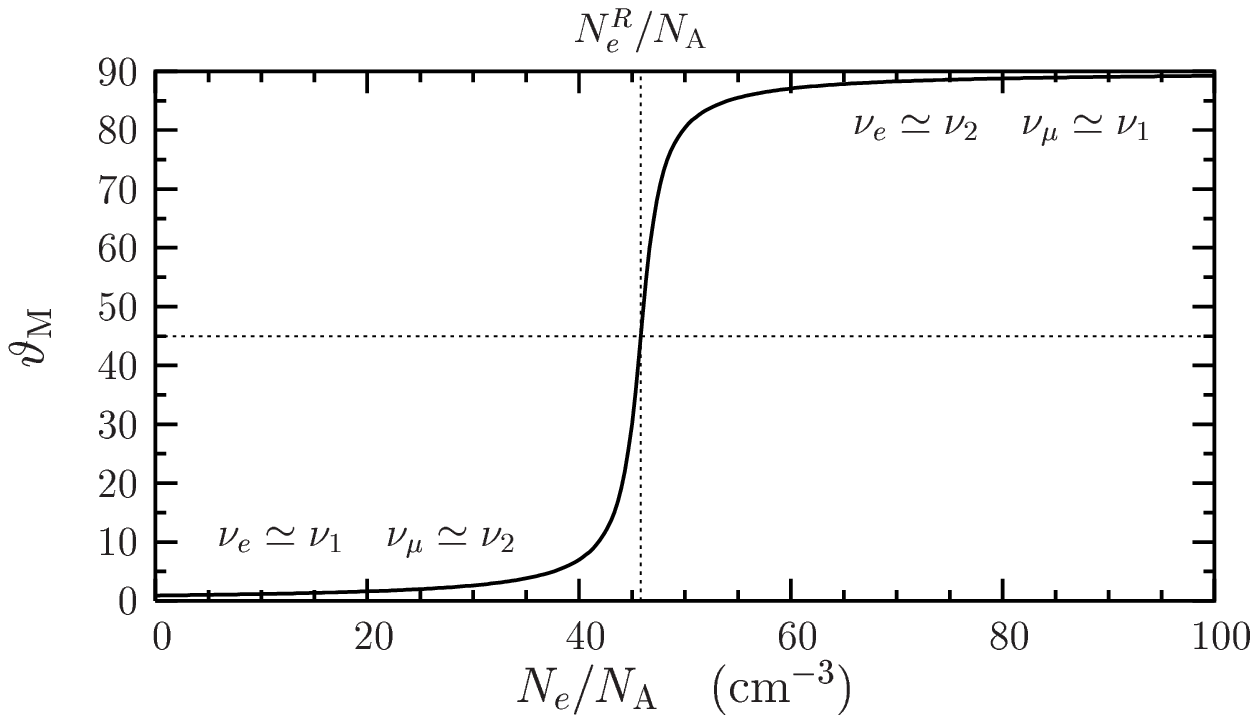}
&
\includegraphics*[bb=119 551 482 754, width=0.48\textwidth]{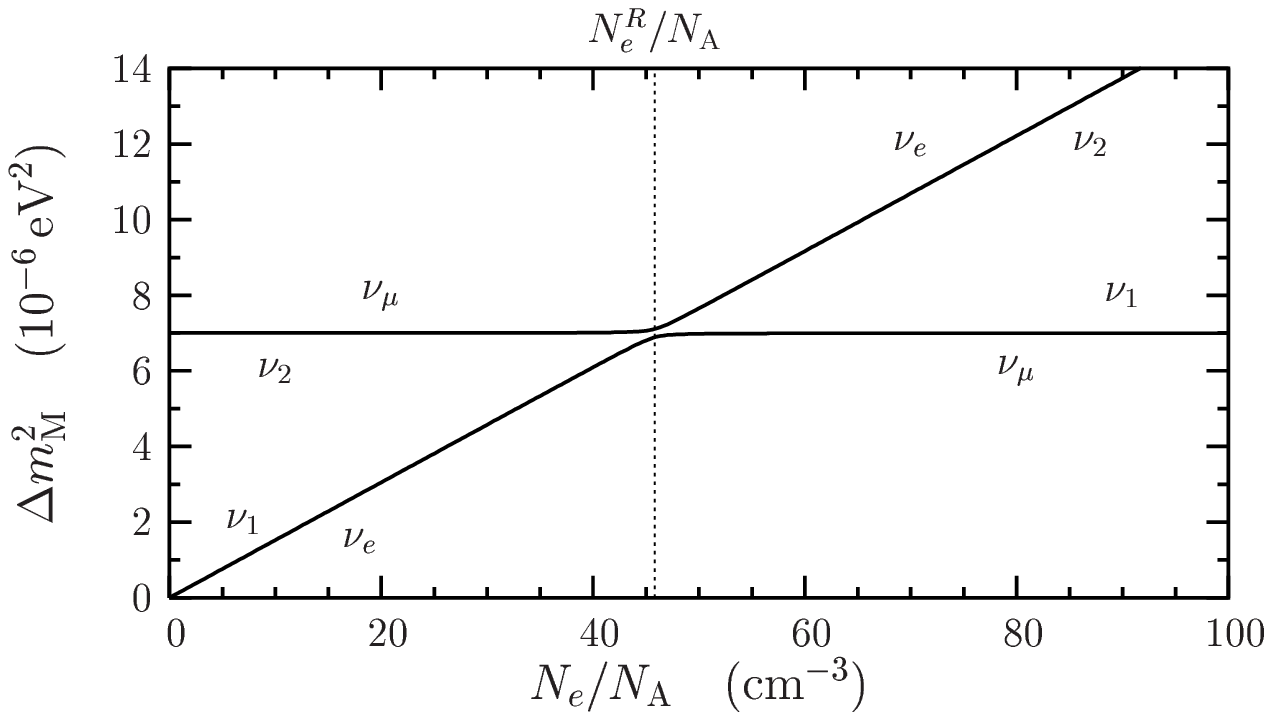}
\end{tabular*}
\end{center}
\caption{ \label{matter}
Effective mixing angle $\vartheta_{\text{M}}$ (left)
and
effective squared-mass difference $\Delta{m}^2_{\text{M}}$ (right)
in matter as functions
of the electron number density $N_e$ divided by the Avogadro number $N_{\text{A}}$,
for
$\Delta{m}^2 = 7 \times 10^{-6} \, \text{eV}^2$,
$\sin^2 2\vartheta = 10^{-3}$.
$
N_e^{\text{R}}
\equiv
\Delta{m}^2 \cos2\vartheta / 2 \sqrt{2} E G_{\text{F}}
$
is the electron number density at the resonance, where
$ \vartheta_{\text{M}} = 45^\circ $.
}
\end{figure}

If the matter density is not constant,
it is necessary to take into account the effect of
$ \text{d}\vartheta_{\text{M}} / \text{d}x $,
\begin{equation}
\dfrac{ \text{d}\vartheta_{\text{M}} }{ \text{d}x }
=
\frac{1}{2}
\,
\frac{
\Delta{m}^2\sin2\vartheta
}{
\left( \Delta{m}^2\cos2\vartheta - A_{\text{CC}} \right)^2
+
\left( \Delta{m}^2\sin2\vartheta \right)^2
}
\,
\dfrac{ \text{d}A_{\text{CC}} }{ \text{d}x }
\,,
\label{148}
\end{equation}
which is maximum at the resonance,
\begin{equation}
\left.
\dfrac{ \text{d}\vartheta_{\text{M}} }{ \text{d}x }
\right|_{\text{R}}
=
\frac{ 1 }{ 2 \, \tan 2\vartheta }
\left.
\dfrac{ \text{d} \ln N_e }{ \text{d}x }
\right|_{\text{R}}
\,.
\label{149}
\end{equation}
This is illustrated in the left panel of Fig.~\ref{matter}
for
$\Delta{m}^2 = 7 \times 10^{-6} \, \text{eV}^2$,
$\sin^2 2\vartheta = 10^{-3}$.
One can see that for
$ N_e \ll N_e^{\text{R}} $
the effective mixing angle is practically equal to the
mixing angle in vacuum,
$ \vartheta_{\text{M}} \simeq \vartheta $,
for
$ N_e \simeq N_e^{\text{R}} $
the effective mixing angle
varies very rapidly with the electron number density,
passing through $45^\circ$ at
$ N_e = N_e^{\text{R}} $
and going rapidly to $90^\circ$
for $ N_e > N_e^{\text{R}} $.

The right panel of Fig.~\ref{matter}
shows the corresponding behavior of
the effective squared-mass difference $\Delta{m}^2_{\text{M}}$,
which is useful in order to understand how the presence of a resonance
can induce an almost complete $\nu_e\to\nu_\mu$ conversion
of solar neutrinos.
If the mixing parameters are such that
at the center of the sun
$ N_e \gg N_e^{\text{R}} $,
the effective mixing angle is practically $90^\circ$
and electron neutrinos are produced as almost pure $\nu_2$.
As the neutrino propagates out of the sun,
it crosses the resonance at $ N_e = N_e^{\text{R}} $,
where the energy gap between $\nu_1$ and $\nu_2$ is minimum.
If the resonance is crossed adiabatically,
the neutrino remains $\nu_2$ and exits the sun as
$\nu_2 = \sin\vartheta \, \nu_e + \cos\vartheta \, \nu_\mu$,
which is almost equal to $\nu_\mu$ if the mixing angle is small,
leading to almost complete $\nu_e\to\nu_\mu$ conversion.
This is the case in which the MSW effect
is most effective and striking, since a large conversion
is achieved in spite of a small mixing angle.

If the resonance is not crossed adiabatically,
$\nu_2\to\nu_1$
transitions occur in an interval around the resonance
and the neutrino emerges out of the sun as a mixture
of $\nu_2$ and $\nu_1$,
leading to partial conversion of
$\nu_e$ into $\nu_\mu$.
Quantitatively, we can write the amplitudes of $\nu_1$ and $\nu_2$
at any point $x$ after resonance crossing as
\begin{align}
\psi_{1}(x)
=
\null & \null
\left[
\cos\vartheta_{\text{M}}^{0}
\exp\left(
i \int_0^{x_{\text{R}}} \frac{\Delta{m}^2_{\text{M}}(x')}{4E} \, \text{d}x'
\right)
\mathcal{A}_{11}^{\text{R}}
+
\sin\vartheta_{\text{M}}^{0}
\exp\left(
- i \int_0^{x_{\text{R}}} \frac{\Delta{m}^2_{\text{M}}(x')}{4E} \, \text{d}x'
\right)
\mathcal{A}_{21}^{\text{R}}
\right]
\nonumber
\\
\null & \null
\times
\exp\left(
i \int_{x_{\text{R}}}^x \frac{\Delta{m}^2_{\text{M}}(x')}{4E} \, \text{d}x'
\right)
\,,
\label{1501}
\\
\psi_{2}(x)
=
\null & \null
\left[
\cos\vartheta_{\text{M}}^{0}
\exp\left(
i \int_0^{x_{\text{R}}} \frac{\Delta{m}^2_{\text{M}}(x')}{4E} \, \text{d}x'
\right)
\mathcal{A}_{12}^{\text{R}}
+
\sin\vartheta_{\text{M}}^{0}
\exp\left(
- i \int_0^{x_{\text{R}}} \frac{\Delta{m}^2_{\text{M}}(x')}{4E} \, \text{d}x'
\right)
\mathcal{A}_{22}^{\text{R}}
\right]
\nonumber
\\
\null & \null
\times
\exp\left(
- i \int_{x_{\text{R}}}^x \frac{\Delta{m}^2_{\text{M}}(x')}{4E} \, \text{d}x'
\right)
\,,
\label{1502}
\end{align}
where
$\mathcal{A}_{kj}^{\text{R}}$
is the amplitude of
$\nu_k\to\nu_j$
transitions in the resonance.

Considering $x$ as the detection point on the earth,
practically in vacuum,
the probability of $\nu_e$ survival is given by
\begin{equation}
\overline{P}_{\nu_e\to\nu_e}(x)
=
|\psi_{ee}(x)|^2
\,,
\qquad
\text{with}
\qquad
\psi_{ee}(x)
=
\cos\vartheta \, \psi_1(x)
+
\sin\vartheta \, \psi_2(x)
\,.
\label{151}
\end{equation}
If $\Delta{m}^2 \gg 10^{-10} \, \text{eV}^2$
all the phases in Eqs.~(\ref{1501}) and (\ref{1502})
are very large and rapidly oscillating as functions of the neutrino energy.
In this case,
the average of the transition probability over the energy resolution
of the detector washes out all interference terms and only
the averaged survival probability
\begin{align}
\overline{P}_{\nu_e\to\nu_e}^{\,\text{sun}}
=
\null & \null
\cos^2\vartheta
\,
\cos^2\vartheta_{\text{M}}^{0}
\,
|\mathcal{A}_{11}^{\text{R}}|^2
+
\cos^2\vartheta
\,
\sin^2\vartheta_{\text{M}}^{0}
\,
|\mathcal{A}_{21}^{\text{R}}|^2
\nonumber
\\
\null & \null
+
\sin^2\vartheta
\,
\cos^2\vartheta_{\text{M}}^{0}
\,
|\mathcal{A}_{12}^{\text{R}}|^2
+
\sin^2\vartheta
\,
\sin^2\vartheta_{\text{M}}^{0}
\,
|\mathcal{A}_{22}^{\text{R}}|^2
\,,
\label{152}
\end{align}
which is independent from the sun--earth distance,
is measurable.
Taking into account that conservation of probability
implies that
\begin{equation}
|\mathcal{A}_{11}^{\text{R}}|^2
=
|\mathcal{A}_{22}^{\text{R}}|^2
=
1 - P_{\text{c}}
\,,
\qquad
|\mathcal{A}_{12}^{\text{R}}|^2
=
|\mathcal{A}_{21}^{\text{R}}|^2
=
P_{\text{c}}
\,,
\label{153}
\end{equation}
where
$P_{\text{c}}$
is the $\nu_1\leftrightarrows\nu_2$
crossing probability at the resonance,
we obtain the so-called Parke formula \cite{Parke:1986jy}
for the averaged $\nu_e$ survival probability:
\begin{equation}
\overline{P}_{\nu_e\to\nu_e}^{\,\text{sun}}
=
\frac{1}{2}
+
\left( \frac{1}{2} - P_{\text{c}} \right)
\cos2\vartheta_{\text{M}}^{0}
\,
\cos2\vartheta
\,.
\label{154}
\end{equation}
This formula has been widely used for the
analysis of solar neutrino data.

\begin{figure}[t]
\begin{center}
\setlength{\tabcolsep}{0cm}
\begin{tabular*}{\textwidth}{l@{\extracolsep{\fill}}r}
\includegraphics*[bb=95 440 517 758, width=0.48\textwidth]{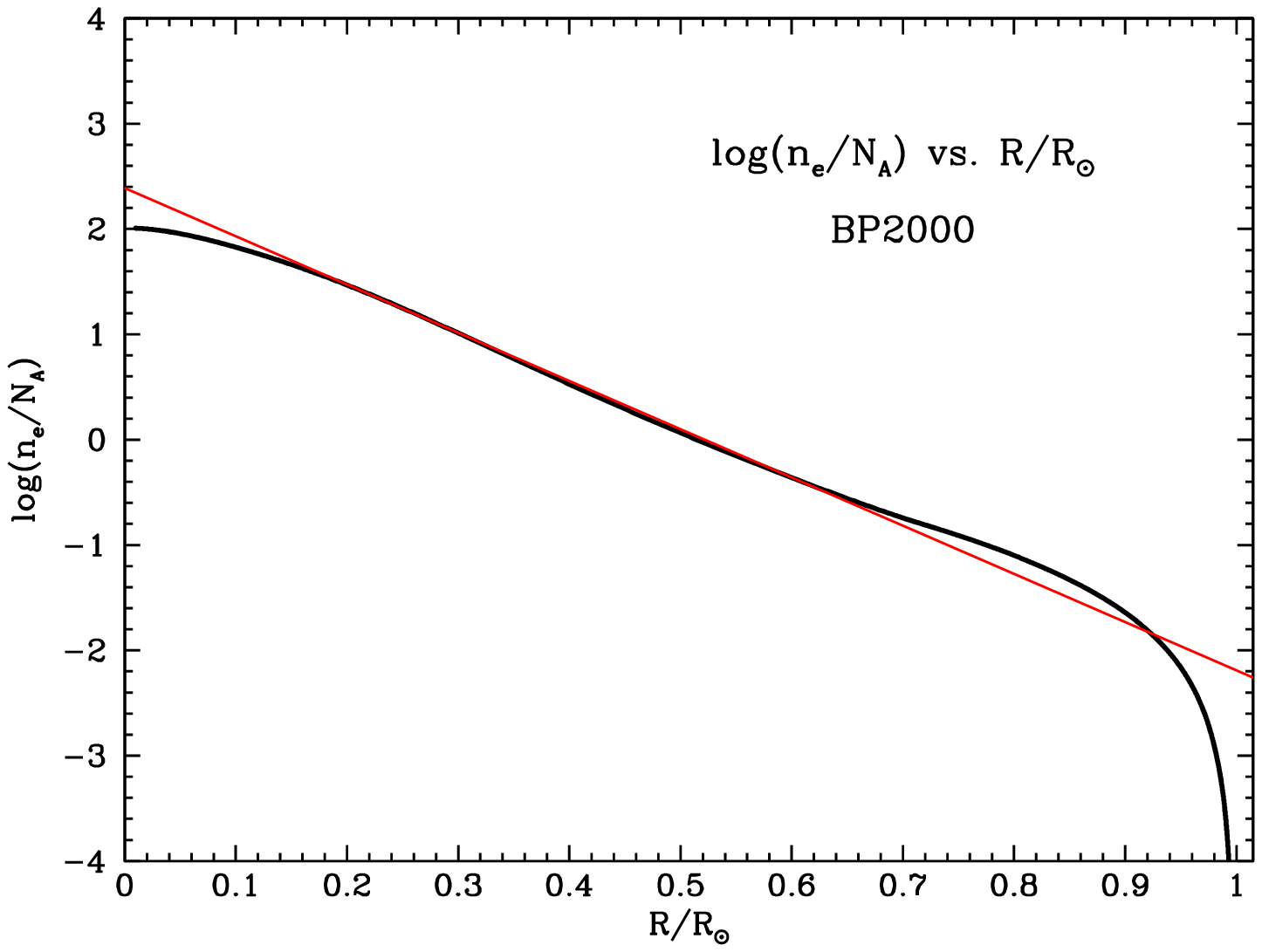}
&
\includegraphics*[bb=110 379 517 768, width=0.48\textwidth]{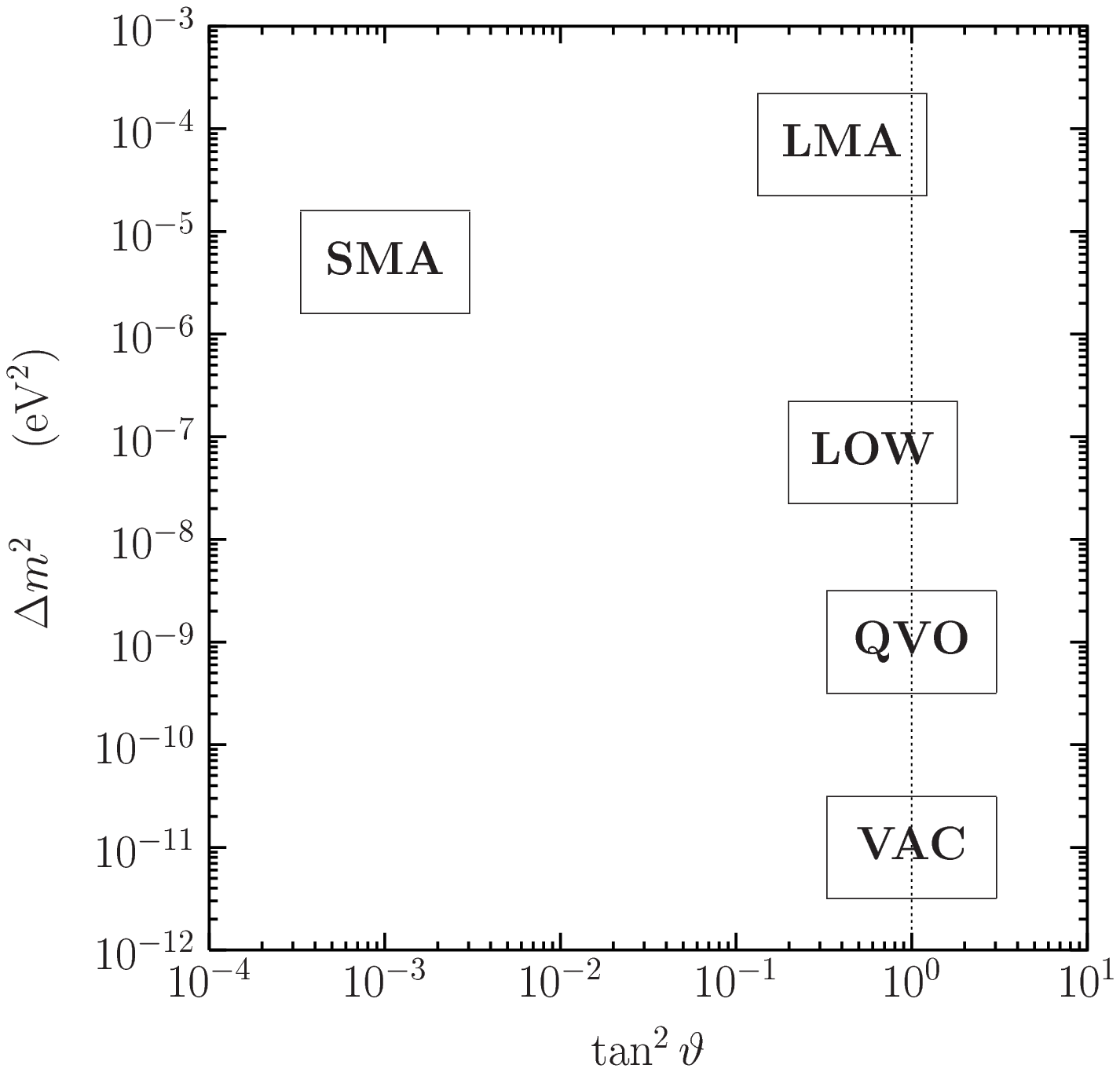}
\end{tabular*}
\end{center}
\caption{ \label{sun}
Left:
Standard Solar Model electron density profile in the sun as a function of the ratio
$R/R_\odot$
\protect\cite{Bahcall:2000nu}.
The straight line represents the approximation in Eq.~(\ref{158}).
Right:
The conventional names for regions in the $\tan^2\vartheta$--$\Delta{m}^2$ plane
obtained from the analysis of solar neutrino data.
The vertical dotted line correspond to maximal mixing.
}
\end{figure}

The main problem in the application of the Parke formula (\ref{154})
is the calculation of the crossing probability.
This probability must involve the energy gap
$ \Delta{m}^2_{\text{M}} / 2 E $
between $\nu_1$ and $\nu_2$
and the off diagonal terms proportional to $ \text{d}\vartheta_{\text{M}} / \text{d}x $
in Eq.~(\ref{145}),
which cause the $\nu_1\leftrightarrows\nu_2$ transitions.
Indeed,
the crossing probability can be written as
\cite{Petcov:1988zj,Krastev:1988ci,Petcov:1988wv,Kuo:1989pn}
\begin{equation}
P_{\text{c}}
=
\frac
{
\exp\left( - \frac{\pi}{2} \gamma F \right)
-
\exp\left( - \frac{\pi}{2} \gamma \frac{F}{\sin^2\vartheta} \right)
}
{
1
-
\exp\left( - \frac{\pi}{2} \gamma \frac{F}{\sin^2\vartheta} \right)
}
\,,
\label{155}
\end{equation}
where
$\gamma$
is the adiabaticity parameter
\begin{equation}
\gamma
=
\left.
\frac{ \Delta{m}^2_{\text{M}} / 2E }{ 2 | \text{d}\vartheta_{\text{M}} / \text{d}x | }
\right|_{\text{R}}
=
\frac
{\Delta{m}^2 \sin^22\vartheta}
{2 E \cos2\vartheta \left|\text{d}\ln N_e/\text{d}x\right|_{\text{R}}}
\,.
\label{156}
\end{equation}
If $\gamma$ is large,
the resonance is crossed adiabatically and $ P_{\text{c}} \ll 1 $,
leading to
\begin{equation}
\overline{P}_{\nu_e\to\nu_e}^{\,\text{sun, adiabatic}}
=
\frac{1}{2}
+
\frac{1}{2}
\,
\cos2\vartheta_{\text{M}}^{0}
\,
\cos2\vartheta
\,.
\label{157}
\end{equation}
The parameter $F$ in Eq.~(\ref{155})
depends on the electron density profile.
The left panel in Figure~\ref{sun} shows the
Standard Solar Model (SSM) electron density profile in the sun
\cite{Bahcall:2000nu},
which is well approximated by the exponential
\begin{equation}
N_e(R)
=
N_e(0) \exp\left( - \frac{R}{R_0} \right)
\,,
\quad
\text{with}
\quad
N_e(0)
=
245 \, N_{\text{A}} / \text{cm}^{3}
\,,
\quad
\text{and}
\quad
R_0 = \frac{R_\odot}{10.54}
\,,
\label{158}
\end{equation}
where $R$ is the distance from the center of the sun and $R_\odot$ is the solar radius.
For an exponential electron density profile
the parameter $F$ is given by
\cite{Petcov:1988zj,Krastev:1988ci,Petcov:1988wv,Pizzochero:1987fj,Toshev:1987jw,Balantekin:1998jp}
\begin{equation}
F = 1 - \tan^2\vartheta
\,.
\label{159}
\end{equation}
For
$\left|\text{d}\ln N_e/\text{d}x\right|_{\text{R}}$
the authors of Ref.~\cite{Lisi:2000su}
suggested the practical prescription,
verified with numerical solutions of the differential evolution equation,
to calculate it numerically from the SSM electron density profile
for
$R \leq 0.904 R_{\odot}$
and take the constant value
$18.9/R_{\odot}$
for
$R > 0.904 R_{\odot}$,
where
the exponential approximation (\ref{158}) breaks down.

For the analysis of solar neutrino data it is also
necessary to take into account the matter effect
along the propagation of neutrinos in the earth during the night
(the so-called ``$\nu_e$ regeneration in the earth''),
which can generate a day-night asymmetry of the rates.
The probability of solar $\nu_e$ survival after crossing the earth is given by
\cite{Mikheev:1987qk,Baltz:1987hn}
\begin{equation}
P_{\nu_e\to\nu_e}^{\,\text{sun+earth}}
=
\overline{P}_{\nu_e\to\nu_e}^{\,\text{sun}}
+
\frac
{
\left(1-2\overline{P}_{\nu_e\to\nu_e}^{\,\text{sun}}\right)
\left(P_{\nu_2\to\nu_e}^{\,\text{earth}}-\sin^2\vartheta\right)
}
{\cos\!2\vartheta}
\,.
\label{160}
\end{equation}
Since the earth density profile is not a smooth function,
the probability $P_{\nu_2\to\nu_e}^{\,\text{earth}}$
must be calculated numerically.
A good approximation is obtained by
approximating the earth density profile with a step function
(see Refs.~\cite{Liu-Maris-Petcov-earth1-97,Petcov-diffractive-98,Akhmedov-parametric-99,Chizhov:1999az,Chizhov:1999he}).
According to Eq.~(\ref{145}),
the effective massive neutrinos propagate as plane waves
in regions of constant density, with a phase
$ \exp\left( \pm i \Delta{m}^2_{\text{M}} \Delta{x} / 4E \right) $,
where $\Delta{x}$ is the width of the step.
At the boundaries of steps the
wave functions of flavor neutrinos are joined,
according to the scheme
\begin{align}
\Psi(x_n)
=
\null & \null
\left[ U_{\text{M}} \, \Phi(x_n-x_{n-1}) \, U_{\text{M}}^\dagger \right]_{(n)}
\left[ U_{\text{M}} \, \Phi(x_{n-1}-x_{n-2}) \, U_{\text{M}}^\dagger \right]_{(n-1)}
\nonumber
\\
\null & \null
\ldots
\left[ U_{\text{M}} \, \Phi(x_2-x_1) \, U_{\text{M}}^\dagger \right]_{(2)}
\left[ U_{\text{M}} \, \Phi(x_1-x_0) \, U_{\text{M}}^\dagger \right]_{(1)}
U \, \Psi(x_0)
\,.
\label{161}
\end{align}
where $x_0$ is the coordinate of the point in which the neutrino enters the earth,
$x_1$, $x_2$, \ldots, $x_n$
are the boundaries of $n$ steps with which the earth density profile is approximated,
$
\Phi(\Delta{x}) = \text{diag}\!\left(
\exp\left( - i \Delta{m}^2_{\text{M}} \Delta{x} / 4E \right)
\,,
\,
\exp\left( i \Delta{m}^2_{\text{M}} \Delta{x} / 4E \right)
\right)
$,
and the notation
$\left[ \ldots \right]_{(i)}$
indicates that all the matter-dependent quantities in the square brackets must be
evaluated with the matter density in the $i^{\text{th}}$ step, that extends
from $x_{i-1}$ to $x_i$.

The right panel in Fig.~\ref{sun}
shows the conventional names for regions in the $\tan^2\vartheta$--$\Delta{m}^2$ plane
obtained from the analysis of solar neutrino data.
The Small Mixing Angle (SMA) region is the one where the mixing angle is very small
and the resonant enhancement of flavor transitions
due to the MSW effect is more efficient.
However, as explained in Section~\ref{Solar neutrino experiments and KamLAND}
there is currently a very strong evidence in favor of the
Large Mixing Angle (LMA) region,
in which both the mixing angle and $\Delta{m}^2$ are large.
Other regions with large mixing are:
the low $\Delta{m}^2$ (LOW) region,
the Quasi-Vacuum-Oscillations (QVO) region,
and
the VACuum Oscillations region (VAC).
In the SMA, LMA and LOW regions vacuum oscillations due to the sun--earth
distance are not observable because the $\Delta{m}^2$
is too high
and interference effects are washed out by the average over
the energy resolution of the detector
(in these cases the Parke formula (\ref{154}) applies).
In the QVO region both matter effects and vacuum oscillations are important
\cite{Friedland:2000cp,Fogli:2000bk,Friedland:2000rn,Lisi:2000su}.
In the VAC region matter effects are negligible and
vacuum oscillations are dominant.

Concluding this Section on the theory of neutrino oscillations,
let us mention that the evolution equation (\ref{134})
allows to prove easily that the Majorana phases in the mixing matrix
do not have any effect on neutrino oscillations in vacuum
\cite{Bilenky-Hosek-Petcov-PLB94-80,Doi-CP-81}
as well as in matter
\cite{Langacker-Petcov-SteigmanToshev-NPB282-88},
because the diagonal matrix of Majorana phases $D(\lambda_{21},\lambda_{31})$
on the right of the mixing matrix
in Eq.~(\ref{046}) cancels
in the product
$ U \Delta{\mathbb{M}}^2 U^{\dagger} $.
Therefore,
the Dirac or Majorana nature of neutrinos cannot
be distinguished in neutrino oscillations.

\section{Neutrino oscillation experiments}
\label{Neutrino oscillation experiments}

In this Section we review the main results of the oscillation experiments which
are connected with the existing model-independent evidences in favor
of oscillations of solar and atmospheric neutrinos
and the interpretation of the experimental data in the framework
of three neutrino mixing, discussed in Section~\ref{Phenomenology of three-neutrino mixing}.
We do not discuss the results of several short-baseline neutrino (SBL)
oscillation experiments,
which have probed scales of $\Delta{m}^2$ bigger than about
$0.1 \, \text{eV}^2$,
that are larger than the scales of
$\Delta{m}^2$
indicated by solar and atmospheric neutrino data.
The SBL experiments whose data give the
most stringent constraints on the different oscillation channels
are listed in Table~\ref{SBL}.

\begin{table}[t]
\begin{center}
\begin{tabular}{|c|c|}
\hline
Experiment
&
Channels
\\
\hline
\hline
Bugey
&
$\bar\nu_e\to\bar\nu_e$
\protect\cite{Bugey}
\\
\hline
CDHS
&
$\overset{\scriptscriptstyle(-)}{\nu}_\mu\to\overset{\scriptscriptstyle(-)}{\nu}_\mu$
\protect\cite{Dydak:1984zq}
\\
\hline
CCFR
&
$\overset{\scriptscriptstyle(-)}{\nu}_\mu\to\overset{\scriptscriptstyle(-)}{\nu}_\mu$
\protect\cite{Stockdale:1985ce},
$\overset{\scriptscriptstyle(-)}{\nu}_\mu\to\overset{\scriptscriptstyle(-)}{\nu}_e$
\protect\cite{Naples:1998va},
$\overset{\scriptscriptstyle(-)}{\nu}_e\to\overset{\scriptscriptstyle(-)}{\nu}_\tau$
\protect\cite{Naples:1998va}
$\overset{\scriptscriptstyle(-)}{\nu}_e\to\overset{\scriptscriptstyle(-)}{\nu}_e$
\protect\cite{Naples:1998va}
\\
\hline
LSND
&
$\bar\nu_\mu\to\bar\nu_e$
\protect\cite{Aguilar:2001ty},
$\nu_\mu\to\nu_e$
\protect\cite{Athanassopoulos:1998er},
\\
\hline
KARMEN
&
$\bar\nu_\mu\to\bar\nu_e$
\protect\cite{Armbruster:2002mp}
\\
\hline
NOMAD
&
$\nu_\mu\to\nu_e$
\protect\cite{hep-ex/0306037}
$\nu_\mu\to\nu_\tau$
\protect\cite{Astier:2001yj},
$\nu_e\to\nu_\tau$
\protect\cite{Astier:2001yj}
\\
\hline
CHORUS
&
$\nu_\mu\to\nu_\tau$
\protect\cite{Eskut:2000de},
$\nu_e\to\nu_\tau$
\protect\cite{Eskut:2000de}
\\
\hline
NuTeV
&
$\overset{\scriptscriptstyle(-)}{\nu}_\mu\to\overset{\scriptscriptstyle(-)}{\nu}_e$
\protect\cite{Avvakumov:2002jj}
\\
\hline
\end{tabular}
\end{center}
\caption{ \label{SBL}
Short-baseline experiments (SBL) whose data give the
most stringent constraints on different oscillation channels.
}
\end{table}

All the SBL experiments in Table~\ref{SBL} did not observe
any indication of neutrino oscillations,
except the LSND experiment \cite{Athanassopoulos:1998er,Aguilar:2001ty}.
A large part
of the region in the
$\sin^2 2\vartheta$--$\Delta{m}^2$
plane allowed by LSND
has been excluded
by the results of other experiments
which are sensitive to similar values of the
neutrino oscillation parameters
(KARMEN \cite{Armbruster:2002mp},
CCFR \cite{Romosan:1997nh},
NOMAD \cite{hep-ex/0306037};
see Ref.~\cite{Church:2002tc} for an accurate combined analysis of
LSND and KARMEN data).
The MiniBooNE experiment
\cite{hep-ex/0210020}
running at Fermilab will tell us the validity
of the LSND indication in the near future.

Some years ago
the oscillations indicated by the LSND experiment
could be accommodated
together with solar and atmospheric neutrino oscillations
in the framework of four-neutrino mixing,
in which there are three light active neutrinos
and one light sterile neutrino
(see Refs.~\cite{BGG-review-98,Barger-Fate-2000,Giunti:2000ur,Peres:2000ic} and references in
Ref.~\cite{Neutrino-Unbound}).
However,
the global fit of recent data in terms of
four-neutrino mixing
is not good
\cite{Maltoni:2003yr},
disfavoring such possibility.
Therefore,
in this review we discuss only three-neutrino mixing,
which cannot explain the LSND indication,
awaiting the response of MiniBooNE
before engaging in wild speculations
(see Refs.~\cite{Bergmann:1998ft,Bueno:2000jy,hep-ph/0204236,hep-ph/0212116,Strumia:2002fw,Sorel:2003hf,hep-ph/0306226,hep-ph/0308299}).

\subsection{Solar neutrino experiments and KamLAND}
\label{Solar neutrino experiments and KamLAND}

At the end of the 60's
the radiochemical Homestake experiment \cite{Cleveland:1998nv}
began the observation of solar neutrinos through the
charged-current reaction
\cite{Pontecorvo-cl-46,Alvarez-cl-49}
\begin{equation}
\nu_e + {}^{37}\text{Cl}
\to
{}^{37}\text{Ar} + e^-
\,,
\label{311}
\end{equation}
with a threshold
$ E_{\text{th}}^{\text{Cl}} = 0.814 \, \text{MeV} $
which allows to observe mainly
$^7\text{Be}$ and $^8\text{B}$ neutrinos
produced,
respectively,
in the reactions
$ e^- + {}^7\text{Be} \to {}^7\text{Li} + \nu_e \, ( E = 0.8631 \, \text{MeV} ) $
and
$ {}^8\text{B} \to {}^8\text{Be}^* + e^+ + \nu_e \, ( E \lesssim 15 \, \text{MeV} ) $
of the thermonuclear $pp$ cycle
that produces energy in the core of the sun
(see Refs.~\cite{Rolfs-Rodney-book-88,Bahcall:1989ks}).

The Homestake experiment is called ``radiochemical''
because the $^{37}\text{Ar}$ atoms
were extracted
every $\sim$35 days
from the detector tank containing
615 tons of tetrachloroethylene ($\text{C}_2\text{Cl}_4$)
through chemical methods
and counted in small proportional counters
which detect the Auger electron produced in the
electron-capture of $^{37}\text{Ar}$.
As all solar neutrino detectors,
the Homestake tank was located deep underground
(1478 m)
in order to have a good shielding from cosmic ray muons.
The Homestake experiment
detected solar electron neutrinos for about 30 years
\cite{Cleveland:1998nv},
measuring a flux which is about one third
of the one predicted
Standard Solar Model (SSM)
\cite{Bahcall:2000nu}:
\begin{equation}
\frac
{ \Phi_{\text{Cl}}^{\text{Hom}} }
{ \Phi_{\text{Cl}}^{\text{SSM}} }
=
0.34 \pm 0.03
\,.
\label{312}
\end{equation}
This deficit was called ``the solar neutrino problem''.

The solar neutrino problem
was confirmed in the late 80's by the real-time
water Cherenkov Kamiokande experiment \cite{Fukuda:1996sz}
(3000 tons of water, 1000 m underground)
which observed solar neutrinos through
the elastic scattering (ES) reaction
\begin{equation}
\nu + e^- \to \nu + e^-
\,,
\label{313}
\end{equation}
which is mainly sensitive to
electron neutrinos,
whose cross section is about six time larger
than the cross section of muon and tau neutrinos.
The experiment is called ``real-time''
because the Cherenkov light produced in water
by the recoil electron in the reaction (\ref{313})
is observed in real time.
The solar neutrino signal is separated statistically
from the background using the fact that
the recoil electron preserves the directionality of the
incoming neutrino.
The energy threshold of the Kamiokande experiment
was 6.75 MeV,
allowing only the detection of $^8\text{B}$ neutrinos.
After 1995 the Kamiokande experiment
has been replaced by the bigger Super-Kamiokande experiment
\cite{Fukuda:2001nj,Fukuda:2002pe,hep-ex/0309011}
(50 ktons of water, 1000 m underground)
which has measured
with high accuracy the flux of solar $^8\text{B}$ neutrinos
with an energy threshold of 4.75 MeV,
obtaining
\cite{Fukuda:2002pe}
\begin{equation}
\frac
{ \Phi_{\text{ES}}^{\text{S-K}} }
{ \Phi_{\text{ES}}^{\text{SSM}} }
=
0.465 \pm 0.015
\,.
\label{314}
\end{equation}

In the early 90's
the
GALLEX \cite{Hampel:1998xg}
(30.3 tons of ${}^{71}\text{Ga}$, 1400 m underground)
and
SAGE \cite{Abdurashitov:2002nt}
(50 tons of ${}^{71}\text{Ga}$, 2000 m underground)
radiochemical experiments
started the observation of solar electron neutrinos through the
charged-current reaction
\cite{Kuzmin-Ga-65}
\begin{equation}
\nu_e + {}^{71}\text{Ga} \to {}^{71}\text{Ge} + e^-
\,,
\label{315}
\end{equation}
which has the low energy threshold of 0.233 MeV,
that allows the detection of the so-called
$pp$ neutrinos
produced in the main reaction
$ p + p \to d + e^+ + \nu_e \, ( E \lesssim 0.42 \, \text{MeV} ) $
of the $pp$ cycle,
besides the $^7\text{Be}$, $^8\text{B}$ and other neutrinos.
After 1997 the GALLEX experiment has been upgraded,
changing its name to GNO \cite{Altmann:2000ft}.
The combined results of the three Gallium experiments
confirm the solar neutrino problem:
\begin{equation}
\frac
{ \Phi_{\text{Ga}} }
{ \Phi_{\text{Ga}}^{\text{SSM}} }
=
0.56 \pm 0.03
\,.
\label{316}
\end{equation}

Although it was difficult to doubt of the Standard Solar Model,
which was well tested by helioseismological measurements
(see Ref.~\cite{Bahcall:1997qw}),
and it was difficult to explain the different suppression
of solar $\nu_e$'s observed in different experiments
with astrophysical mechanisms,
a definitive model-independent proof
that the solar neutrino problem is due to neutrino physics
was lacking until
the real-time heavy-water Cherenkov detector
SNO \cite{Ahmad:2001an,Ahmad:2002jz,nucl-ex/0309004}
(1 kton of $\text{D}_2\text{O}$, 2073 m underground)
observed
solar $^8\text{B}$ neutrinos through
the charged-current (CC) reaction
\begin{equation}
\nu_e + d \to p + p + e^-
\,,
\label{302}
\end{equation}
with
$ E_{\text{th,CC}}^{\text{SNO}} = 8.2 \, \text{MeV} $
and the neutral-current (NC) reaction
\begin{equation}
\nu + d \to p + n + \nu
\,,
\label{301}
\end{equation}
with
$ E_{\text{th,NC}}^{\text{SNO}} = 2.2 \, \text{MeV} $,
besides
the ES reaction (\ref{313})
with
$ E_{\text{th,ES}}^{\text{SNO}} = 7.0 \, \text{MeV} $.
The observation of solar neutrinos
through the CC and NC reactions has provided the breakthrough
for the definitive solution of the
solar neutrino problem in favor of new neutrino physics.
The charged-current reaction is very
important because it allows to measure
with high statistics the energy spectrum
of solar $\nu_e$'s.
The neutral current reaction
is extremely important for the measurement of the total flux
of active $\nu_e$, $\nu_\mu$ and $\nu_\tau$,
which interact with the same cross section.

In June 2001 the combination of the first SNO CC data \cite{Ahmad:2001an}
and the high-precision Super-Kamiokande ES data \cite{Fukuda:2001nj}
allowed to extract a model-independent
indication of the oscillations of solar electron neutrinos into
active $\nu_\mu$'s and/or $\nu_\tau$'s
\cite{Ahmad:2001an}
(see also Refs.~\cite{Fogli:2001vr,Giunti:2001ws}).
In April 2002
the observation of solar neutrinos
through the NC
and CC
reactions
allowed the SNO experiment \cite{Ahmad:2002jz}
to solve definitively the long-standing solar neutrino problem
in favor of the existence of $\nu_e \to \nu_\mu, \nu_\tau$
transitions.
In this first phase \cite{Ahmad:2002jz},
called ``$\text{D}_2\text{O}$ phase'',
the neutron produced in the neutral-current reaction (\ref{301})
was detected by observing the photon produced in
the reaction
\begin{equation}
n + d \to {}^3\text{H} + \gamma \, ( E_\gamma = 6.25 \, \text{MeV})
\,.
\label{303}
\end{equation}
In September 2003 the SNO collaboration released the data obtained
in the second phase \cite{nucl-ex/0309004},
called ``salt phase'',
in which 2 tons of salt has been added to the heavy water in the SNO detector,
allowing the detection
of the neutron produced in the neutral-current reaction (\ref{301})
by observing the photons produced in
the reaction
\begin{equation}
n + {}^{35}\text{Cl} \to {}^{36}\text{Cl} + \text{several $\gamma$'s} \,
( E_\gamma^{\text{tot}} = 8.6 \, \text{MeV})
\,.
\label{304}
\end{equation}
The better signature given by several photons
and the higher cross-section of reaction (\ref{304})
with respect to reaction (\ref{303})
have allowed the SNO collaboration to measure with good precision
the total flux of active neutrinos
coming from $^8\text{B}$ decay in the core of the sun \cite{nucl-ex/0309004}:
\begin{equation}
\Phi_{\text{NC}}^{\text{SNO}}
=
5.21 \pm 0.47
\times 10^6 \, \text{cm}^{-2} \, \text{s}^{-1}
\,,
\label{305}
\end{equation}
which is in good agreement with
the value predicted by the Standard Solar Model (SSM)
\cite{Bahcall:2000nu},
\begin{equation}
\Phi_{^8\text{B}}^{\text{SSM}}
=
5.05 \, {}^{+1.01}_{-0.81}
\times 10^6 \, \text{cm}^{-2} \, \text{s}^{-1}
\,.
\label{306}
\end{equation}
On the other hand,
the flux of electron neutrinos coming from $^8\text{B}$ decay
measured through the
CC reaction (\ref{302})
is only \cite{nucl-ex/0309004}
\begin{equation}
\Phi_{\text{CC}}^{\text{SNO}}
=
1.59 {}^{+0.10}_{-0.11}
\times 10^6 \, \text{cm}^{-2} \, \text{s}^{-1}
\,.
\label{307}
\end{equation}
The fact that the ratio \cite{nucl-ex/0309004}
\begin{equation}
\frac{ \Phi_{\text{CC}}^{\text{SNO}} }{ \Phi_{\text{NC}}^{\text{SNO}} }
=
0.306 \pm 0.035
\label{308}
\end{equation}
differs from unity by
about 19 standard deviations
is a very convincing proof that
solar electron neutrinos have transformed into muon and/or
tau neutrinos on their way to the earth. 

\begin{figure}[t]
\begin{center}
\begin{tabular}{cc}
\begin{minipage}[t]{0.45\textwidth}
\begin{center}
\includegraphics*[bb=2 2 275 256, width=0.90\textwidth]{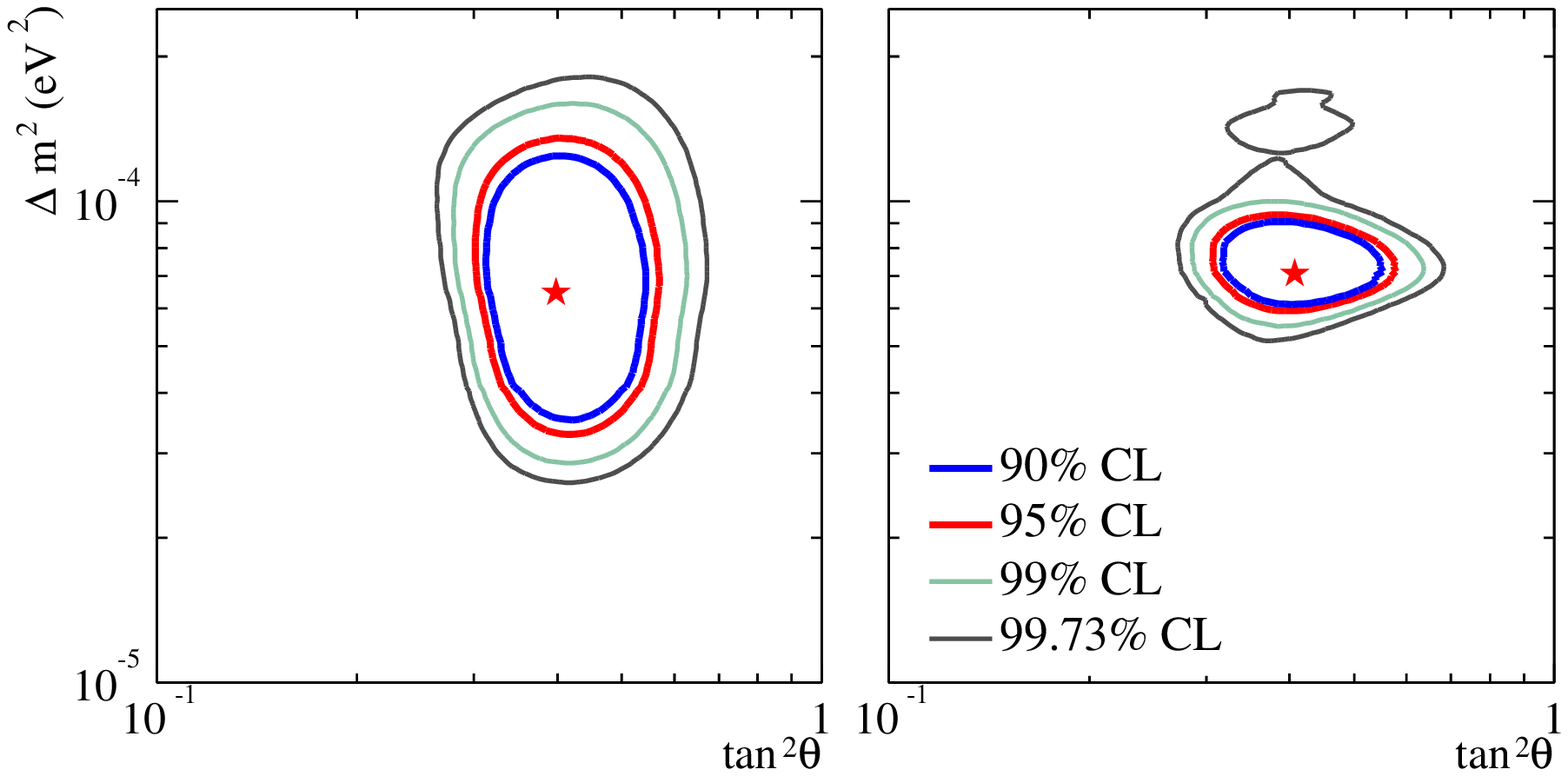}
\end{center}
\end{minipage}
&
\begin{minipage}[t]{0.45\textwidth}
\begin{center}
\includegraphics*[bb=6 7 555 525, width=0.90\textwidth]{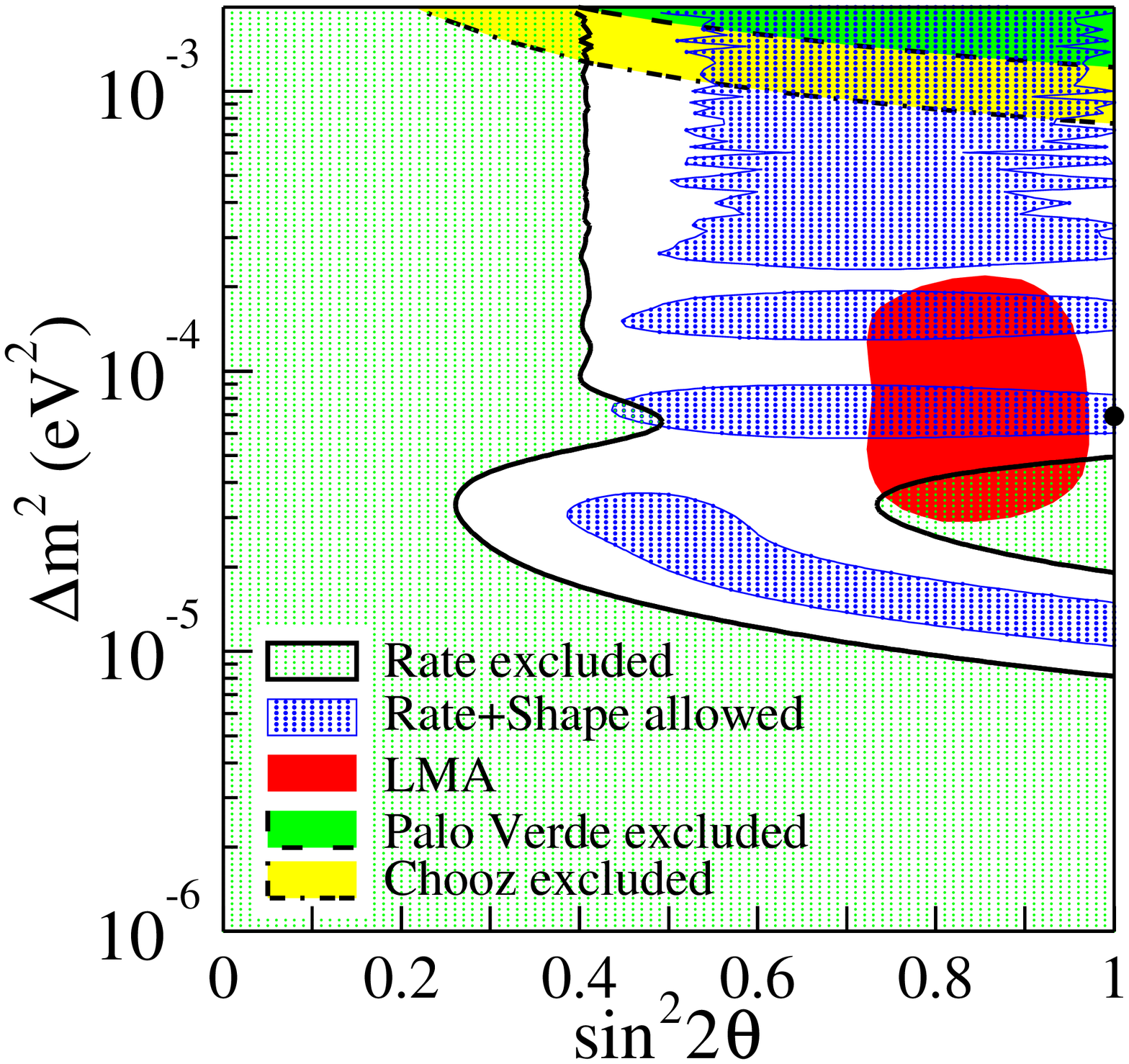}
\end{center}
\end{minipage}
\end{tabular}
\end{center}
\caption{ \label{sno-0309004-f05+kamland-0212021-f06}
Left:
Allowed regions of neutrino oscillation parameters
obtained from the global analysis of
solar neutrino data
\protect\cite{nucl-ex/0309004}.
The best-fit point is marked by a star.
Right:
KamLAND
excluded regions of neutrino oscillation parameters
for the rate analysis and allowed regions for the combined rate and
energy spectrum analysis at 95\% C.L \protect\cite{hep-ex/0212021}.
At the top are the 95\% C.L. excluded region from CHOOZ \cite{Apollonio:1999ae}
and Palo Verde \cite{Boehm:2001ik} experiments, respectively.
The dark area is the 95\% C.L. LMA allowed
region obtained in Ref.~\cite{Fogli:2002pt}.
The thick dot indicates the best fit of
KamLAND data.
}
\end{figure}

The result of the global analysis of all solar neutrino data
in terms of the simplest
hypothesis of two-neutrino oscillations
favors the so-called Large Mixing Angle (LMA)
region with effective two-neutrino mixing parameters
$
\Delta{m}^2_{\text{SUN}}
\sim
7 \times 10^{-5} \, \text{eV}^2
$
and
$ \tan^2 \vartheta_{\text{SUN}} \sim 0.4 $,
as shown in the left panel in Fig.~\ref{sno-0309004-f05+kamland-0212021-f06},
taken from Ref.~\cite{nucl-ex/0309004}.

\begin{figure}[t]
\begin{center}
\begin{tabular}{cc}
\begin{minipage}[t]{0.45\textwidth}
\begin{center}
\includegraphics*[bb=2 2 46 256, height=6cm]{fig/sno-0309004-f05.eps}
\hspace{-0.27cm}
\includegraphics*[bb=289 2 518 256, height=6cm]{fig/sno-0309004-f05.eps}
\end{center}
\end{minipage}
&
\begin{minipage}[t]{0.45\textwidth}
\begin{center}
\includegraphics*[bb=21 256 278 548, height=6cm]{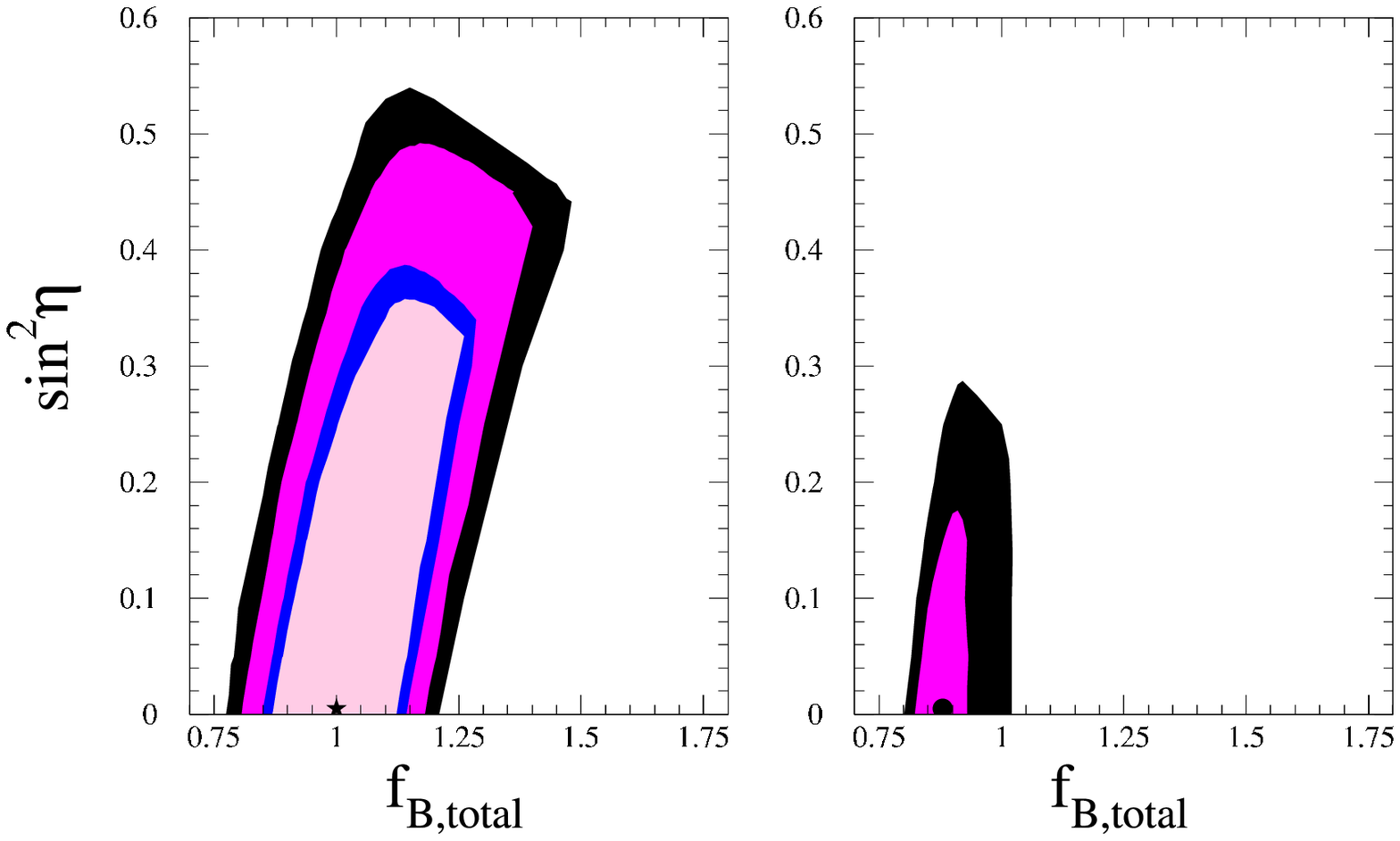}
\end{center}
\end{minipage}
\end{tabular}
\end{center}
\caption{ \label{sno-0309004-f05+eps-bahcall-0212147-f06}
Left:
Allowed
regions
obtained from the global analysis of
solar and KamLAND data
\cite{nucl-ex/0309004}.
Right:
Allowed
90\%, 95\%, 99\%, 99.73\% C.L.
regions obtained from the global analysis of
solar and KamLAND data \protect\cite{hep-ph/0212147}.
The best-fit points are marked by stars.
}
\end{figure}

A spectacular proof of the correctness of the LMA region
has been obtained at the end of 2002 in
the KamLAND long-baseline $\bar\nu_e$ disappearance experiment
\cite{hep-ex/0212021},
in which the suppression
\begin{equation}
\frac
{N^{\text{KamLAND}}_{\text{observed}}}
{N^{\text{KamLAND}}_{\text{expected}}}
=
0.611 \pm 0.094
\,.
\label{320}
\end{equation}
of the $\bar\nu_e$ flux
produced by nuclear reactors at an average distance of about 180 km
was observed.
The right panel in Fig.~\ref{sno-0309004-f05+kamland-0212021-f06}
shows the regions of oscillation parameters allowed by KamLAND,
compared with the allowed LMA region obtained in
Ref.~\cite{Fogli:2002pt} in 2002
after the release of the data of the first $\text{D}_2\text{O}$ phase
of the SNO experiment
\cite{Ahmad:2002jz}.
From the right panel in Fig.~\ref{sno-0309004-f05+kamland-0212021-f06}
one can see that the LMA region and the KamLAND allowed regions
overlap in two subregions at
$
\Delta{m}^2_{\text{SUN}}
\simeq
7 \times 10^{-5} \, \text{eV}^2
$
and
$
\Delta{m}^2_{\text{SUN}}
\simeq
1.5 \times 10^{-4} \, \text{eV}^2
$.
Therefore,
the combined fit of 2002 solar neutrino data and KamLAND data
yielded two allowed LMA subregions.
The 2003 SNO salt phase data lead to a
restriction of the LMA region allowed by solar neutrino data,
which favors the lower LMA subregion,
as shown in the left panel in Fig.~\ref{sno-0309004-f05+eps-bahcall-0212147-f06}
which depicts
the most updated allowed region of the two-neutrino oscillation parameters
obtained from the global analysis of solar and KamLAND neutrino data.
The effective two-neutrino mixing parameters
are constrained at 99.73\% C.L. ($3\sigma$)
in the ranges
\cite{hep-ph/0309130}
\begin{eqnarray}
&
5.4 \times 10^{-5} \, \text{eV}^2
<
\Delta{m}^2_{\text{SUN}}
<
9.4 \times 10^{-5} \, \text{eV}^2
\,,
&
\label{321}
\\
&
0.30
<
\tan^2 \vartheta_{\text{SUN}}
<
0.64
\,,
&
\label{322}
\end{eqnarray}
with best-fit values
\cite{hep-ph/0309130}
\begin{equation}
\Delta{m}^{2\,\text{bf}}_{\text{SUN}}
=
6.9 \times 10^{-5} \, \text{eV}^2
\,,
\qquad
\tan^2 \vartheta_{\text{SUN}}^{\text{bf}}
=
0.43
\,.
\label{323}
\end{equation}
Maximal mixing
is excluded at a confidence level equivalent to $5.4\sigma$
\cite{nucl-ex/0309004}.

Transitions of solar $\nu_e$'s into sterile states are disfavored by
the data.
The right panel in Fig.~\ref{sno-0309004-f05+eps-bahcall-0212147-f06}
shows the allowed regions in the
$\text{f}_{\text{B,total}}$--$\sin^2\eta$ plane
obtained in Ref.~\cite{hep-ph/0212147} before the release of the SNO salt data,
where
$\text{f}_{\text{B,total}}=\Phi_{^8\text{B}}/\Phi_{^8\text{B}}^{\text{SSM}}$
is the ratio of the $^8\text{B}$ solar neutrino flux
and its value predicted by the Standard Solar Model (SSM)
\cite{Bahcall:2000nu}.
The parameter $\sin^2\eta$ quantifies the fraction of
solar $\nu_e$'s that transform into sterile $\nu_s$:
$\nu_e \to \cos\eta \, \nu_a + \sin\eta \, \nu_s $,
where $\nu_a$ are active neutrinos.
From the right panel of Fig.~\ref{sno-0309004-f05+eps-bahcall-0212147-f06}
it is clear that there is a correlation between
$\text{f}_{\text{B,total}}$
and
$\sin^2\eta$,
which is due to the constraint on the total flux of
$^8\text{B}$ active neutrinos given by the SNO neutral-current measurement:
disappearance into sterile states is possible only if the
$^8\text{B}$ solar neutrino flux
is larger than the SSM prediction.
The allowed ranges for
$\Phi_{^8\text{B}}$
and
$\sin^2\eta$
are
\cite{hep-ph/0212147}
\begin{equation}
\Phi_{^8\text{B}} = 1.00 \pm 0.06 \, \Phi_{^8\text{B}}^{\text{SSM}}
\,,
\quad
\sin^2\eta < 0.52
\,.
\label{eps-014}
\end{equation}
The allowed interval for
$\Phi_{^8\text{B}}$
shows a remarkable agreement of the
data with the SSM,
independently from possible $\nu_e\to\nu_s$ transitions.
The recent SNO salt data
do not allow to improve significantly
the bound on $\sin^2\eta$
\cite{hep-ph/0309174}.

In the future it is expected that
the KamLAND experiment will allow to
reach a relatively high accuracy in the determination of
$\Delta{m}^2_{\text{SUN}}$
\cite{Inoue:2003qs},
whereas new low-energy solar neutrino experiments
or a new dedicated reactor neutrino experiment
are needed in order to improve significantly our knowledge
of the solar effective mixing angle $\vartheta_{\text{SUN}}$
\cite{hep-ph/0302243,Bahcall:2003ce,hep-ph/0306017}.

\subsection{Atmospheric neutrino experiments and K2K}
\label{Atmospheric neutrino experiments and K2K}

Atmospheric neutrinos are produced by cosmic rays
(mainly protons) which interact with
the atmosphere producing pions, which decay into muon and neutrinos,
\begin{equation}
\pi^+ \to \mu^+ + \nu_\mu
\,,
\qquad
\pi^- \to \mu^- + \bar\nu_\mu
\,.
\label{351}
\end{equation}
At low energy the muons decay before hitting the ground into
electrons and neutrinos,
\begin{equation}
\mu^+ \to e^+ + \nu_e + \bar\nu_\mu
\,,
\qquad
\mu^- \to e^- + \bar\nu_e + \nu_\mu
\,.
\label{352}
\end{equation}
Hence, the predicted ratio of
$\nu_\mu+\bar\nu_\mu$
and
$\nu_e+\bar\nu_e$
is about 2 at neutrino energy $ E \lesssim 1 \text{GeV} $.
At higher energies the ratio increases,
but it can be calculated with reasonable accuracy
(about 5\%).
On the other hand,
the calculation of the absolute value of the atmospheric neutrino flux
suffers from a large uncertainty
(20\% or 30\%)
due to the uncertainty of the absolute value of the cosmic ray flux
and
the uncertainties of the cross sections of cosmic ray interactions
with the nuclei in the atmosphere
(see Ref.~\cite{Gaisser:2002jj}).
Therefore,
the traditional way that has been followed for testing
the atmospheric neutrino flux calculation
is to measure the ratio of ratios
\begin{equation}
R \equiv
\frac{
\left[ N(\nu_\mu+\bar\nu_\mu) / N(\nu_e+\bar\nu_e) \right]_{\text{data}}
}{
\left[ N(\nu_\mu+\bar\nu_\mu) / N(\nu_e+\bar\nu_e) \right]_{\text{theo}}
}
\,,
\label{353}
\end{equation}
where the subscripts ``data'' and ``theo''
indicate, respectively,
the measured and calculated
ratio.
If nothing happens to neutrinos on their way to the detector
the ratio of ratios should be equal to one.

Atmospheric neutrinos are observed through high-energy charged-current
interactions in which the flavor, direction and energy of the neutrino
are strongly correlated with
the measured flavor, direction and energy of the produced charged lepton.

In 1988 the
Kamiokande \cite{Hirata:1988uy} and IMB \cite{Bionta:1988an} experiments
measured a ratio of ratios significantly lower than one.
The current values of $R$ measured in the Super-Kamiokande experiment
are
\cite{Toshito:2001dk}
\begin{eqnarray}
&&
R^{\text{S-K}}(E < 1.33 \, \text{GeV}) = 0.638 \pm 0.053
\,,
\label{3541}
\\
&&
R^{\text{S-K}}(E > 1.33 \, \text{GeV}) = 0.675 \pm 0.087
\,.
\label{3542}
\end{eqnarray}
The boundary of 1.33 GeV
has been chosen by the Super-Kamiokande Collaboration
for historical reasons connected with proton decay search.

Also the Soudan-2 experiment \cite{hep-ex/0307069}
observed a ratio of ratios significantly lower than one,
\begin{equation}
R^{\text{Soudan-2}} = 0.69 \pm 0.12
\,,
\label{355}
\end{equation}
and the MACRO experiment \cite{Ambrosio:2003yz}
measured a disappearance of upward-going muons.

Although the values (\ref{3541}), (\ref{3542}) and (\ref{355}) of the ratio of ratios
suggest an evidence of an atmospheric neutrino anomaly
probably due to neutrino oscillations,
they are not completely model-independent.

\begin{figure}[t]
\begin{center}
\begin{tabular}{cc}
\begin{minipage}[t]{0.45\textwidth}
\begin{center}
\includegraphics*[bb=8 6 526 516, width=\textwidth]{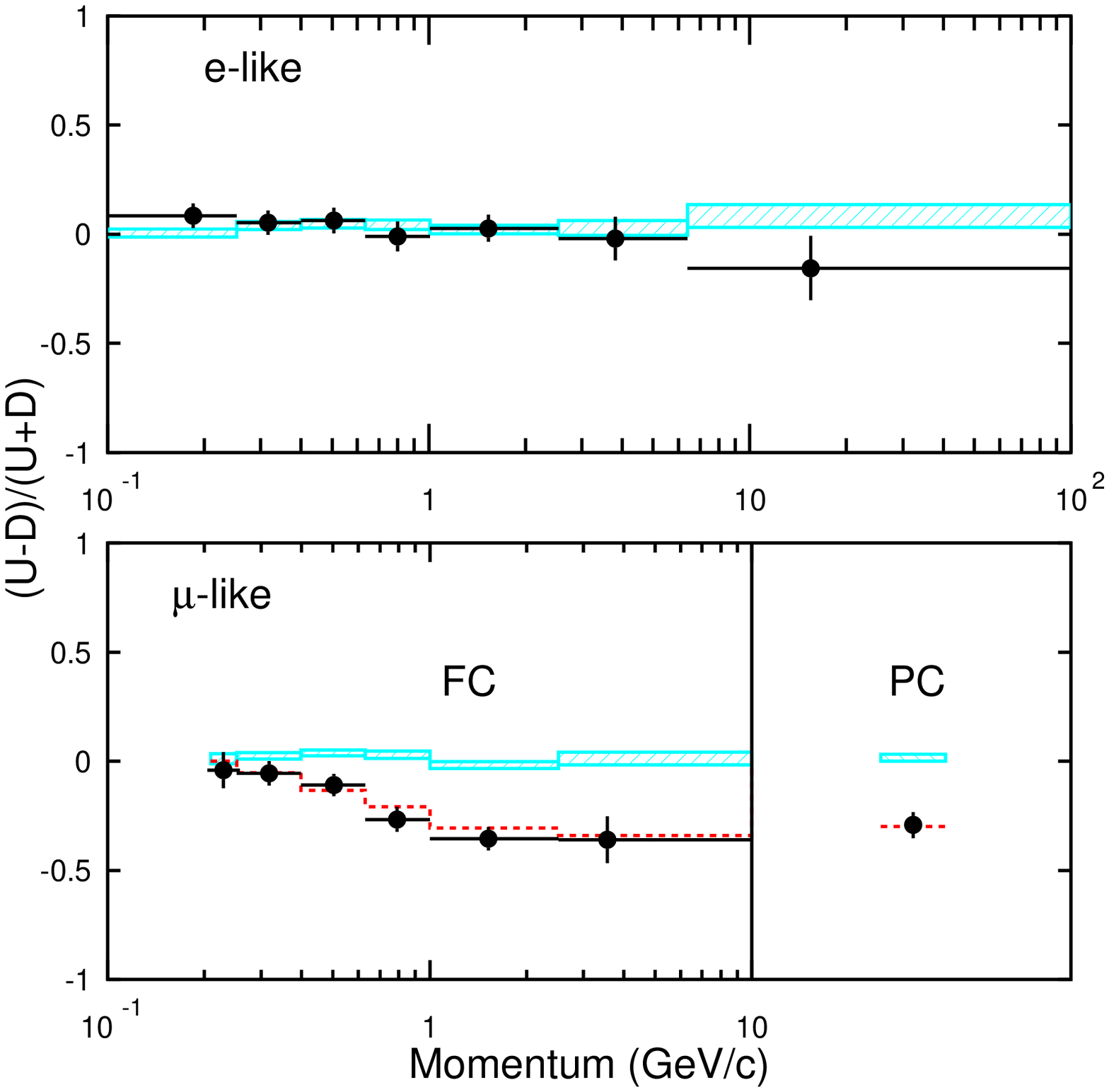}
\end{center}
\end{minipage}
&
\begin{minipage}[t]{0.45\textwidth}
\begin{center}
\includegraphics*[bb=2 19 534 548, width=\textwidth]{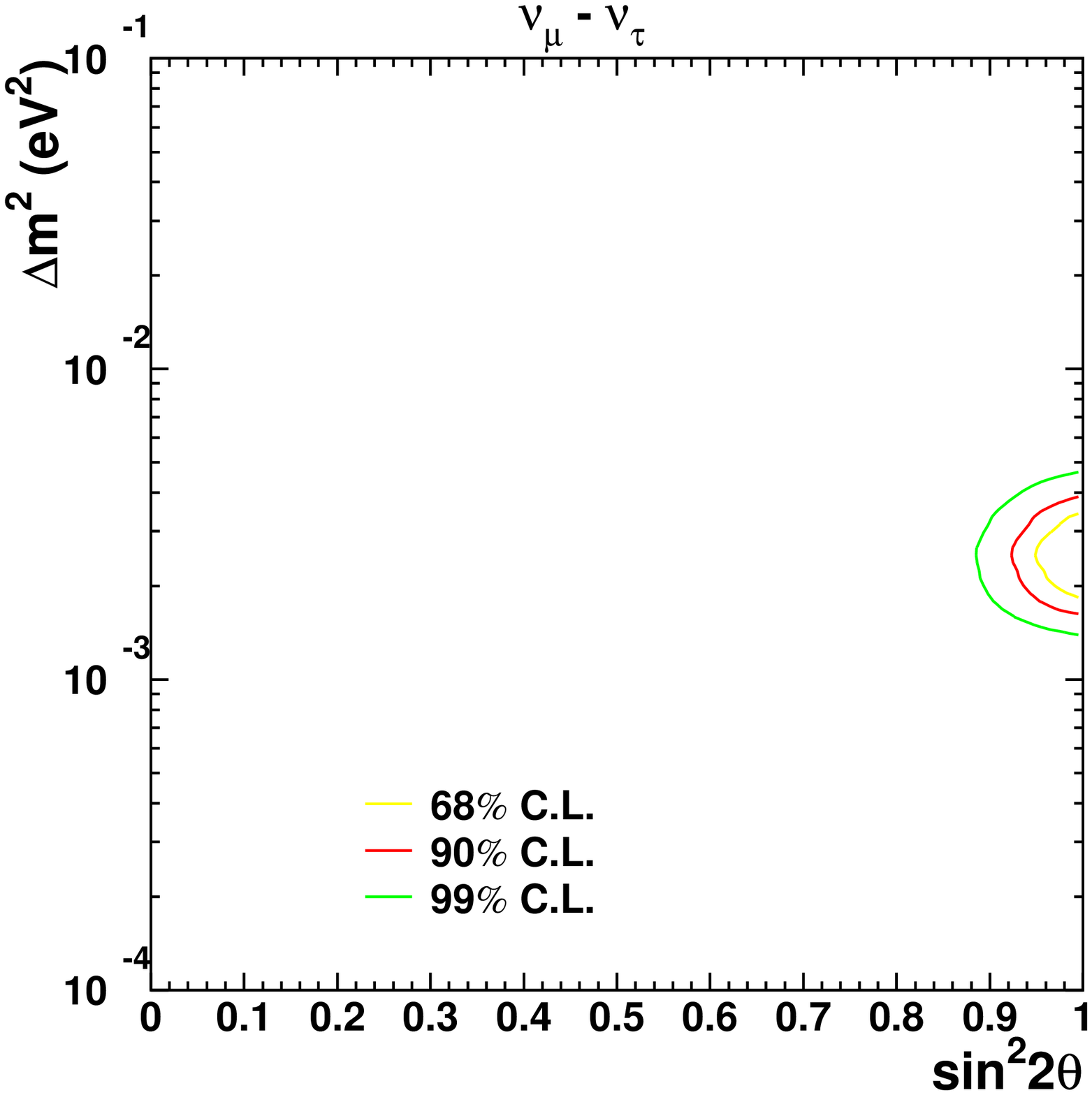}
\end{center}
\end{minipage}
\end{tabular}
\end{center}
\caption{ \label{sk-scholberg-9905016-f04+sk-wilkes-0212035-combined-allowed}
Left:
Up-down Super-Kamiokande asymmetry as a function of
momentum for $e$-like and $\mu$-like events
generated,
respectively,
by atmospheric $\nu_e$, $\bar\nu_e$ and
$\nu_\mu$, $\bar\nu_\mu$
\protect\cite{Scholberg:1999ar}.
The division of
$\mu$-like into fully contained (FC) and partially contained (PC)
is explained, for example, in Ref.~\cite{Kajita:2000mr}.
The hatched region shows the theoretical expectation without neutrino oscillations.
The dashed line for $\mu$-like events
represents the fit of the data in the case of two-generation
$\nu_\mu\to\nu_\tau$ oscillations
with
$ \Delta{m}^2 = 3.5 \times 10^{-3} \, \text{eV}^2 $
and
$ \sin^22\vartheta = 1.0$.
Right:
Allowed region contours for $\nu_\mu \to \nu_\tau$
oscillations obtained by the Super-Kamiokande experiment
\cite{hep-ex/0212035}.
%Allowed regions obtained in different experiments
%for the effective two-neutrino mixing parameters
%that generate
%atmospheric $\nu_\mu\to\nu_\tau$ oscillations
%\protect\cite{Kearns:2002jh}.
}
\end{figure}

\begin{figure}[t]
\begin{center}
\begin{tabular}{cc}
\begin{minipage}[t]{0.45\textwidth}
\begin{center}
\includegraphics*[bb=131 558 452 778, width=\textwidth]{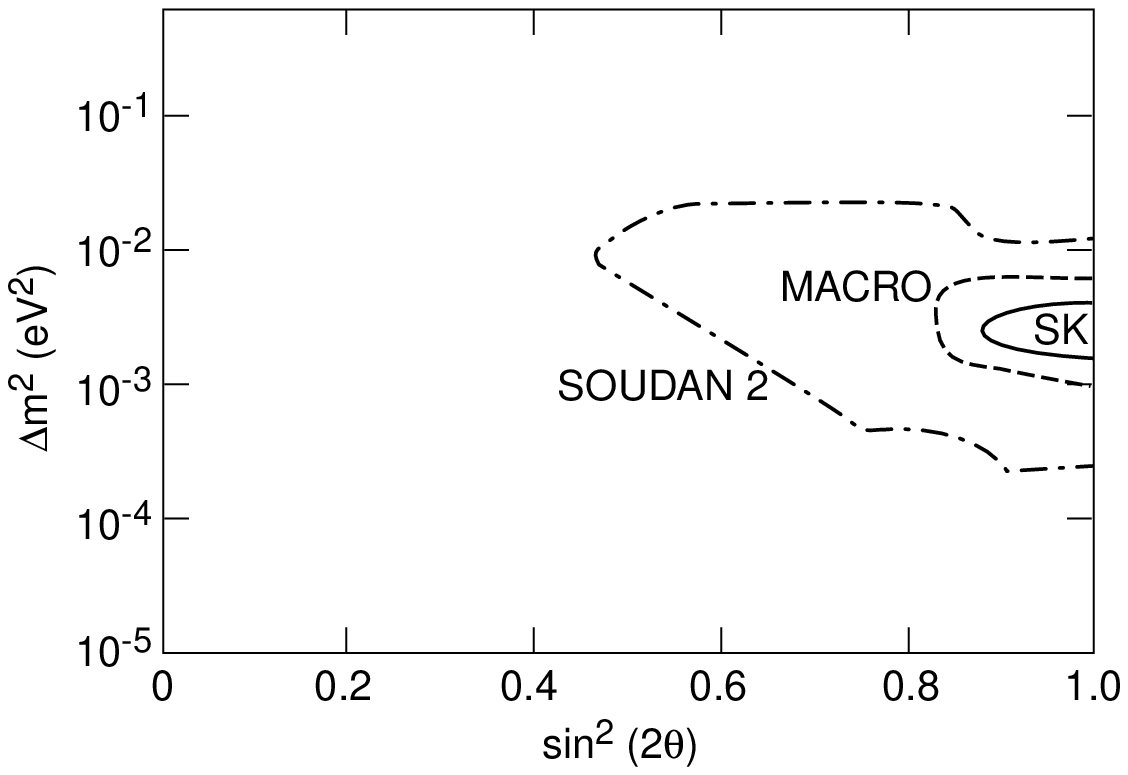}
\end{center}
\end{minipage}
&
\begin{minipage}[t]{0.45\textwidth}
\begin{center}
\includegraphics*[bb=20 145 555 650, width=\textwidth]{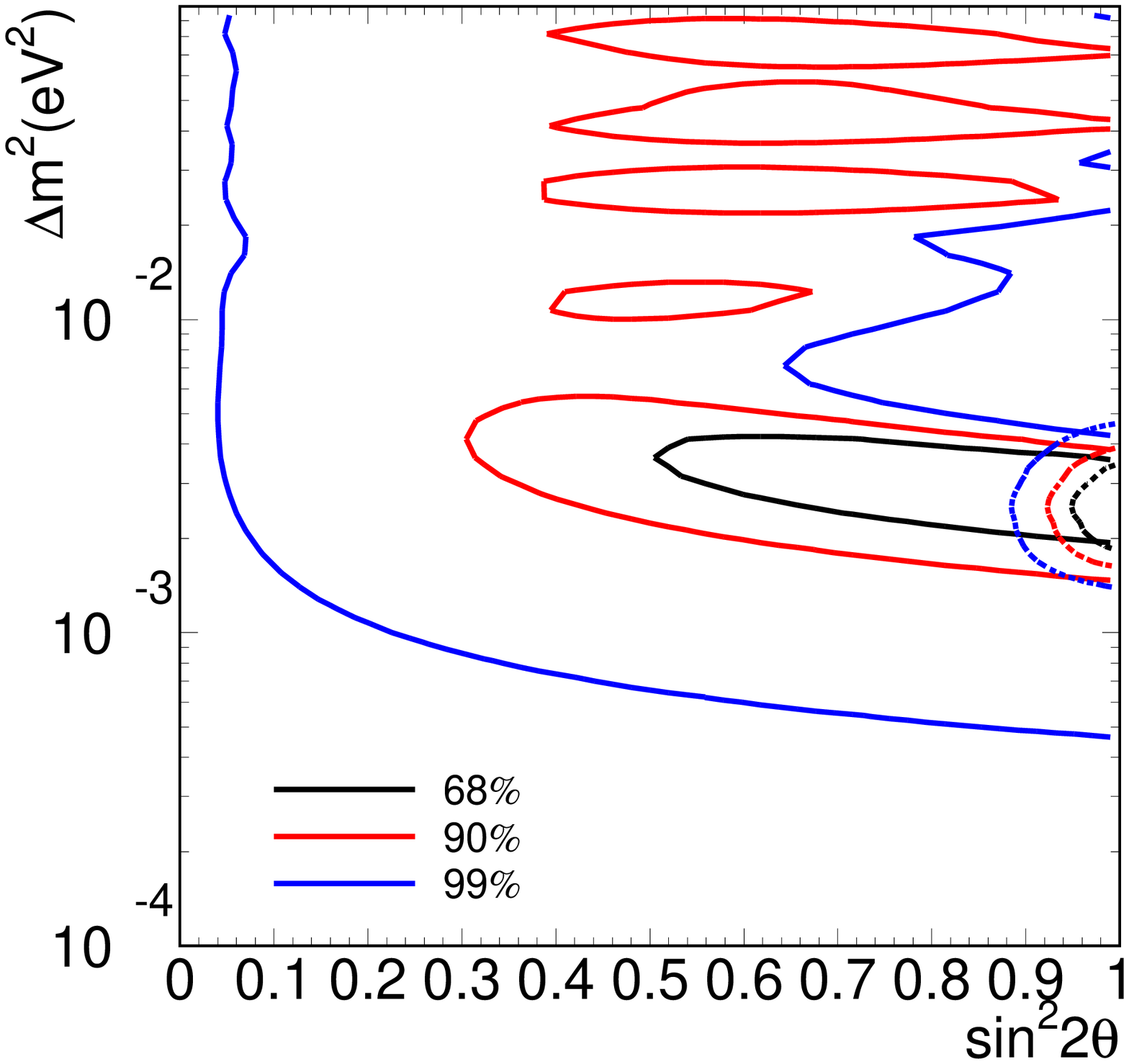}
\end{center}
\end{minipage}
\end{tabular}
\end{center}
\caption{ \label{giacomelli-0201032-f10+k2k-0210030-f09}
Left:
90\% C.L. allowed region contours for $\nu_\mu \to \nu_\tau$
oscillations obtained by the Super-Kamiokande, MACRO and Soudan-2 experiments
\cite{Giacomelli-0201032}.
Right:
Allowed region contours
for $\nu_\mu$ disappearance obtained in the K2K experiment
confronted with the allowed regions for $\nu_\mu \to \nu_\tau$
oscillations obtained in the Super-Kamiokande experiment
\cite{hep-ex/0210030}.
}
\end{figure}

The breakthrough in atmospheric neutrino research
occurred in 1998,
when the Super-Kamiokande Collaboration
\cite{Fukuda:1998mi}
discovered the up-down asymmetry of high-energy events generated
by atmospheric $\nu_\mu$'s,
providing a model independent proof
of atmospheric $\nu_\mu$ disappearance.
Indeed,
on the basis of simple geometrical arguments
the fluxes of upward-going and downward-going
high-energy events generated
by atmospheric $\nu_\mu$'s
should be equal if nothing happens to
neutrinos on their way from the production in the atmosphere
to the detector
(see Ref.~\cite{Kayser:2002ed}).
The last published value of the measured up-down asymmetry is
\cite{Scholberg:1999ar}
\begin{equation}
A_{\nu_\mu}^{\text{up-down}}(\text{SK})
=
\left(
\frac
{ N_{\nu_\mu}^{\text{up}} - N_{\nu_\mu}^{\text{down}} }
{ N_{\nu_\mu}^{\text{up}} + N_{\nu_\mu}^{\text{down}} }
\right)
=
- 0.31 \pm 0.04
\,,
\label{356}
\end{equation}
showing a $7\sigma$ evidence of disappearance
of atmospheric high-energy upward-going
muon neutrinos.
These neutrinos travel a distance from about 2650 to about 12780 km
($ 0.2 < \cos\theta < 1 $, where $\theta$ is the nadir angle of the neutrino trajectory),
whereas the downward-going neutrinos
travel a distance from about 20 to about 100 km
($ -1 < \cos\theta < -0.2 $).
Therefore,
the simplest explanation of
the atmospheric neutrino data
is neutrino oscillations.
The left panel in
Fig.~\ref{sk-scholberg-9905016-f04+sk-wilkes-0212035-combined-allowed}
shows the Super-Kamiokande up-down asymmetry as a function of
momentum for $e$-like and $\mu$-like events
generated,
respectively,
by atmospheric $\nu_e$, $\bar\nu_e$ and
$\nu_\mu$, $\bar\nu_\mu$.
One can see that there is a clear deficit of high-energy
upward going muon neutrinos
with respect to downward-going ones,
which do not have time to oscillate.
On the other hand,
there is no up-down asymmetry at
low energies because also most of the downward-going muon neutrinos
have time to oscillate and a possible asymmetry
is washed out by a poor correlation between
the directions of the incoming neutrino and the observed
charged lepton
(the average angle between the two directions
is $55^\circ$ at $p = 400$ MeV and $20^\circ$ at 1.5
GeV \cite{Fukuda:1998mi}).

At the end of 2002
the long-baseline K2K experiment
\cite{Ahn:2002up}
confirmed the neutrino oscillation
interpretation of the atmospheric neutrino anomaly
observing the disappearance of accelerator $\nu_\mu$'s
with average energy energy $\overline{E} \simeq 1.3 \, \text{GeV}$
traveling 250 km from KEK to the Super-Kamiokande detector
(only 56 of the $80.1 {}^{+6.2}_{-5.4}$
expected events
were observed).

\begin{figure}[t]
\begin{center}
\begin{tabular}{cc}
\begin{minipage}[t]{0.45\textwidth}
\begin{center}
\includegraphics*[bb=42 299 86 509, height=7cm]{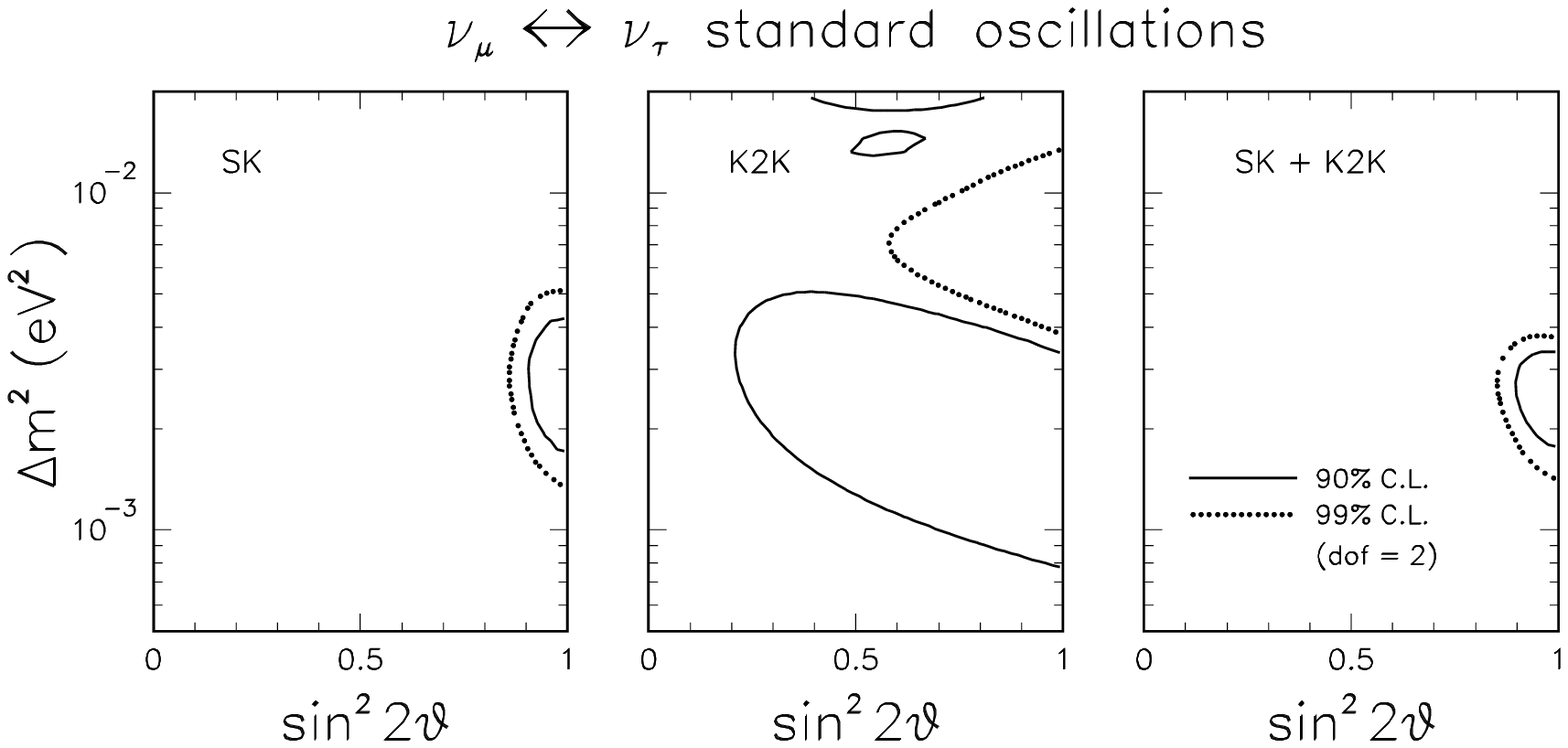}
\includegraphics*[bb=398 299 536 509, height=7cm]{fig/fogli-0303064-f01.eps.pdf.eps}
\end{center}
\end{minipage}
&
\begin{minipage}[t]{0.45\textwidth}
\begin{center}
\includegraphics*[bb=23 20 541 800, height=7cm]{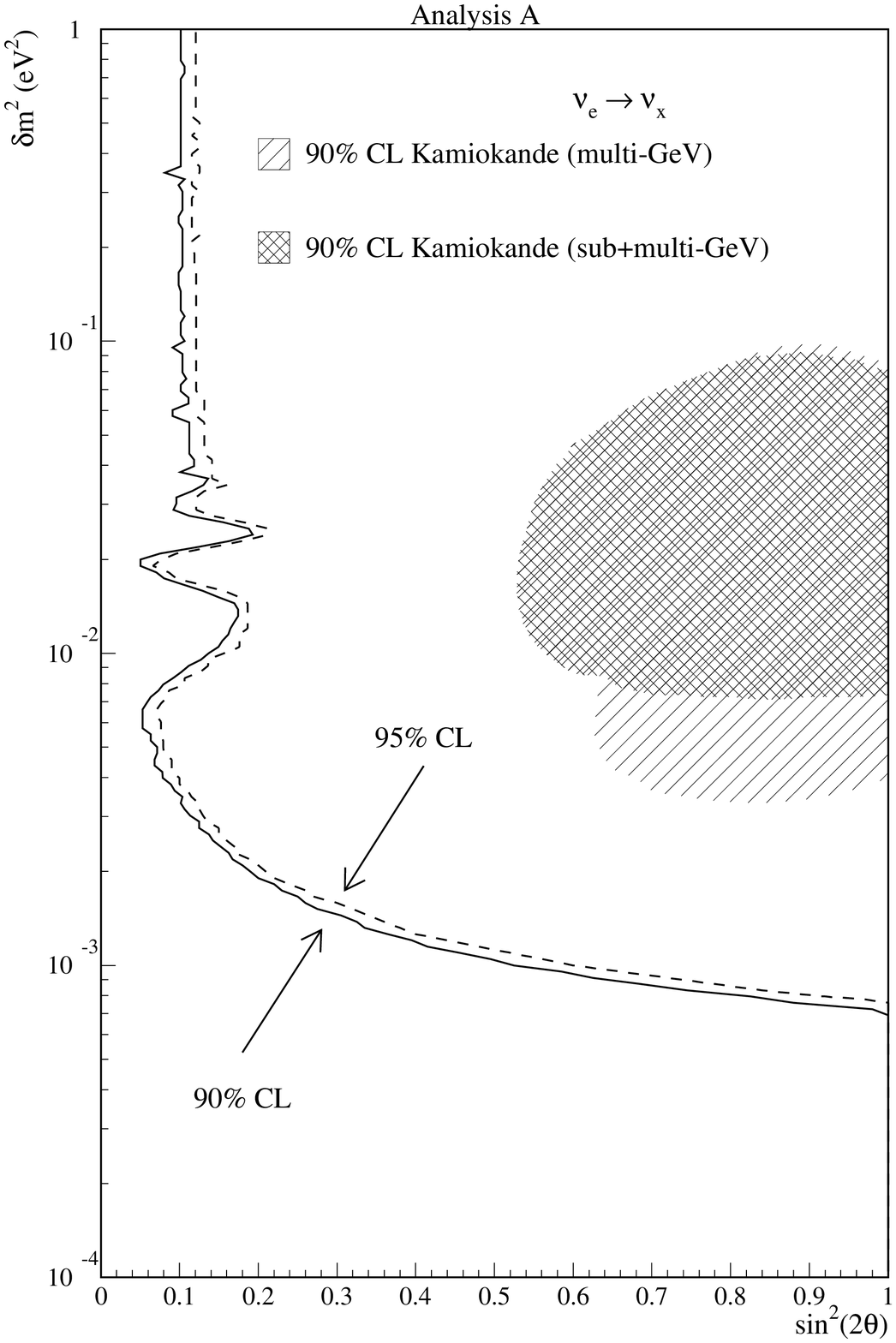}
\end{center}
\end{minipage}
\end{tabular}
\end{center}
\caption{ \label{fogli-0303064-f01+chooz-9907037-f09}
Left:
Allowed region
obtained from the analysis of
Super-Kamiokande atmospheric and K2K data
in terms of
$\nu_\mu\to\nu_\tau$
oscillations \protect\cite{hep-ph/0303064}.
Right:
CHOOZ exclusion curves \cite{Apollonio:1999ae}
confronted with the Kamiokande allowed regions
\cite{Fukuda:1994mc}.
}
\end{figure}

The Super-Kamiokande atmospheric neutrino data and the data of the K2K experiment
are well fitted by
$\nu_\mu \to \nu_\tau$
transitions with the effective two-neutrino mixing parameters
constrained in the ranges
\cite{hep-ph/0303064}
\begin{eqnarray}
&
1.4 \times 10^{-3} \, \text{eV}^2
<
\Delta{m}^2_{\text{ATM}}
<
5.1 \times 10^{-3} \, \text{eV}^2
\,,
&
\label{361}
\\
&
\sin^2 2 \vartheta_{\text{ATM}} > 0.86
\,,
&
\label{362}
\end{eqnarray}
at 99.73\% C.L. ($3\sigma$),
with best-fit values
\begin{equation}
\Delta{m}^{2\,\text{bf}}_{\text{ATM}}
=
2.6 \times 10^{-3} \, \text{eV}^2
\,,
\qquad
\sin^2 2 \vartheta_{\text{ATM}}^{\text{bf}} = 1
\,.
\label{363}
\end{equation}
Hence,
the best-fit effective atmospheric mixing
is maximal.
The right panel in Fig.~\ref{sk-scholberg-9905016-f04+sk-wilkes-0212035-combined-allowed}
shows the region in the
$\sin^2 2 \vartheta_{\text{ATM}}$--$\Delta{m}^2_{\text{ATM}}$
plane for $\nu_\mu\to\nu_\tau$ oscillations
allowed by Super-Kamiokande data
\cite{hep-ex/0212035}.
The left panel in Fig.~\ref{giacomelli-0201032-f10+k2k-0210030-f09}
shows the
90\% C.L. allowed regions for $\nu_\mu \to \nu_\tau$
oscillations obtained in the MACRO and Soudan-2 experiments
confronted with the corresponding region obtained
in the Super-Kamiokande experiment
\cite{Giacomelli-0201032}.
The right panel in Fig.~\ref{giacomelli-0201032-f10+k2k-0210030-f09}
shows the allowed regions
for $\nu_\mu$ disappearance obtained in the K2K experiment
confronted with the allowed regions for $\nu_\mu \to \nu_\tau$
oscillations obtained in the Super-Kamiokande experiment
\cite{hep-ex/0210030}.
The left panel in Fig.~\ref{fogli-0303064-f01+chooz-9907037-f09}
shows the
allowed region
obtained in Ref.~\cite{hep-ph/0303064} from the combined analysis of
Super-Kamiokande atmospheric and K2K data.

Transitions of
atmospheric $\nu_\mu$'s into $\nu_e$'s
or sterile states
are disfavored.
The fraction
$\sin^2\xi$ of atmospheric $\nu_\mu$'s
that transform into sterile $\nu_s$
($\nu_\mu \to \cos\xi \, \nu_\tau + \sin\xi \, \nu_s $)
is limited at 90\% C.L. by \cite{Nakaya:2002ki}
\begin{equation}
\sin^2\xi < 0.19
\,.
\label{eps-0041}
\end{equation}

In the next years the
MINOS \cite{Diwan:2002pu} experiment
will measure with improved precision
the disappearance of muon neutrinos
over a long-baseline of about 730 km.
The
OPERA
\cite{Guler:2000bd}
and ICARUS
\cite{Arneodo:2001tx}
experiments
belonging to the
CERN to Gran Sasso program
(CNGS) \cite{hep-ex/0209082}
are aimed at a direct measurement of
$\nu_\mu\to\nu_\tau$ oscillation
over a similar long-baseline of about 730 km.

\subsection{The reactor experiment CHOOZ}
\label{The reactor experiment CHOOZ}

CHOOZ was a long-baseline
reactor $\nu_e$ disappearance experiment
\cite{Apollonio:1998xe,Apollonio:1999ae,Apollonio:2003gd}
which did not observe any disappearance of electron neutrinos
at a distance of about 1 km from the source.
In spite of such negative result,
the CHOOZ experiment is very important,
because it shows that the oscillations of electron neutrinos at the atmospheric
scale of $\Delta{m}^2$
are small or zero.
This constraint is particularly important in the framework of three-neutrino
mixing,
as will be discussed in Section~\ref{Phenomenology of three-neutrino mixing}.
Therefore,
we briefly review the results of the CHOOZ experiment.

The CHOOZ detector consisted in 5 tons of liquid scintillator
in which neutrinos were revealed through the inverse $\beta$-decay reaction\footnote{
The inverse $\beta$-decay reaction
(\ref{581})
has been used by all experiments aimed at the detection
of reactor electron antineutrinos,
starting from the Cowan and Reines experiment
in 1953 \cite{Reines:1953pu}, in which neutrinos were detected for the first time.
The same reaction is used in the KamLAND experiment discussed in
Section~\ref{Solar neutrino experiments and KamLAND}.
}
\begin{equation}
\bar\nu_e + p \to n + e^+
\,,
\label{581}
\end{equation}
with a threshold
$ E_{\text{th}} = 1.8 \, \text{MeV} $.
The neutrino energy is measured through the positron energy:
$ E = E_{e^+} - 1.8 \, \text{MeV} $.
The detector was located at a distance of about 1 km
from the Chooz power station, which has two pressurized-water reactors.

The ratio of observed and expected number of events
in the CHOOZ experiment is
\begin{equation}
\frac
{N^{\text{CHOOZ}}_{\text{observed}}}
{N^{\text{CHOOZ}}_{\text{expected}}}
=
1.01 \pm 0.04
\,,
\label{382}
\end{equation}
showing no indication of any
electron antineutrino disappearance.
The right panel in Fig.~\ref{fogli-0303064-f01+chooz-9907037-f09}
\cite{Apollonio:1999ae}
shows the CHOOZ exclusion curves
confronted with the Kamiokande allowed regions
for $\nu_\mu\to\nu_e$ transitions
\cite{Fukuda:1994mc}.
The area on the right of the exclusion curves is excluded.
Since the Kamiokande allowed region lies in the excluded area,
the disappearance of muon neutrinos observed in Kamiokande
(and IMB, Super-Kamiokande, Soudan-2 and MACRO)
cannot be due to $\nu_\mu\to\nu_e$ transitions.
Indeed, $\nu_\mu\to\nu_e$ transitions
are also disfavored by Super-Kamiokande data,
which prefer the $\nu_\mu\to\nu_\tau$ channel \cite{Nakaya:2002ki}
(therefore, the Super-Kamiokande collaboration
did not calculate an allowed region for $\nu_\mu\to\nu_e$ transitions
and the CHOOZ collaboration correctly compared
their exclusion curve with the regions
allowed by the results of the Kamiokande experiment).

The results of the CHOOZ experiment
have been confirmed,
albeit with lower accuracy,
by the Palo Verde experiment \cite{Boehm:2001ik}.

\section{Phenomenology of three-neutrino mixing}
\label{Phenomenology of three-neutrino mixing}

The solar and atmospheric evidences of neutrino oscillations
are nicely accommodated in the minimal framework of three-neutrino mixing,
in which the three flavor neutrinos
$\nu_e$,
$\nu_\mu$,
$\nu_\tau$
are unitary linear combinations of
three neutrinos
$\nu_1$,
$\nu_2$,
$\nu_3$
with masses
$m_1$,
$m_2$,
$m_3$,
according to Eq.~(\ref{044}).
As explained in Section~\ref{Neutrino masses and mixing}
this scenario is theoretically motivated by the see-saw mechanism,
which also predicts that
massive neutrinos are Majorana particles.

\begin{figure}[t]
\begin{center}
\begin{tabular}{lcr}
\includegraphics*[bb=181 466 428 775, width=0.25\textwidth]{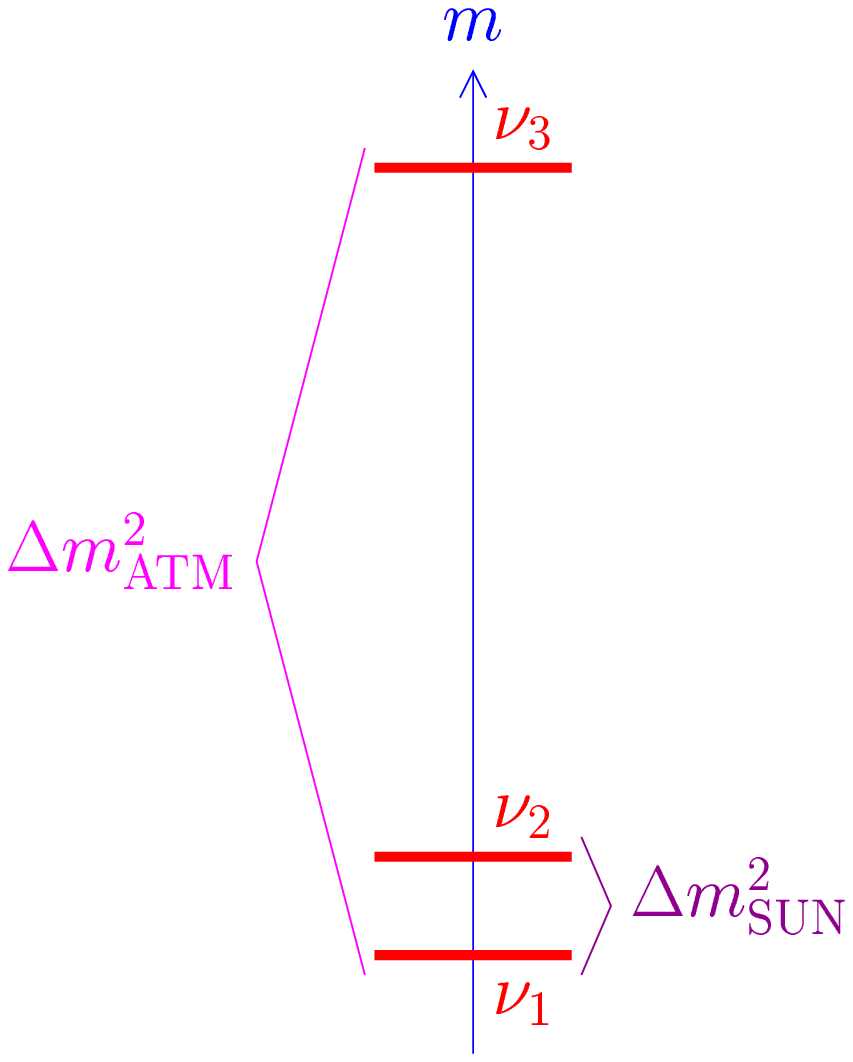}
&
\hspace{0.15\textwidth}
&
\includegraphics*[bb=183 466 432 775, width=0.25\textwidth]{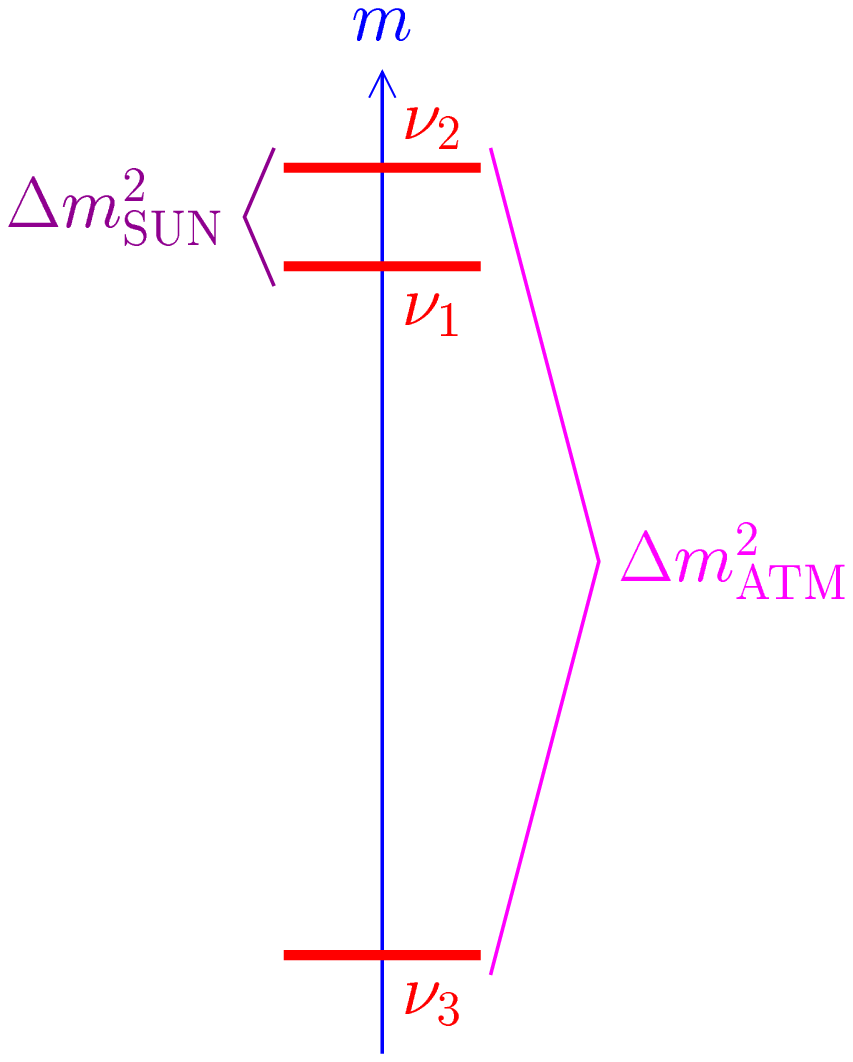}
\\
\textsf{normal}
&
&
\textsf{inverted}
\end{tabular}
\end{center}
\caption{ \label{eps-3nu}
The two three-neutrino schemes allowed by the hierarchy
$\Delta{m}^2_{\text{SUN}} \ll \Delta{m}^2_{\text{ATM}}$.
}
\end{figure}

\subsection{Three-neutrino mixing schemes}
\label{Three-neutrino mixing schemes}

\begin{figure*}
\begin{minipage}[t]{0.47\textwidth}
\begin{center}
\includegraphics*[bb=120 427 463 750, width=0.99\textwidth]{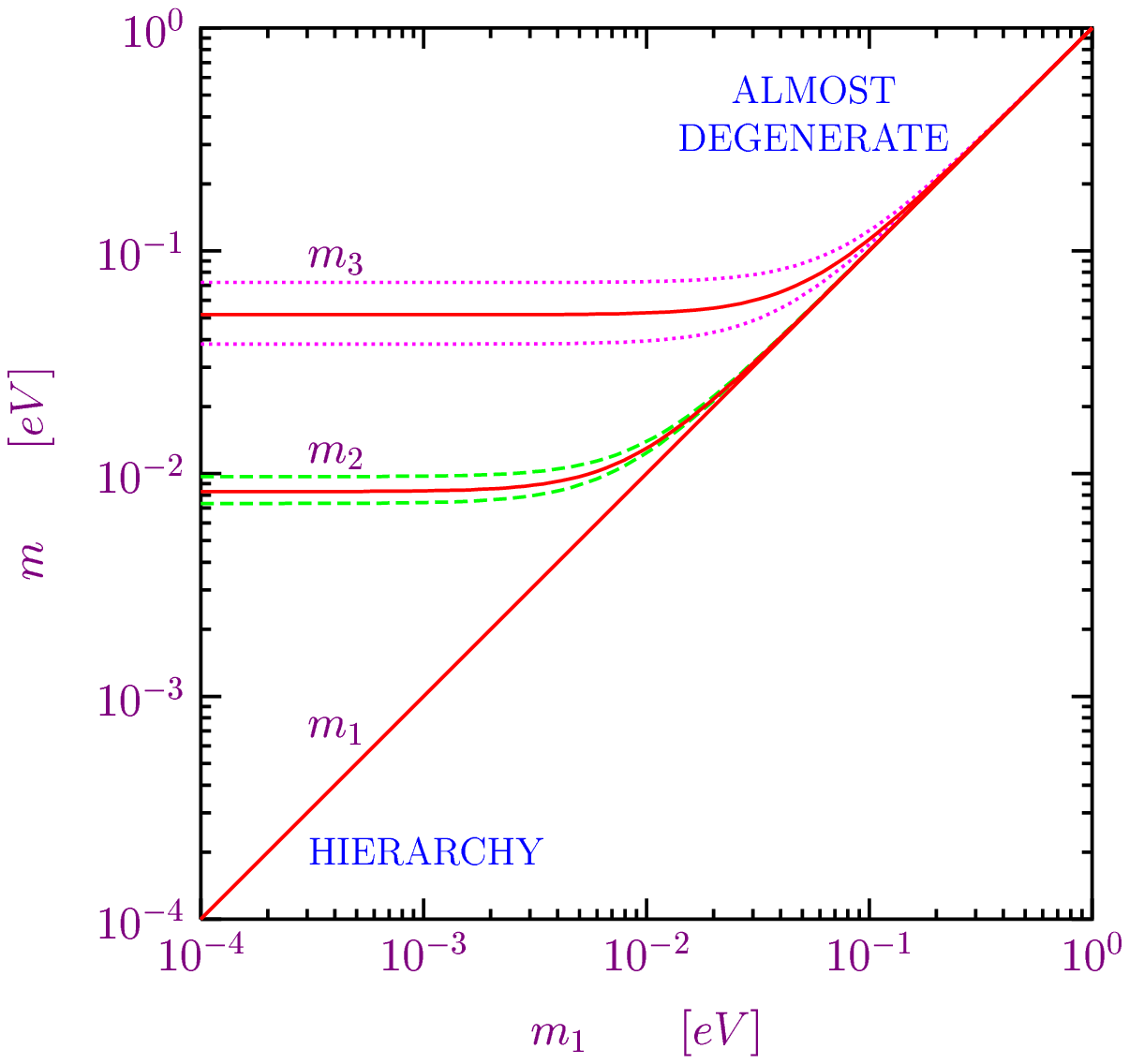}
\end{center}
\end{minipage}
\hfill
\begin{minipage}[t]{0.47\textwidth}
\begin{center}
\includegraphics*[bb=120 427 463 750, width=0.99\textwidth]{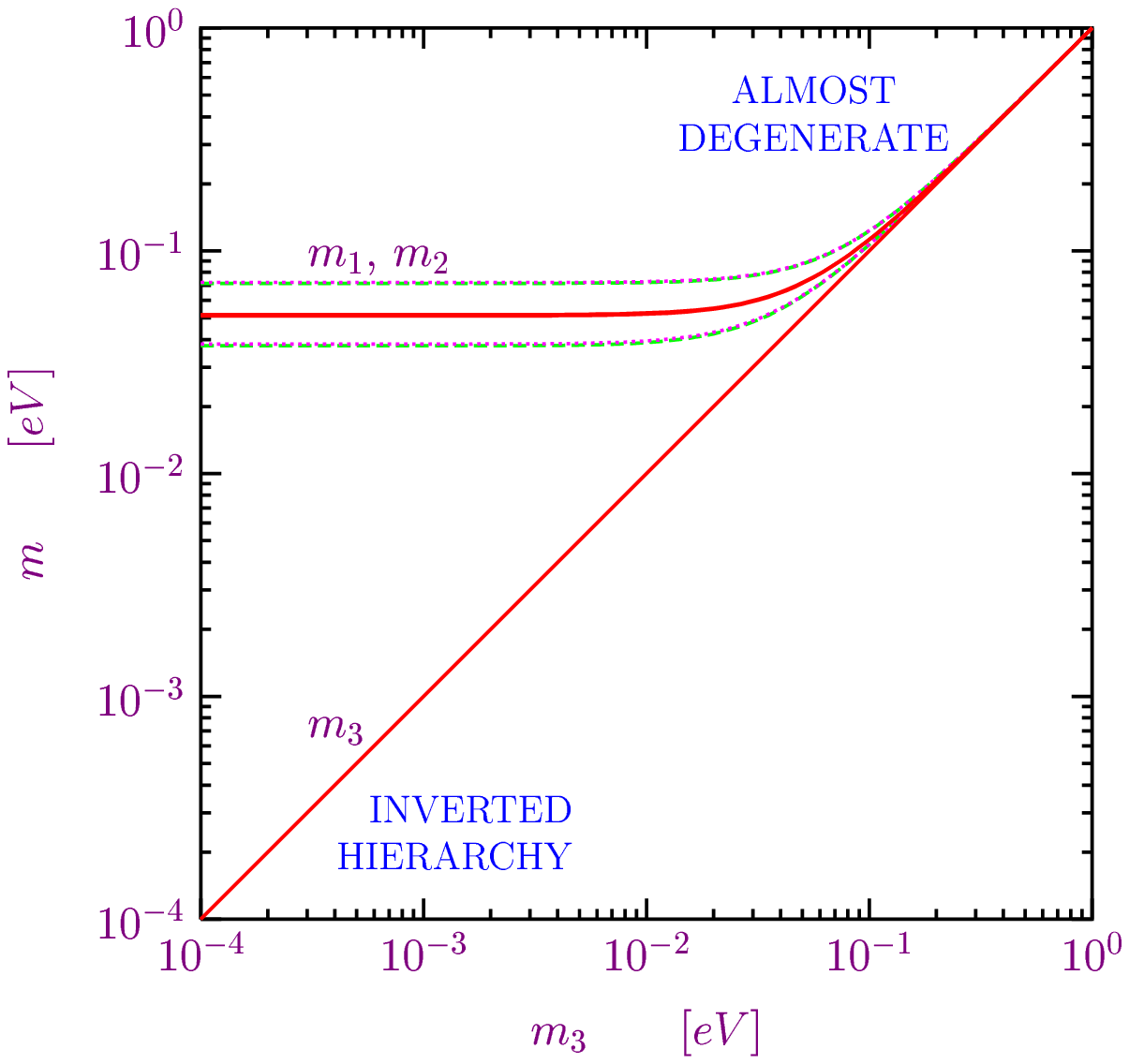}
\end{center}
\end{minipage}
\caption{ \label{3ma}
Allowed ranges for the neutrino masses as functions
of the lightest mass $m_1$ and $m_3$ in the normal and inverted
three-neutrino scheme, respectively.
}
\end{figure*}

Figure~\ref{eps-3nu}
shows the two three-neutrino schemes
allowed by the observed hierarchy
of squared-mass differences,
$\Delta{m}^2_{\text{SUN}} \ll \Delta{m}^2_{\text{ATM}}$,
with
the massive neutrinos labeled in order to have
\begin{equation}
\Delta{m}^2_{\text{SUN}}
=
\Delta{m}^2_{21}
\,,
\quad
\Delta{m}^2_{\text{ATM}}
\simeq
|\Delta{m}^2_{31}|
\simeq
|\Delta{m}^2_{32}|
\,.
\label{eps-006}
\end{equation}
The two schemes
in Fig.~\ref{eps-3nu} are usually called
``normal''
and
``inverted'',
because in the normal scheme the smallest
squared-mass difference is generated by the two lightest neutrinos
and a natural neutrino mass hierarchy can be realized\footnote{
The absolute scale of neutrino masses
is not determined by the observation of
neutrino oscillations,
which
depend
only on the differences of the squares of neutrino masses.
}
if
$m_1 \ll m_2$,
whereas in the inverted scheme the smallest
squared-mass difference is generated
by the two heaviest neutrinos,
which are almost degenerate for any value of the lightest neutrino mass $m_3$.
This is shown in Fig.~\ref{3ma},
where we have depicted the allowed ranges
(between the dashed and dotted lines)
for the neutrino masses
obtained from the allowed values of
$\Delta{m}^2_{\text{SUN}}$
in Eq.~(\ref{321})
and
$\Delta{m}^2_{\text{ATM}}$
in Eq.~(\ref{361}),
as functions of the lightest mass
in the normal and inverted schemes.
The solid lines correspond to the best fit values of
$\Delta{m}^2_{\text{SUN}}$
and
$\Delta{m}^2_{\text{ATM}}$
in Eqs.~(\ref{323}) and (\ref{363}),
respectively.
One can see that at least two neutrinos have masses
larger than about
$7 \times 10^{-3} \, \text{eV}$.

In the case of three-neutrino mixing there are no light
sterile neutrinos,
in agreement with the absence of any indication in favor
of active--sterile transitions in
both solar and atmospheric neutrino experiments.
Let us however emphasize
that three-neutrino mixing cannot explain the
indications in favor of short-baseline
$\bar\nu_\mu\to\bar\nu_e$
transitions observed in the LSND experiment
\cite{Aguilar:2001ty},
which are presently under investigation in the
MiniBooNE experiment
\cite{hep-ex/0210020}.

Let us now discuss the current information on
the three-neutrino mixing matrix $U$.
In solar neutrino experiments
$\nu_\mu$ and $\nu_\tau$ are
indistinguishable,
because the energy is well below $\mu$ and $\tau$ production
and $\nu_\mu$, $\nu_\tau$ can be detected only through flavor-blind
neutral-current interactions.
Hence,
solar neutrino oscillations,
as well as the oscillations in the KamLAND experiment,
depend only on the absolute value of the elements in the first row
of the mixing matrix,
$|U_{e1}|$,
$|U_{e2}|$,
$|U_{e3}|$
which regulates $\nu_e$ and $\bar\nu_e$ disappearance.
Indeed,
the survival probability of solar electron neutrinos
can be written as
\cite{Shi:1992zw}
\begin{equation}
\overline{P}_{\nu_e\to\nu_e}^{\,\text{sun}}
=
\left(1-|U_{e3}|^{2}\right)^{2}
\overline{P}_{\nu_e\to\nu_e}^{\,\text{sun, (1,2)}}
+
|U_{e3}|^{4}
\,,
\label{201}
\end{equation}
where
$ \overline{P}_{\nu_e\to\nu_e}^{\,\text{sun, (1,2)}} $
is the two-neutrino survival probability in matter (\ref{154})
calculated with the charged-current matter potential
$V_{\text{CC}}$
multiplied by
$(1-|U_{e3}|^2)$
and
$ \vartheta = \vartheta_{12} $
in the parameterization (\ref{046})
of the mixing matrix.

The hierarchy $\Delta{m}^2_{\text{SUN}} \ll \Delta{m}^2_{\text{ATM}}$
implies that neutrino oscillations generated by
$\Delta{m}^2_{\text{ATM}}$
depend only on the absolute value of the elements in the last column
of the mixing matrix,
$|U_{e3}|$,
$|U_{\mu3}|$,
$|U_{\tau3}|$
because $m_1$ and $m_2$ are indistinguishable.
Indeed,
taking into account also the matter effects in the earth,
the evolution equation of the neutrino amplitudes
is given by Eq.~(\ref{133}) with
$ \Delta{m}^2_{k1} \simeq \Delta{m}^2_{31} \delta_{k3} $,
leading to
\begin{equation}
i \, \frac{\text{d}}{\text{d}x} \,
\psi_{\alpha\beta}(x)
=
\sum_\rho
\left(
U_{\beta 3}
\,
\frac{\Delta{m}^2_{31}}{2E}
\,
U_{\rho 3}^*
+
\delta_{\beta e} \, \delta_{\rho e} \, V_{\text{CC}}
\right)
\psi_{\alpha\rho}(x)
\,,
\label{202}
\end{equation}
which clearly depends only on the elements
$U_{e3}$,
$U_{\mu3}$ and
$U_{\tau3}$
of the mixing matrix.
In order to further demonstrate that only the absolute values
$|U_{e3}|$,
$|U_{\mu3}|$,
$|U_{\tau3}|$
are relevant,
we notice that
in the parameterization (\ref{046})
of the mixing matrix we have
$ U_{\beta 3} = |U_{\beta 3}| e^{-i\varphi_{13}\delta_{\beta e}} $,
and
Eq.~(\ref{202})
can be written as
\begin{equation}
i \, \frac{\text{d}}{\text{d}x} \,
\psi_{\alpha\beta}(x)
=
e^{-i\varphi_{13}\delta_{\beta e}}
\sum_\rho
\left(
|U_{\beta 3}|
\,
\frac{\Delta{m}^2_{31}}{2E}
\,
|U_{\rho 3}|
+
\delta_{\beta e} \, \delta_{\rho e} \, V_{\text{CC}}
\right)
e^{i\varphi_{13}\delta_{\rho e}}
\,
\psi_{\alpha\rho}(x)
\,.
\label{203}
\end{equation}
Since the flavor transition probabilities
depend on the squared absolute value of the
flavor amplitudes
(see Eq.~(\ref{132})),
we can change arbitrarily the phases of the flavor amplitudes.
Making the change of phase
\begin{equation}
\psi_{\alpha\beta}(x)
\to
e^{-i\varphi_{13}\delta_{\beta e}}
\,
\psi_{\alpha\beta}(x)
\,,
\label{204}
\end{equation}
we obtain the evolution equation
\begin{equation}
i \, \frac{\text{d}}{\text{d}x} \,
\psi_{\alpha\beta}(x)
=
\sum_\rho
\left(
|U_{\beta 3}|
\,
\frac{\Delta{m}^2_{31}}{2E}
\,
|U_{\rho 3}|
+
\delta_{\beta e} \, \delta_{\rho e} \, V_{\text{CC}}
\right)
\psi_{\alpha\rho}(x)
\,,
\label{205}
\end{equation}
which depends\footnote{
A simpler way to obtain the same result
is to adopt a parameterization of the mixing matrix
in which the Dirac phase is associated with the mixing angle
$\vartheta_{12}$,
which does not contribute to the evolution equation (\ref{202})
(see Ref.~\cite{GKM-atm-98}).
}
only on
$|U_{e3}|$,
$|U_{\mu3}|$ and
$|U_{\tau3}|$.

The only connection between
solar and atmospheric oscillations
is due to $|U_{e3}|$.
Therefore,
any information on the value of $|U_{e3}|$
is of crucial importance.

The key experiment for the determination of
$|U_{e3}|$
has been the
CHOOZ long-baseline reactor
$\bar\nu_e$ disappearance experiment
\cite{Apollonio:1998xe,Apollonio:1999ae,Apollonio:2003gd},
which did not observe any disappearance at a distance of about 1 km from the reactor source
(see Section~\ref{The reactor experiment CHOOZ}).
The negative result of the CHOOZ experiment,
confirmed by the Palo Verde experiment \cite{Boehm:2001ik},
implies that the oscillations of electron neutrinos
at the atmospheric scale
are very small or even zero.
The CHOOZ bound on
the effective mixing angle
(see Refs.~\cite{Bilenky:1998tw,BGG-review-98})
\begin{equation}
\sin^2 2\vartheta_{\text{CHOOZ}}
=
4 \, |U_{e3}|^2 \left( 1 - |U_{e3}|^2 \right)
\label{251}
\end{equation}
implies that
$|U_{e3}|$
is small:
\begin{equation}
|U_{e3}|^2 < 5 \times 10^{-2}
\,,
\label{252}
\end{equation}
at 99.73\% C.L.
\cite{Fogli:2002pb}.
Therefore,
solar and atmospheric neutrino oscillations are practically decoupled
\cite{Bilenky:1998tw}
and the effective mixing angles in
solar, atmospheric and CHOOZ experiments
can be related to the elements of the three-neutrino mixing matrix by
(see also Ref.~\cite{hep-ph/0212142})
\begin{equation}
\sin^2\vartheta_{\text{SUN}}
=
\frac{|U_{e2}|^2}{1-|U_{e3}|^2}
\,,
\qquad
\sin^2\vartheta_{\text{ATM}}
=
|U_{\mu3}|^2
\,,
\qquad
\sin^2\vartheta_{\text{CHOOZ}}
=
|U_{e3}|^2
\,.
\label{eps-011}
\end{equation}
Taking into account the best-fit values of
$\vartheta_{\text{SUN}}$ and $\vartheta_{\text{ATM}}$
in Eqs.~(\ref{323}) and (\ref{363}),
respectively,
and
\begin{equation}
\sin^2\vartheta_{\text{CHOOZ}}^{\text{bf}}
=
0
\,,
\label{254}
\end{equation}
the best-fit value for the mixing matrix $U$ is
\begin{equation}
U_{\text{bf}}
\simeq
\left( \begin{smallmatrix}
 0.84 &  0.55 & 0.00 \\
-0.39 &  0.59 & 0.71 \\
 0.39 & -0.59 & 0.71
\end{smallmatrix} \right)
\,.
\label{eps-012}
\end{equation}
Using the 99.73\% C.L. allowed ranges for
$\vartheta_{\text{SUN}}$, $\vartheta_{\text{ATM}}$ and $\vartheta_{\text{CHOOZ}}$
given by Eqs.~(\ref{322}), (\ref{362}) and (\ref{252}),
respectively,
we have reconstructed the allowed ranges for the
elements of the mixing matrix:
\begin{equation}
|U|
\simeq
\left( \begin{smallmatrix}
0.76-0.88 & 0.47-0.62 & 0.00-0.22 \\
0.09-0.62 & 0.29-0.79 & 0.55-0.85 \\
0.11-0.62 & 0.32-0.80 & 0.51-0.83
\end{smallmatrix} \right)
\,.
\label{eps-013}
\end{equation}
Such mixing matrix,
with all elements large except $U_{e3}$,
is called ``bilarge''.
It is very different from the quark mixing matrix,
in which mixing is very small.
This difference is an important
piece of information for our understanding
of the physics beyond the Standard Model,
which presumably involves some sort of quark-lepton unification.

An important open problem
is the determination of the
absolute values of neutrino masses.
The most sensitive known ways to probe the
absolute values of neutrino masses
are
the observation of the end-point part of
the electron spectrum in Tritium $\beta$-decay,
the observation of large-scale structures
in the early universe
and
the search for neutrinoless double-$\beta$ decay,
if neutrinos are Majorana particles
(we do not consider here the interesting possibility
to determine neutrino masses through the
observation of supernova neutrinos;
see Ref.~\cite{Bilenky:2002aw} and references therein).

\begin{figure*}
\begin{minipage}[t]{0.47\textwidth}
\begin{center}
\includegraphics*[bb=119 427 463 750, width=0.99\textwidth]{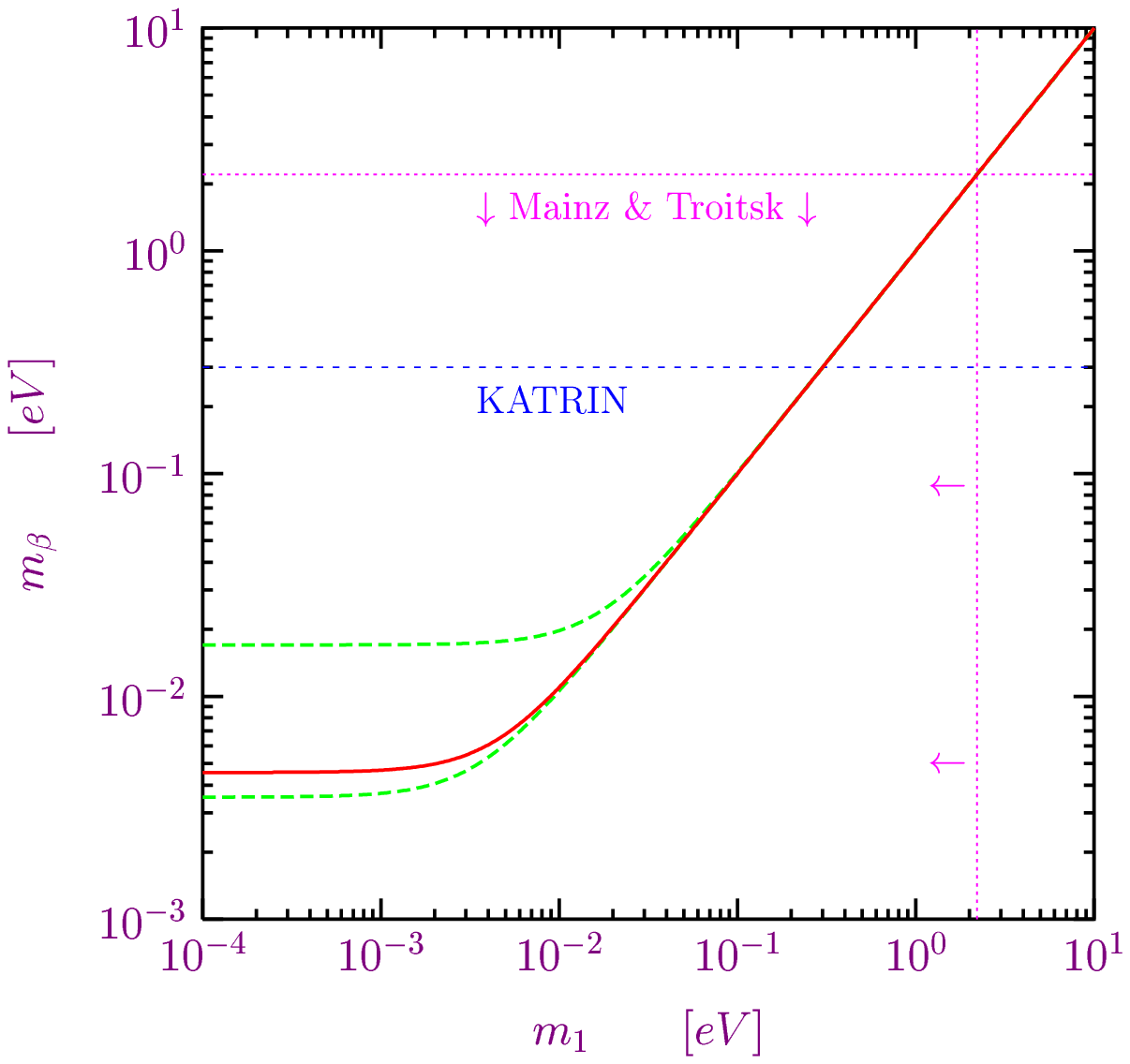}
\end{center}
\end{minipage}
\hfill
\begin{minipage}[t]{0.47\textwidth}
\begin{center}
\includegraphics*[bb=119 427 463 750, width=0.99\textwidth]{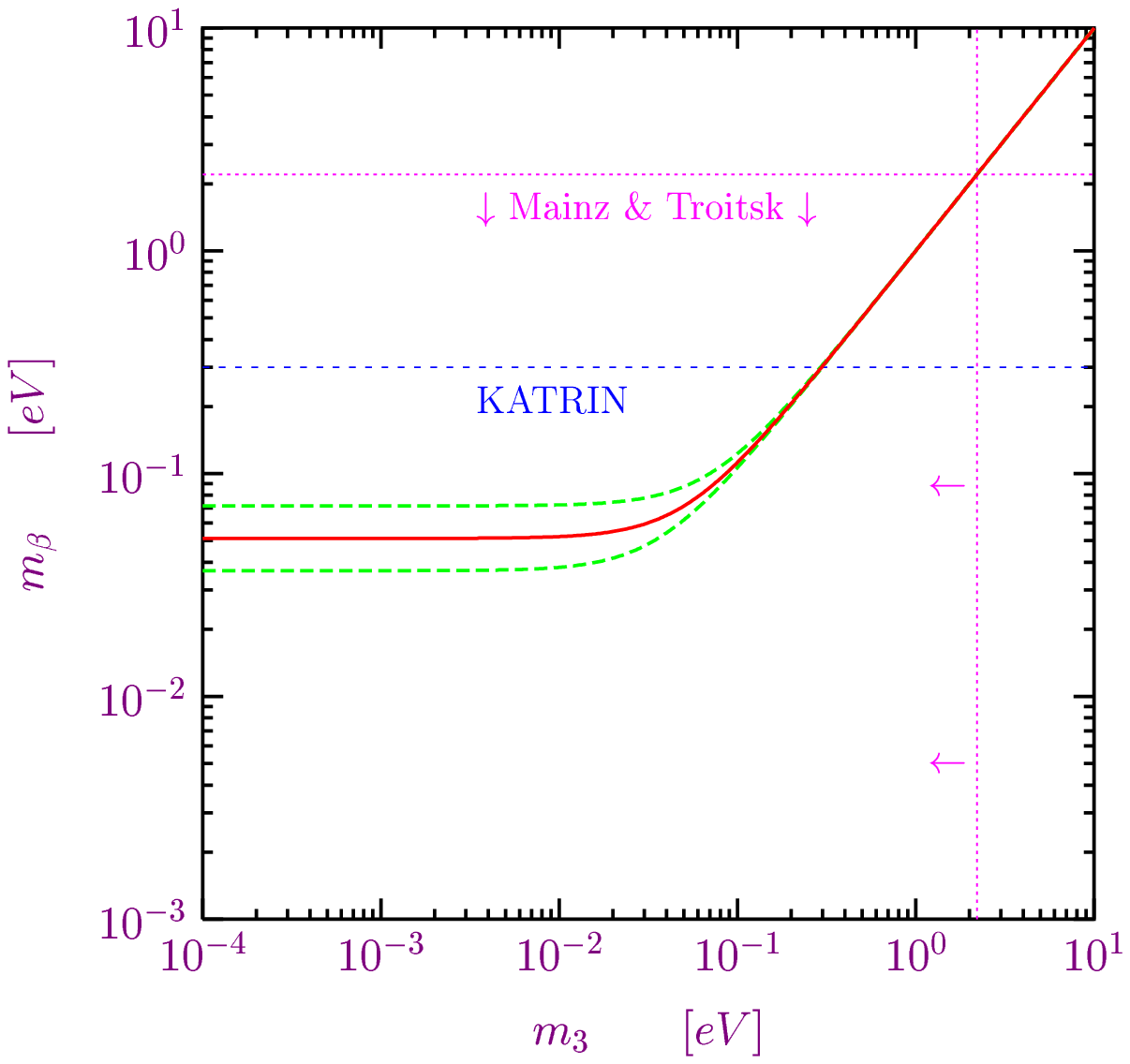}
\end{center}
\end{minipage}
\caption{ \label{mb}
Effective neutrino mass $m_\beta$
in Tritium $\beta$-decay experiments as a function
of the lightest mass $m_1$ and $m_3$ in the normal and inverted
three-neutrino scheme, respectively.
}
\end{figure*}

\subsection{Tritium $\beta$-decay}
\label{Tritium beta-decay}

Up to now,
no indication of a neutrino mass has been found in
Tritium $\beta$-decay experiments,
leading to the 95\% C.L. upper limit
\cite{hep-ex/0210050}
\begin{equation}
m_\beta < 2.2 \, \text{eV}
\label{599}
\end{equation}
on the effective mass
\begin{equation}
m_\beta = \sqrt{ \sum_k |U_{ek}|^2 m_k^2 }
\,,
\label{598}
\end{equation}
obtained in the Mainz \cite{Weinheimer:1999tn}
and Troitsk \cite{Lobashev:1999tp} experiments.
After 2007, the KATRIN experiment
\cite{hep-ex/0109033}
will explore $m_\beta$ down to about
$0.2-0.3 \, \text{eV}$.
Figure~\ref{mb} shows the allowed range (between the dashed lines)
for $m_\beta$
obtained from the 99.73\% C.L. allowed values of the oscillation parameters
in Eqs.~(\ref{321}), (\ref{322}), (\ref{361}), (\ref{362}),
as a function of the lightest mass
in the normal and inverted three-neutrino schemes.
The solid line corresponds to the best fit values of the oscillation parameters
in Eqs.~(\ref{323}) and (\ref{363}).
One can see that in the normal scheme with a mass hierarchy
$m_\beta$
has a value between about
$3 \times 10^{-3} \, \text{eV}$
and
$2 \times 10^{-2} \, \text{eV}$,
whereas in the inverted scheme
$m_\beta$
is larger than about
$3 \times 10^{-2} \, \text{eV}$.
Therefore,
if in the future it will be possible to constraint
$m_\beta$
to be smaller than about
$3 \times 10^{-2} \, \text{eV}$,
a normal hierarchy of neutrino masses will be established.

From Figs.~\ref{3ma} and \ref{mb}
it is clear that the bound (\ref{599})
can be saturated only if the three neutrino masses are almost
degenerate.
In this case the dependence of $m_k$ on the index $k$
can be neglected in Eq.~(\ref{598}), leading to
$ m_\beta \simeq m_k $.
Therefore,
the upper limit for each mass is the same as the one on
$m_\beta$ in Eq.~(\ref{599}): at 95\% C.L.
\begin{equation}
m_k < 2.2 \, \text{eV}
\,.
\label{597}
\end{equation}

\subsection{Cosmological bounds on neutrino masses}
\label{Cosmological bounds on neutrino masses}

In the early hot universe
neutrinos are in equilibrium in the primeval plasma through the weak interaction reactions
$
\nu \bar\nu \leftrightarrows e^+ e^-
$,
$
\overset{\scriptscriptstyle(-)}{\nu} e \leftrightarrows \overset{\scriptscriptstyle(-)}{\nu} e
$,
$
\overset{\scriptscriptstyle(-)}{\nu} N \leftrightarrows \overset{\scriptscriptstyle(-)}{\nu} N
$,
$
\nu_e n \leftrightarrows p e^-
$,
$
\bar\nu_e p \leftrightarrows n e^+
$,
$
n \leftrightarrows p e^- \bar\nu_e
$.
As the universe expands and cools,
the rate of weak interactions decreases.
When the temperature of the Universe
goes below
$T \simeq 1 \, \text{MeV}$,
the mean neutrino free path becomes larger than the
horizon\footnote{
The horizon is the distance traveled by light from the beginning of the universe.
}
and
neutrinos practically cease to interact with the plasma.
At $T \lesssim 0.5 \, \text{MeV}$
electron and positron in the plasma annihilate into photons,
increasing the photon temperature $T_\gamma$ with
respect to the neutrino temperature $T_\nu$
by a factor $(11/4)^{1/3}$,
easily calculated from entropy conservation
(see Refs.~\cite{Mohapatra:1998rq,CWKim-book,Dolgov:2002wy}).
From the well measured temperature of the
Cosmic Microwave Background Radiation (CMBR)
$ T_\gamma = 2.725 \pm 0.001 \, \text{K} $,
we infer the neutrino temperature
$ T_\nu = 1.945 \pm 0.001 \, \text{K}$,
and
$ k \, T_\nu = 1.676 \pm 0.001 \times 10^{-4} \, \text{eV} $.
As we have seen in Section~\ref{Three-neutrino mixing schemes},
at least two neutrinos have masses
larger than about
$7 \times 10^{-3} \, \text{eV}$
(see Fig.~\ref{3ma}).
Hence,
at least two massive neutrinos
in the present relic neutrino background are non-relativistic.
The number density of relic non-relativistic neutrinos
can be calculated from the Fermi-Dirac distribution
(see Refs.~\cite{Mohapatra:1998rq,CWKim-book,Dolgov:2002wy}):
\begin{equation}
n_{\nu_k,\bar\nu_k}
=
\frac{3}{2} \, \frac{\zeta(3)}{\pi^2} T_\nu^3
\simeq
0.1827 \, T_\nu^3
\simeq
112 \, \text{cm}^{-3}
\,,
\label{601}
\end{equation}
with
$ \zeta(3) \simeq 1.202 $.
Their contribution to the present density of the universe
(normalized to the critical density
$
\rho_c
=
3 H^2 / 8 \pi G_{\text{N}}
$,
where
$H$ is the Hubble constant
and
$G_{\text{N}}$
is the Newton constant)
is given by
\begin{equation}
\Omega_k
=
\frac{ n_{\nu_k,\bar\nu_k} \, m_k }{ \rho_c }
\simeq
\frac{1}{h^2}
\frac{ m_k }{ 94.14 \, \text{eV} }
\,,
\label{602}
\end{equation}
where $h$ is the value of the Hubble constant
in units of $100 \, \text{km} \, \text{s}^{-1} \, \text{Mpc}^{-1}$
(the current determination of $h$
from a global fit of cosmological data is $h = 0.71 {}^{+0.04}_{-0.03}$
\cite{Spergel:2003cb}).
The total contribution of relic neutrinos to
the present density of the universe is given by
\cite{Gershtein:1966gg,Cowsik:1972gh}
\begin{equation}
\Omega_\nu \, h^2
=
\frac{ \sum_k m_k }{ 94.14 \, \text{eV} }
\,.
\label{603}
\end{equation}
It is clear that,
just from the need to avoid overclosing the Universe,
the sum of neutrino masses has to be lighter than
about 100 eV.
If one further takes $h \lesssim 0.8$
and
$\Omega_\nu \lesssim 0.1$,
as indicated by astronomical data,
one gets a quite restrictive upper bound of about 6 eV
for the sum of neutrino masses.

An even stronger bound on the sum of neutrino masses
follows from more sophisticated calculations of
structure formation in the early universe.
Neutrinos with masses of the order of 1 eV or lighter
constitute what is called ``hot dark matter'',
which is dark matter that is now non-relativistic,
but was relativistic at the time of matter-radiation equality,
when the contribution of matter and radiation to the density of the universe
was equal.
Since the radiation energy density scales as $T^4$ and matter
energy density scale as $T^3$,
matter started to dominate the density of the universe
and structures started to form
after matter-radiation equality.
However,
hot dark matter particles did not participate to the
beginning of structure formation at matter-radiation equality,
but streamed freely until they become non relativistic.
Hence, neutrinos contribute only to the formation of
structures with size given by the
free-streaming distance traveled by neutrinos until
they become non relativistic.
The formation of structures on smaller scales
is suppressed with respect to
a universe without hot dark matter.
The absence of such suppression in
the present astronomical observations of
large scale structures (LSS) in the universe
allow to put a strong upper bound
on the sum of neutrino masses
(see Refs.~\cite{Hu:1998mj,Wang:2001gy,Hannestad:2002cn}).

The recent high-precision CMBR data of the WMAP satellite
\cite{Bennett:2003bz}
combined with the LSS data of the 2dF Galaxy Redshift Survey (2dFGRS)
\cite{Colless:2003wz}
and other astronomical data
(see Ref.~\cite{Spergel:2003cb})
allowed the WMAP collaboration to derive the impressive bound
\begin{equation}
\Omega_\nu h^2 < 0.0076
\,,
\label{604}
\end{equation}
with 95\% confidence,
which, using Eq.~(\ref{603}), yields
\begin{equation}
\sum_k m_k < 0.71 \, \text{eV}
\,.
\label{605}
\end{equation}

From the smallness of the squared-mass differences
implied by solar and atmospheric neutrino data
(see Eqs.~(\ref{321}), (\ref{361}) and (\ref{eps-006})),
it is clear that the bound (\ref{605}) can be saturated
only if the three neutrino masses are almost degenerate.
Therefore,
the upper limit on each neutrino mass
is one third of the bound in Eq.(\ref{605}):
\begin{equation}
m_k < 0.23 \, \text{eV}
\,.
\label{606}
\end{equation}
This impressive limit is one order of magnitude
more restrictive than
the limit (\ref{597})
obtained in Tritium experiments,
reaching already the level of sensitivity of the
future KATRIN experiment.
Let us emphasize,
however,
that the KATRIN experiment is important in order to probe
the neutrino masses in a model-independent way.
Indeed,
the cosmological bound relies on several assumptions
on the cosmological model
and
some controversial data
(see the discussion in Ref.~\cite{astro-ph/0310133}
and Ref.~\cite{astro-ph/0304237} for an alternative model).
Using only the WMAP and 2dFGRS data, the author of Ref.~\cite{astro-ph/0303076}
found the 95\% confidence limit
\begin{equation}
\sum_k m_k < 2.12 \, \text{eV}
\,.
\label{607}
\end{equation}
Adding also the Hubble Space Telescope
determination of $h$,
the authors of Ref.~\cite{astro-ph/0303089}
obtained the 95\% confidence limit
\begin{equation}
\sum_k m_k < 1.1 \, \text{eV}
\,.
\label{608}
\end{equation}

\begin{figure*}
\begin{minipage}[t]{0.47\textwidth}
\begin{center}
\includegraphics*[bb=120 427 463 750, width=0.99\textwidth]{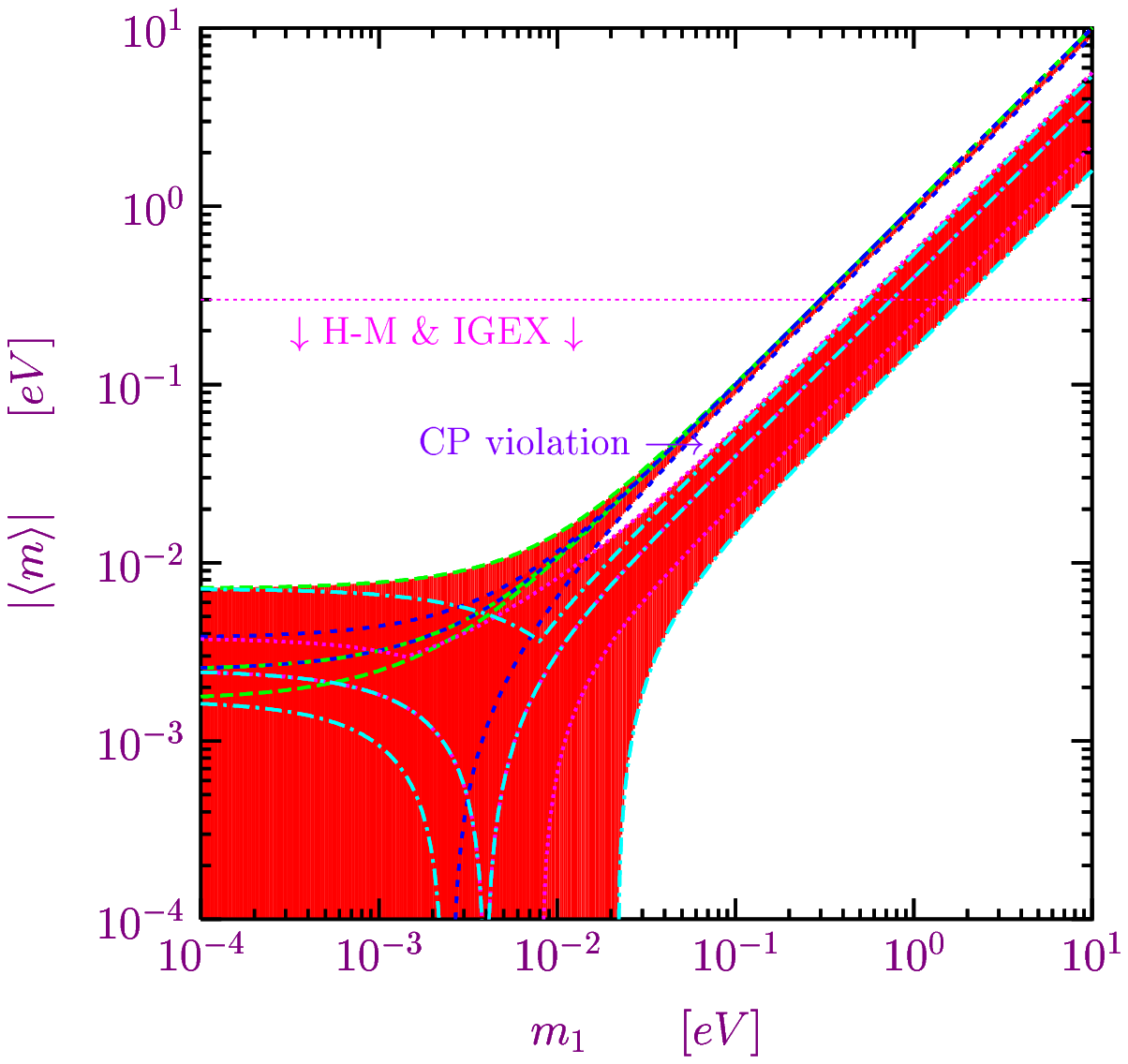}
\end{center}
\end{minipage}
\hfill
\begin{minipage}[t]{0.47\textwidth}
\begin{center}
\includegraphics*[bb=120 427 463 750, width=0.99\textwidth]{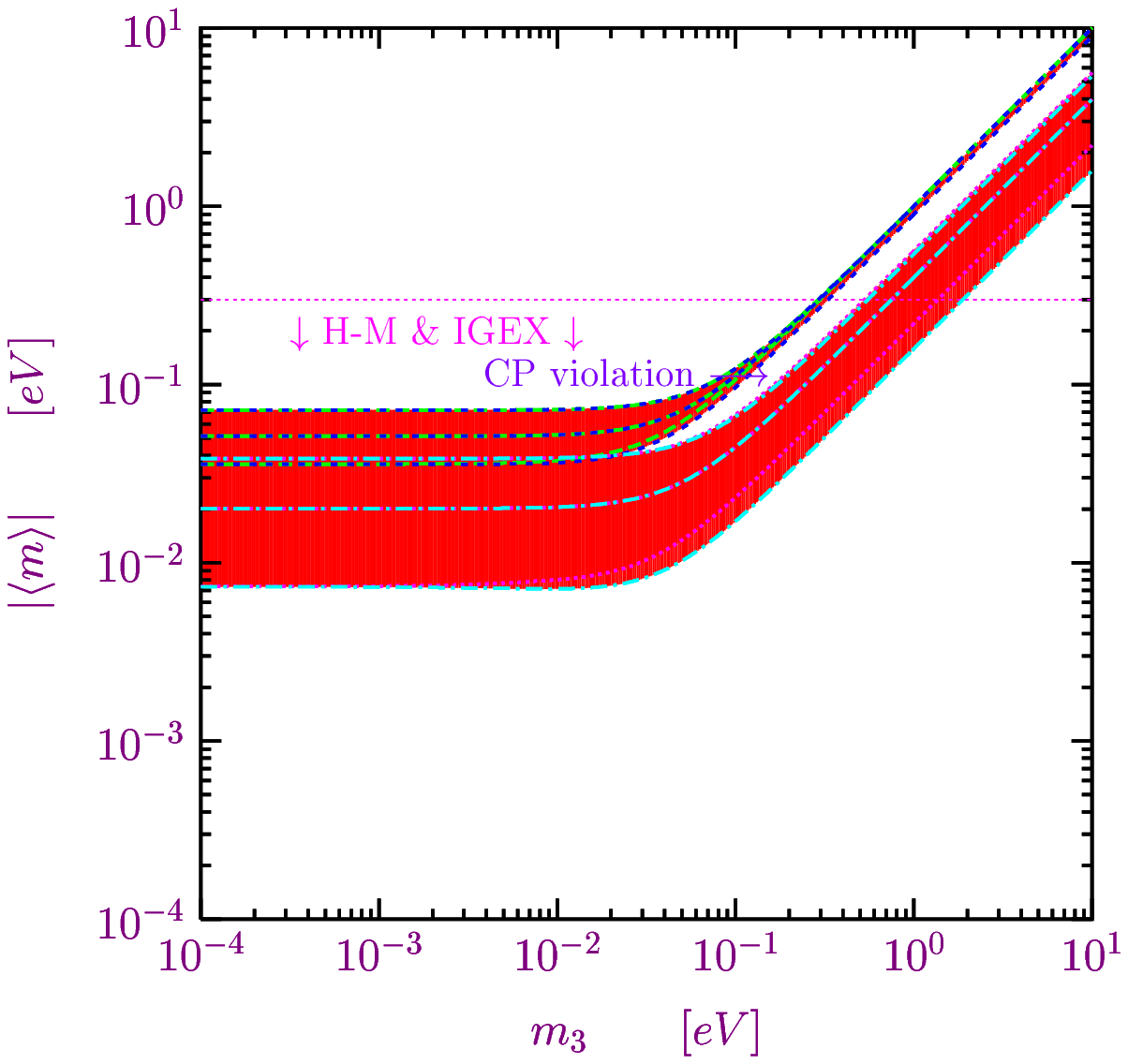}
\end{center}
\end{minipage}
\caption{ \label{eps-db}
Effective Majorana mass $|\langle{m}\rangle|$
in neutrinoless double-$\beta$ decay experiments as a function
of the lightest mass $m_1$ and $m_3$ in the normal and inverted
three-neutrino scheme, respectively.
}
\end{figure*}

\subsection{Neutrinoless double-$\beta$ decay}
\label{Neutrinoless double-beta decay}

A very important open problem in neutrino physics
is the Dirac or Majorana nature of neutrinos.
From the theoretical point of view it
is expected that neutrinos are Majorana particles,
with masses generated by the see-saw mechanism
(see Section~\ref{The see-saw mechanism})
or by effective Lagrangian terms
in which heavy degrees of freedom have been integrated out
(see Section~\ref{Effective dimension-five operator} and Ref.~\cite{Altarelli:2003vk}).

The best known way to search for Majorana neutrino masses
is neutrinoless double-$\beta$ decay,
whose amplitude is proportional to the effective Majorana mass
(see Refs.~\cite{Mohapatra:1998rq,CWKim-book,Bilenky:1987ty,BGG-review-98,Elliott:2002xe})
\begin{equation}
|\langle{m}\rangle|
=
\bigg|
\sum_k
U_{ek}^2 \, m_k
\bigg|
\,.
\label{eps-021}
\end{equation}
The present experimental upper limit on $|\langle{m}\rangle|$,
between about 0.3 eV and 1.3 eV, has been
obtained in the Hei\-del\-berg-Moscow \cite{Klapdor-Kleingrothaus:2001yx}
and IGEX experiments \cite{Aalseth:2002rf}.
The large uncertainty is due to the difficulty
of calculating the nuclear matrix element in the decay.
Figure~\ref{eps-db} shows the allowed range for $|\langle{m}\rangle|$
obtained from the 99.73\% C.L. allowed values of the oscillation parameters
in Eqs.~(\ref{321}), (\ref{322}), (\ref{361}), (\ref{362}),
as a function of the lightest mass
in the normal and inverted three-neutrino schemes
(see also Refs.~\cite{Pascoli:2002qm,hep-ph/0310003,hep-ph/0304276}).
If CP is conserved,
$|\langle{m}\rangle|$
is constrained to lie in the shadowed region.
Finding $|\langle{m}\rangle|$ in an unshaded strip
would signal CP violation.
One can see that in the normal scheme large cancellations
between the three mass contributions are possible
and
$|\langle{m}\rangle|$
can be arbitrarily small.
On the other hand,
the cancellations in the inverted scheme are limited,
because $\nu_1$ and $\nu_2$,
with which the electron neutrino has large mixing,
are almost degenerate and much heavier than $\nu_3$.
Since the solar mixing angle is less than maximal,
a complete cancellation between the contributions of $\nu_1$ and $\nu_2$
is excluded,
leading to a lower bound of about
$7 \times 10^{-3} \, \text{eV}$
for $|\langle{m}\rangle|$
in the inverted scheme
(see also Ref.~\cite{hep-ph/0309114}).
If in the future
$|\langle{m}\rangle|$
will be found to be smaller than
about
$7 \times 10^{-3} \, \text{eV}$,
it will be established that either neutrinos have a mass hierarchy
or they are Dirac particles.
Many neutrinoless double-$\beta$ decay experiments are planned for the future
(see Refs.~\cite{Cremonesi:2002is,Bilenky:2002aw,Zdesenko-RMP-02,Gratta:LP03}).
The most sensitive may be able to probe such small
values of $|\langle{m}\rangle|$.

\section{Future prospects}
\label{Future prospects}

As reviewed in Section~\ref{Neutrino oscillation experiments},
in recent years
neutrino oscillations have been established in a model-independent way
in solar ($\sim 19 \sigma$) and atmospheric ($\sim 7 \sigma$)
neutrino experiments.
An impressive proof of neutrino oscillations has also been obtained in the KamLAND
reactor experiment ($\sim 4 \sigma$).
As we have discussed in Section~\ref{Phenomenology of three-neutrino mixing},
all these experimental results are nicely accommodated
in the framework of three-neutrino mixing.
Only the
LSND anomaly \cite{Aguilar:2001ty} does not fit in this scheme,
as already noticed.
Its interpretation in terms of neutrino oscillation
is presently under investigation in the MiniBooNE experiment
\cite{hep-ex/0210020}.

Putting aside for the moment the controversial LSND anomaly,
the immediate future prospects of neutrino oscillation research deal with
the precise determination of the parameters of the
three-neutrino mixing matrix.
The actual value of $|U_{e3}|$, the real unknown, will be
searched for in the near future
with the long baseline accelerator neutrino programs
MINOS
\cite{Diwan:2002pu,hep-ex/0210005}
in the United States,
OPERA 
\cite{Komatsu:2002sz} 
and ICARUS 
\cite{Rubbia:2002rb}
in Europe
(CERN to Gran Sasso \cite{hep-ex/0209082})
and JHF 
\cite{Itow-ICHEP02}
in Japan.
All these projects use conventional neutrino beams.

However,
the evolution of neutrino physics demands new schemes to produce intense,
collimated and pure neutrino beams.
New possibilities have been studied in the last few years:
neutrino beams from a Neutrino Factory, Beta-Beams and Super-Beams,
that here we briefly summarize.

The current Neutrino Factory concept implies
the production, collection, and storage of muons to produce
very intense beams of muon and
electron neutrinos with equal fluxes
through the decays (\ref{352}).
Research and development addressing the feasibility
of a Neutrino Factory are currently in progress. 
Review studies on the physics reach of the Neutrino Factory
option are given in Refs.~\cite{hep-ph/0210192,Albright:2000xi}.
Incidentally, we notice that anomalous muon decays, due to non standard
weak interactions, if responsible of the observed
LSND effect, could be easily tested at a Neutrino Factory
with a short-baseline experiment,
using a 10 ton detector capable of charge discrimination
\cite{Bueno:2000jy}.

The Beta-Beam concept,
first proposed in Ref.~\cite{Zucchelli:2001gp},
is based on the acceleration and storage of
radioactive ions.
The $\beta$-decay of these radioactive ions
can produce a very intense beam of electron neutrinos or antineutrinos
with
perfectly known energy spectrum.
The physics reach of a CERN based Beta-Beam and
of a Super-Beam+Beta-Beam combination is studied in Ref.~\cite{hep-ex/0302007}.
In this study a Super-Beam is a very intense $\nu_{\mu}$ beam
that has a Super Proton Linac (SPL) as injector
delivering $10^{23}$ protons on target per year
with energy of 2.2 GeV (see also Ref.~\cite{Blondel:2001jk}).
The sensitivity to the Dirac CP-violating phase $\varphi_{13}$ in the neutrino mixing matrix
(\ref{046})
reachable in a Super-Beam+Beta-Beam combination
results to be comparable to the sensitivity reachable
in a Neutrino Factory if $\sin^2{\vartheta_{13}} \gtrsim 10^{-4}$.
Comparisons between the sensitivities of different projects
already approved or planned are presented in Ref.~\cite{Harris:LP03}.

A next-generation neutrino oscillation experiment using reactor antineutrinos 
could
give important information on the size of the mixing angle $\vartheta_{13}$
\cite{Minakata:2002jv,hep-ex/0306031}.
Reactor experiments can give a clean measure of the mixing angle without
ambiguities associated with the size of the other mixing angles, matter 
effects,
and effects due to CP violation. The key question is whether a 
next-generation
experiment can reach the needed sensitivity goals to make a measurement 
of
$\sin^{2}2\vartheta_{13}$ at the $10^{-2}$ level \cite{hep-ph/0303232,hep-ex/0306031}.

However, the search for $|U_{13}|$ and CP violation in the lepton sector
does not cover all the items in today neutrino physics.
Let us emphasize that still several fundamental
characteristics of neutrinos are unknown.
Among them, the Dirac or Majorana nature of neutrinos,
the absolute scale of neutrino masses, the distinction
between the normal and inverted schemes and the electromagnetic properties
of neutrinos.

In our opinion the most important question in today neutrino physics is:
are massive neutrinos Majorana particles?
This question can be resolved with an affirmative answer
if neutrinoless beta decay is observed
\cite{Schechter:1982bd,Takasugi:1984xr}.
Many experimental proposals exist that will increase dramatically
the sensitivity of the neutrinoless double-$\beta$ decay search,
reaching $|\langle{m}\rangle| \sim 10^{-2}$ eV
(see Refs.~\cite{Cremonesi:2002is,Bilenky:2002aw,Zdesenko-RMP-02,Gratta:LP03}).

Besides masses, neutrinos can have magnetic moments
(see Refs.~\cite{Bilenky:1987ty,Mohapatra:1998rq,CWKim-book,hep-ph/0307149}).
In the case of a Dirac neutrino,
a large enough magnetic moment could lead to spin precession
in a transverse magnetic field
\cite{Cisneros:1971nq,Voloshin:1986ty,Okun:1986na,Okun:1986hi},
generating transitions between
a neutrino with negative helicity and one with positive helicity,
which is practically sterile.
A Majorana neutrino cannot have a magnetic moment,
but different Majorana neutrinos can have transition magnetic moments,
which could lead to Spin-Flavor Precession (SFP).
However, SFP is suppressed in vacuum by the mass difference
of different neutrinos.
In 1988 Akhmedov \cite{Akhmedov:1988nc,Akhmedov:1988uk}
and
Lim and Marciano \cite{Lim:1988tk}
discovered that the mass difference can be compensated by the
matter potential for neutrinos propagating in a medium,
leading to the so-called
Resonant Spin-Flavor Precession (RSFP)
mechanism,
which is analogous to the MSW effect.

The RSFP mechanism was proposed as a possible explanation of the
solar neutrino problem and was considered a viable possibility for many years.
Now we know that the solar neutrino problem
is due to neutrino oscillations
(see Section~\ref{Solar neutrino experiments and KamLAND}),
but RSFP could still be a subdominant mechanism (see Ref.~\cite{Akhmedov:2002mf}).
A possible signature of RSFP
would be a periodic variability of the solar neutrino flux
due to temporal variations of the magnetic field or to
the inclination of the solar equator with respect to the ecliptic.
A recent analysis of Super-Kamiokande data
put severe limits to possible periodic modulations of the $^8\text{B}$ solar neutrino flux
\cite{Yoo:2003rc}
(see, however, Ref.~\cite{hep-ph/0309191} for a different point of view).

If massive neutrinos are Majorana particles,
a combination of the RSFP mechanism and oscillations can induce
besides
$\nu_e\to\nu_\mu,\nu_\tau,\bar\nu_\mu,\bar\nu_\tau$
transitions
also
$\nu_e\to\bar\nu_e$
transitions\footnote{
Here $\bar\nu_e$ is the conventional
name for a neutrino state with almost exact positive helicity,
which can induce the creation of a positron
upon scattering with matter
(see the remark at the end of Section~\ref{Three-neutrino mixing}).
}
of solar neutrinos \cite{Akhmedov:1989df,Akhmedov:1991uk,Akhmedov:1993ta},
which can be observed in solar neutrino detectors
through the inverse $\beta$-decay process (\ref{581}).
However,
the Super-Kamiokande experiment
did not find any indication of $\bar\nu_e$'s coming from the sun
and established an upper limit of $8 \times 10^{-3}$ for the
averaged probability of $\nu_e\to\bar\nu_e$ conversion
in the energy range 8-20 MeV
(assuming the initial Standard Solar Model $^8\text{B}$ $\nu_e$ flux)
\cite{Gando:2002ub}.

In any case,
even if at the moment there seems to be no indication of
an effect of the RSFP mechanism in solar neutrino data,
it is important to pursue this line of research,
because magnetic moments are important properties
and the existence of large neutrino magnetic moments
could give crucial indications on the physics beyond the Standard Model.

Direct measurements of neutrino magnetic moments are also
planned:
a proposal on the direct detection of antineutrino-electron scattering with
an artificial tritium source
\cite{Neganov:2001bn}
will search for a neutrino magnetic moment about
two orders of magnitude smaller than
the present-day laboratory upper limit,
reaching a sensitivity of about
$10^{-12} \, \mu_{\text{B}}$,
where $\mu_{\text{B}}$ is the Bohr magneton.

\section{Conclusions}
\label{Conclusions}

The recent years have been extraordinarily fruitful
for neutrino physics,
yielding model-independent proofs
of solar and atmospheric neutrino oscillations,
which have been confirmed, respectively, by the reactor experiment
KamLAND and the accelerator experiment K2K.
Taking into account the negative result of the CHOOZ
long-baseline reactor $\bar\nu_e$ disappearance experiment,
the global fit of solar, KamLAND, atmospheric and K2K data
have provided
important information on the neutrino mixing parameters
in the framework of three-neutrino mixing,
which is predicted by the natural versions of the see-saw mechanism.

The only experimental result that cannot be explained in the framework of
three-neutrino mixing is the controversial
indication in favor of short-baseline
$\bar\nu_\mu\to\bar\nu_e$
transitions observed in the LSND experiment.
It is very important that the MiniBooNE experiment
running at Fermilab will check the validity
of the LSND indication in the near future.
If MiniBooNE will obtain a positive result,
the investigation of new possibilities
(as four-neutrino mixing,
CPT violation, etc.)
will become imperative.
These new phenomena would be very interesting
for our understanding of the physics beyond the Standard Model.

Even if the values of some parameters of three-neutrino mixing
are determined with a precision that was unthinkable a few years ago,
still several fundamental characteristics of neutrinos
remain unknown.
Among them the most important are:
the Dirac or Majorana nature of neutrinos,
the absolute scale of neutrino masses,
the distinction between the normal and inverted schemes,
the value of $|U_{e3}|$,
the existence of CP violation in the lepton sector,
the number of light neutrinos
and
the electromagnetic properties of neutrinos.
Several existing experiments and future projects are aimed
at the exploration of these characteristics,
which
are very important for our understanding
of neutrino physics.
Their determination is likely
to shed some light on
the new physics beyond the Standard Model.
Therefore,
we think that interesting years lie ahead in neutrino physics research.

%\begin{center}
%\textbf{Acknowledgments}
%\end{center}

%\bibliographystyle{h-elsevier2}%{physrev3}%{myplainnat}
%\input{bibtex/bib.tex}


\begin{thebibliography}{100}

\bibitem{Pontecorvo:1957cp}
B. Pontecorvo,
Sov. Phys. JETP 6 (1957) 429
[Zh. Eksp. Teor. Fiz. 33 (1957) 549].

\bibitem{Pontecorvo-58}
B. Pontecorvo,
Sov. Phys. JETP 7 (1958) 172
[Zh. Eksp. Teor. Fiz. 34 (1958) 247].

\bibitem{Fukuda:1998mi}
Super-Kamiokande, Y. Fukuda et~al.,
Phys. Rev. Lett. 81 (1998) 1562,
\href{http://arxiv.org/abs/hep-ex/9807003}{hep-ex/9807003}.

\bibitem{Hirata:1988uy}
Kamiokande-II, K.S. Hirata et~al.,
Phys. Lett. B205 (1988) 416.

\bibitem{Bionta:1988an}
IMB, R.M. Bionta et~al.,
Phys. Rev. D38 (1988) 768.

\bibitem{Ahn:2002up}
K2K, M.H. Ahn et~al.,
Phys. Rev. Lett. 90 (2003) 041801,
\href{http://arxiv.org/abs/hep-ex/0212007}{hep-ex/0212007}.

\bibitem{Ahmad:2001an}
SNO, Q.R. Ahmad et~al.,
Phys. Rev. Lett. 87 (2001) 071301,
\href{http://arxiv.org/abs/nucl-ex/0106015}{nucl-ex/0106015}.

\bibitem{Fukuda:2001nj}
Super-Kamiokande, S. Fukuda et~al.,
Phys. Rev. Lett. 86 (2001) 5651,
\href{http://arxiv.org/abs/hep-ex/0103032}{hep-ex/0103032}.

\bibitem{Cleveland:1998nv}
Homestake, B.T. Cleveland et~al.,
Astrophys. J. 496 (1998) 505.

\bibitem{Ahmad:2002jz}
SNO, Q.R. Ahmad et~al.,
Phys. Rev. Lett. 89 (2002) 011301,
\href{http://arxiv.org/abs/nucl-ex/0204008}{nucl-ex/0204008}.

\bibitem{nucl-ex/0309004}
SNO, S. Ahmed et~al.,
\href{http://arxiv.org/abs/nucl-ex/0309004}{nucl-ex/0309004}.

\bibitem{hep-ex/0212021}
KamLAND, K. Eguchi et~al.,
Phys. Rev. Lett. 90 (2003) 021802,
\href{http://arxiv.org/abs/hep-ex/0212021}{hep-ex/0212021}.

\bibitem{Mohapatra:1998rq}
R.N. Mohapatra and P.B. Pal,
{Massive neutrinos in physics and astrophysics} (World Sci. Lect.
  Notes Phys. 60, 1998).

\bibitem{CWKim-book}
C.W. Kim and A. Pevsner,
{Neutrinos in physics and astrophysics} (Harwood Academic Press,
  Chur, Switzerland, 1993),
Contemporary Concepts in Physics, Vol. 8.

\bibitem{Bahcall:1989ks}
J.N. Bahcall,
{Neutrino Astrophysics} (Cambridge University Press, 1989).

\bibitem{Bilenky:1978nj}
S.M. Bilenky and B. Pontecorvo,
Phys. Rept. 41 (1978) 225.

\bibitem{Bilenky:1987ty}
S.M. Bilenky and S.T. Petcov,
Rev. Mod. Phys. 59 (1987) 671.

\bibitem{Mikheev:1987qk}
S.P. Mikheev and A.Y. Smirnov,
Sov. Phys. Usp. 30 (1987) 759.

\bibitem{Kuo:1989qe}
T.K. Kuo and J. Pantaleone,
Rev. Mod. Phys. 61 (1989) 937.

\bibitem{Castellani:1997cm}
V. Castellani et~al.,
Phys. Rept. 281 (1997) 309,
\href{http://arxiv.org/abs/astro-ph/9606180}{astro-ph/9606180}.

\bibitem{BGG-review-98}
S.M. Bilenky, C. Giunti and W. Grimus,
Prog. Part. Nucl. Phys. 43 (1999) 1,
\href{http://arxiv.org/abs/hep-ph/9812360}{hep-ph/9812360}.

\bibitem{Altarelli:1999gu}
G. Altarelli and F. Feruglio,
Phys. Rept. 320 (1999) 295,
\href{http://arxiv.org/abs/hep-ph/9905536}{hep-ph/9905536}.

\bibitem{Kajita:2000mr}
T. Kajita and Y. Totsuka,
Rev. Mod. Phys. 73 (2001) 85.

\bibitem{Jung:2001dh}
C.K. Jung et~al.,
Ann. Rev. Nucl. Part. Sci. 51 (2001) 451.

\bibitem{Altmann:2001eu}
M.F. Altmann, R.L. Mossbauer and L.J.N. Oberauer,
Rept. Prog. Phys. 64 (2001) 97.

\bibitem{Bilenkii:2001yh}
S.M. Bilenky and C. Giunti,
Int. J. Mod. Phys. A16 (2001) 3931,
\href{http://arxiv.org/abs/hep-ph/0102320}{hep-ph/0102320}.

\bibitem{Beuthe:2001rc}
M. Beuthe,
Phys. Rept. 375 (2003) 105,
\href{http://arxiv.org/abs/hep-ph/0109119}{hep-ph/0109119}.

\bibitem{Gaisser:2002jj}
T.K. Gaisser and M. Honda,
Ann. Rev. Nucl. Part. Sci. 52 (2002) 153,
\href{http://arxiv.org/abs/hep-ph/0203272}{hep-ph/0203272}.

\bibitem{Giacomelli-0201032}
G. Giacomelli, M. Giorgini and M. Spurio,
\href{http://arxiv.org/abs/hep-ex/0201032}{hep-ex/0201032}.

\bibitem{hep-ph/0202058}
M. Gonzalez-Garcia and Y. Nir,
Rev. Mod. Phys. 75 (2003) 345,
\href{http://arxiv.org/abs/hep-ph/0202058}{hep-ph/0202058}.

\bibitem{Elliott:2002xe}
S.R. Elliott and P. Vogel,
Ann. Rev. Nucl. Part. Sci. 52 (2002) 115,
\href{http://arxiv.org/abs/hep-ph/0202264}{hep-ph/0202264}.

\bibitem{Dolgov:2002wy}
A.D. Dolgov,
Phys. Rept. 370 (2002) 333,
\href{http://arxiv.org/abs/hep-ph/0202122}{hep-ph/0202122}.

\bibitem{Kayser:2002qs}
B. Kayser,
\href{http://arxiv.org/abs/hep-ph/0211134}{hep-ph/0211134}.

\bibitem{Miramonti-Reseghetti-2002}
L. Miramonti and F. Reseghetti,
La Rivista del Nuovo Cimento 25 (2002) 1,
\href{http://arxiv.org/abs/hep-ex/0302035}{hep-ex/0302035}.

\bibitem{Bilenky:2002aw}
S.M. Bilenky et~al.,
Phys. Rept. 379 (2003) 69,
\href{http://arxiv.org/abs/hep-ph/0211462}{hep-ph/0211462}.

\bibitem{hep-ph/0307149}
W. Grimus,
\href{http://arxiv.org/abs/hep-ph/0307149}{hep-ph/0307149}.

\bibitem{hep-ph/0301276}
C. Giunti and M. Laveder,
\href{http://arxiv.org/abs/hep-ph/0301276}{hep-ph/0301276}.

\bibitem{Altarelli:2003vk}
G. Altarelli and F. Feruglio,
\href{http://arxiv.org/abs/hep-ph/0306265}{hep-ph/0306265}.

\bibitem{Neutrino-Unbound}
C. Giunti and M. Laveder,
Neutrino Unbound,
\href{http://www.nu.to.infn.it}{http://www.nu.to.infn.it}.

\bibitem{Glashow-SM-61}
S.L. Glashow,
Nucl. Phys. 22 (1961) 579.

\bibitem{Weinberg-SM-67}
S. Weinberg,
Phys. Rev. Lett. 19 (1967) 1264.

\bibitem{Salam-SM-68}
A. Salam,
(1969),
Proc. of the 8$^{\mathrm{th}}$ Nobel Symposium on \textit{Elementary
  particle theory, relativistic groups and analyticity}, Stockholm, Sweden,
  1968, edited by N. Svartholm, p.367-377.

\bibitem{Landau-57}
L. Landau,
Nucl. Phys. 3 (1957) 127.

\bibitem{Lee-Yang-57}
T.D. Lee and C.N. Yang,
Phys. Rev. 105 (1957) 1671.

\bibitem{Salam-57}
A. Salam,
Nuovo Cim. 5 (1957) 299.

\bibitem{Majorana:1937vz}
E. Majorana,
Nuovo Cim. 14 (1937) 171.

\bibitem{Yanagida-SeeSaw-1979}
T. Yanagida,
(1979),
Proc. of the Workshop on Unified Theory and the Baryon Number of the
  Universe, KEK, Japan.

\bibitem{GellMann-Ramond-Slansky-SeeSaw-1979}
M. Gell-Mann, P. Ramond and R. Slansky,
(1979),
In 'Supergravity', p.~315, edited by F. van Nieuwenhuizen and D.
  Freedman, North Holland, Amsterdam.

\bibitem{Mohapatra:1980ia}
R.N. Mohapatra and G. Senjanovic,
Phys. Rev. Lett. 44 (1980) 912.

\bibitem{Weinberg:1979sa}
S. Weinberg,
Phys. Rev. Lett. 43 (1979) 1566.

\bibitem{Weinberg:1980bf}
S. Weinberg,
Phys. Rev. D22 (1980) 1694.

\bibitem{Weldon:1980gi}
H.A. Weldon and A. Zee,
Nucl. Phys. B173 (1980) 269.

\bibitem{PDG}
Particle Data Group, K. Hagiwara et~al.,
Phys. Rev. D66 (2002) 010001.

\bibitem{Bulanov:2003ka}
S.S. Bulanov et~al.,
\href{http://arxiv.org/abs/hep-ph/0301268}{hep-ph/0301268}.

\bibitem{hep-ph/0210153}
K. Belotsky et~al.,
Phys. Rev. D68 (2003) 054027,
\href{http://arxiv.org/abs/hep-ph/0210153}{hep-ph/0210153}.

\bibitem{Murnaghan-book-62}
F.D. Murnaghan,
{The unitary and rotation groups} (Spartan Articles, Washington D.C.,
  1962).

\bibitem{Schechter-Valle-COMMENT-80}
J. Schechter and J.W.F. Valle,
Phys. Rev. D21 (1980) 309.

\bibitem{Schechter-Valle-MASSES-80}
J. Schechter and J.W.F. Valle,
Phys. Rev. D22 (1980) 2227.

\bibitem{GKM-atm-98}
C. Giunti, C.W. Kim and M. Monteno,
Nucl. Phys. B521 (1998) 3,
\href{http://arxiv.org/abs/hep-ph/9709439}{hep-ph/9709439}.

\bibitem{Giunti:2002pp}
C. Giunti and M. Tanimoto,
Phys. Rev. D66 (2002) 113006,
\href{http://arxiv.org/abs/hep-ph/0209169}{hep-ph/0209169}.

\bibitem{Hampel:1998xg}
GALLEX, W. Hampel et~al.,
Phys. Lett. B447 (1999) 127.

\bibitem{Abdurashitov:2002nt}
SAGE, J.N. Abdurashitov et~al.,
J. Exp. Theor. Phys. 95 (2002) 181,
\href{http://arxiv.org/abs/astro-ph/0204245}{astro-ph/0204245}.

\bibitem{Altmann:2000ft}
GNO, M. Altmann et~al.,
Phys. Lett. B490 (2000) 16,
\href{http://arxiv.org/abs/hep-ex/0006034}{hep-ex/0006034}.

\bibitem{Bugey}
Bugey, Y. Declais et~al.,
Nucl. Phys. B434 (1995) 503.

\bibitem{CHOOZ-99}
CHOOZ, M. Apollonio et~al.,
Phys. Lett. B466 (1999) 415,
\href{http://arxiv.org/abs/hep-ex/9907037}{hep-ex/9907037}.

\bibitem{Fukuda:2002pe}
Super-Kamiokande, S. Fukuda et~al.,
Phys. Lett. B539 (2002) 179,
\href{http://arxiv.org/abs/hep-ex/0205075}{hep-ex/0205075}.

\bibitem{hep-ex/0309011}
Super-Kamiokande, M. Smy et~al.,
\href{http://arxiv.org/abs/hep-ex/0309011}{hep-ex/0309011}.

\bibitem{Giunti:1992cb}
C. Giunti, C.W. Kim and U.W. Lee,
Phys. Rev. D45 (1992) 2414.

\bibitem{Giunti:2000kw}
C. Giunti and C.W. Kim,
Found. Phys. Lett. 14 (2001) 213,
\href{http://arxiv.org/abs/hep-ph/0011074}{hep-ph/0011074}.

\bibitem{Giunti:2003ax}
C. Giunti,
\href{http://arxiv.org/abs/hep-ph/0302026}{hep-ph/0302026}.

\bibitem{Dydak:1984zq}
CDHS, F. Dydak et~al.,
Phys. Lett. B134 (1984) 281.

\bibitem{Naples:1998va}
CCFR/NuTeV, D. Naples et~al.,
Phys. Rev. D59 (1999) 031101,
\href{http://arxiv.org/abs/hep-ex/9809023}{hep-ex/9809023}.

\bibitem{Eskut:2000de}
CHORUS, E. Eskut et~al.,
Phys. Lett. B497 (2001) 8.

\bibitem{hep-ex/0306037}
NOMAD, P. Astier et~al.,
Phys. Lett. B570 (2003) 19,
\href{http://arxiv.org/abs/hep-ex/0306037}{hep-ex/0306037}.

\bibitem{Aguilar:2001ty}
LSND, A. Aguilar et~al.,
Phys. Rev. D64 (2001) 112007,
\href{http://arxiv.org/abs/hep-ex/0104049}{hep-ex/0104049}.

\bibitem{Armbruster:2002mp}
KARMEN, B. Armbruster et~al.,
Phys. Rev. D65 (2002) 112001,
\href{http://arxiv.org/abs/hep-ex/0203021}{hep-ex/0203021}.

\bibitem{Apollonio:2003gd}
CHOOZ, M. Apollonio et~al.,
Eur. Phys. J. C27 (2003) 331,
\href{http://arxiv.org/abs/hep-ex/0301017}{hep-ex/0301017}.

\bibitem{Boehm:2001ik}
Palo Verde, F. Boehm et~al.,
Phys. Rev. D64 (2001) 112001,
\href{http://arxiv.org/abs/hep-ex/0107009}{hep-ex/0107009}.

\bibitem{Diwan:2002pu}
MINOS, M.V. Diwan,
eConf C0209101 (2002) TH08,
\href{http://arxiv.org/abs/hep-ex/0211026}{hep-ex/0211026}.

\bibitem{hep-ex/0209082}
D. Duchesneau,
eConf C0209101 (2002) TH09,
\href{http://arxiv.org/abs/hep-ex/0209082}{hep-ex/0209082}.

\bibitem{Fukuda:1994mc}
Kamiokande, Y. Fukuda et~al.,
Phys. Lett. B335 (1994) 237.

\bibitem{Becker-Szendy:1992hq}
IMB, R. Becker-Szendy et~al.,
Phys. Rev. D46 (1992) 3720.

\bibitem{hep-ex/0307069}
Soudan 2, M. Sanchez et~al.,
\href{http://arxiv.org/abs/hep-ex/0307069}{hep-ex/0307069}.

\bibitem{Ambrosio:2003yz}
MACRO, M. Ambrosio et~al.,
Phys. Lett. B566 (2003) 35,
\href{http://arxiv.org/abs/hep-ex/0304037}{hep-ex/0304037}.

\bibitem{Fukuda:1996sz}
Kamiokande, Y. Fukuda et~al.,
Phys. Rev. Lett. 77 (1996) 1683.

\bibitem{Wolfenstein:1978ue}
L. Wolfenstein,
Phys. Rev. D17 (1978) 2369.

\bibitem{Giunti:1992sx}
C. Giunti, C.W. Kim and U.W. Lee,
Phys. Lett. B274 (1992) 87.

\bibitem{Langacker:1983ih}
P. Langacker, J.P. Leveille and J. Sheiman,
Phys. Rev. D27 (1983) 1228.

\bibitem{Mikheev:1986wj}
S.P. Mikheev and A.Y. Smirnov,
Nuovo Cim. C9 (1986) 17.

\bibitem{Bethe:1986ej}
H.A. Bethe,
Phys. Rev. Lett. 56 (1986) 1305.

\bibitem{Parke:1986jy}
S.J. Parke,
Phys. Rev. Lett. 57 (1986) 1275.

\bibitem{Petcov:1988zj}
S.T. Petcov,
Phys. Lett. B200 (1988) 373.

\bibitem{Krastev:1988ci}
P.I. Krastev and S.T. Petcov,
Phys. Lett. B207 (1988) 64
[Erratum-ibid. B214 (1988) 661].

\bibitem{Petcov:1988wv}
S.T. Petcov,
Phys. Lett. B214 (1988) 139.

\bibitem{Kuo:1989pn}
T.K. Kuo and J. Pantaleone,
Phys. Rev. D39 (1989) 1930.

\bibitem{Bahcall:2000nu}
J.N. Bahcall, M.H. Pinsonneault and S. Basu,
Astrophys. J. 555 (2001) 990,
\href{http://arxiv.org/abs/astro-ph/0010346}{astro-ph/0010346}.

\bibitem{Pizzochero:1987fj}
P. Pizzochero,
Phys. Rev. D36 (1987) 2293.

\bibitem{Toshev:1987jw}
S. Toshev,
Phys. Lett. B196 (1987) 170.

\bibitem{Balantekin:1998jp}
A.B. Balantekin,
Phys. Rev. D58 (1998) 013001,
\href{http://arxiv.org/abs/hep-ph/9712304}{hep-ph/9712304}.

\bibitem{Lisi:2000su}
E. Lisi et~al.,
Phys. Rev. D63 (2001) 093002,
\href{http://arxiv.org/abs/hep-ph/0011306}{hep-ph/0011306}.

\bibitem{Baltz:1987hn}
A.J. Baltz and J. Weneser,
Phys. Rev. D35 (1987) 528.

\bibitem{Liu-Maris-Petcov-earth1-97}
Q.Y. Liu, M. Maris and S.T. Petcov,
Phys. Rev. D56 (1997) 5991,
\href{http://arxiv.org/abs/hep-ph/9702361}{hep-ph/9702361}.

\bibitem{Petcov-diffractive-98}
S.T. Petcov,
Phys. Lett. B434 (1998) 321,
\href{http://arxiv.org/abs/hep-ph/9805262}{hep-ph/9805262}.

\bibitem{Akhmedov-parametric-99}
E.K. Akhmedov,
Nucl. Phys. B538 (1999) 25,
\href{http://arxiv.org/abs/hep-ph/9805272}{hep-ph/9805272}.

\bibitem{Chizhov:1999az}
M.V. Chizhov and S.T. Petcov,
Phys. Rev. Lett. 83 (1999) 1096,
\href{http://arxiv.org/abs/hep-ph/9903399}{hep-ph/9903399}.

\bibitem{Chizhov:1999he}
M.V. Chizhov and S.T. Petcov,
Phys. Rev. D63 (2001) 073003,
\href{http://arxiv.org/abs/hep-ph/9903424}{hep-ph/9903424}.

\bibitem{Friedland:2000cp}
A. Friedland,
Phys. Rev. Lett. 85 (2000) 936,
\href{http://arxiv.org/abs/hep-ph/0002063}{hep-ph/0002063}.

\bibitem{Fogli:2000bk}
G.L. Fogli et~al.,
Phys. Rev. D62 (2000) 113004,
\href{http://arxiv.org/abs/hep-ph/0005261}{hep-ph/0005261}.

\bibitem{Friedland:2000rn}
A. Friedland,
Phys. Rev. D64 (2001) 013008,
\href{http://arxiv.org/abs/hep-ph/0010231}{hep-ph/0010231}.

\bibitem{Bilenky-Hosek-Petcov-PLB94-80}
S.M. Bilenky, J. Hosek and S.T. Petcov,
Phys. Lett. B94 (1980) 495.

\bibitem{Doi-CP-81}
M. Doi et~al.,
Phys. Lett. B102 (1981) 323.

\bibitem{Langacker-Petcov-SteigmanToshev-NPB282-88}
P. Langacker et~al.,
Nucl. Phys. B282 (1987) 589.

\bibitem{Stockdale:1985ce}
CCFR, I.E. Stockdale et~al.,
Z. Phys. C27 (1985) 53.

\bibitem{Athanassopoulos:1998er}
LSND, C. Athanassopoulos et~al.,
Phys. Rev. C58 (1998) 2489,
\href{http://arxiv.org/abs/nucl-ex/9706006}{nucl-ex/9706006}.

\bibitem{Astier:2001yj}
NOMAD, P. Astier et~al.,
Nucl. Phys. B611 (2001) 3,
\href{http://arxiv.org/abs/hep-ex/0106102}{hep-ex/0106102}.

\bibitem{Avvakumov:2002jj}
NuTeV, S. Avvakumov et~al.,
Phys. Rev. Lett. 89 (2002) 011804,
\href{http://arxiv.org/abs/hep-ex/0203018}{hep-ex/0203018}.

\bibitem{Romosan:1997nh}
CCFR/NuTeV, A. Romosan et~al.,
Phys. Rev. Lett. 78 (1997) 2912,
\href{http://arxiv.org/abs/hep-ex/9611013}{hep-ex/9611013}.

\bibitem{Church:2002tc}
E.D. Church et~al.,
Phys. Rev. D66 (2002) 013001,
\href{http://arxiv.org/abs/hep-ex/0203023}{hep-ex/0203023}.

\bibitem{hep-ex/0210020}
MiniBooNE, A. Bazarko,
\href{http://arxiv.org/abs/hep-ex/0210020}{hep-ex/0210020}.

\bibitem{Barger-Fate-2000}
V. Barger et~al.,
Phys. Lett. B489 (2000) 345,
\href{http://arxiv.org/abs/hep-ph/0008019}{hep-ph/0008019}.

\bibitem{Giunti:2000ur}
C. Giunti and M. Laveder,
JHEP 02 (2001) 001,
\href{http://arxiv.org/abs/hep-ph/0010009}{hep-ph/0010009}.

\bibitem{Peres:2000ic}
O.L.G. Peres and A.Y. Smirnov,
Nucl. Phys. B599 (2001) 3,
\href{http://arxiv.org/abs/hep-ph/0011054}{hep-ph/0011054}.

\bibitem{Maltoni:2003yr}
M. Maltoni et~al.,
\href{http://arxiv.org/abs/hep-ph/0305312}{hep-ph/0305312}.

\bibitem{Bergmann:1998ft}
S. Bergmann and Y. Grossman,
Phys. Rev. D59 (1999) 093005,
\href{http://arxiv.org/abs/hep-ph/9809524}{hep-ph/9809524}.

\bibitem{Bueno:2000jy}
A. Bueno et~al.,
JHEP 06 (2001) 032,
\href{http://arxiv.org/abs/hep-ph/0010308}{hep-ph/0010308}.

\bibitem{hep-ph/0204236}
K.S. Babu and S. Pakvasa,
\href{http://arxiv.org/abs/hep-ph/0204236}{hep-ph/0204236}.

\bibitem{hep-ph/0212116}
G. Barenboim, L. Borissov and J. Lykken,
\href{http://arxiv.org/abs/hep-ph/0212116}{hep-ph/0212116}.

\bibitem{Strumia:2002fw}
A. Strumia,
Phys. Lett. B539 (2002) 91,
\href{http://arxiv.org/abs/hep-ph/0201134}{hep-ph/0201134}.

\bibitem{Sorel:2003hf}
M. Sorel, J. Conrad and M. Shaevitz,
\href{http://arxiv.org/abs/hep-ph/0305255}{hep-ph/0305255}.

\bibitem{hep-ph/0306226}
M. Gonzalez-Garcia, M. Maltoni and T. Schwetz,
\href{http://arxiv.org/abs/hep-ph/0306226}{hep-ph/0306226}.

\bibitem{hep-ph/0308299}
V. Barger, D. Marfatia and K. Whisnant,
\href{http://arxiv.org/abs/hep-ph/0308299}{hep-ph/0308299}.

\bibitem{Pontecorvo-cl-46}
B. Pontecorvo,
(1946),
Chalk River Report PD 205.

\bibitem{Alvarez-cl-49}
L.W. Alvarez,
(1949),
University of California Radiation Laboratory Report UCRL 328.

\bibitem{Rolfs-Rodney-book-88}
C.E. Rolfs and W.S. Rodney,
{Cauldrons in the Cosmos} (The University of Chicago Press, 1988).

\bibitem{Kuzmin-Ga-65}
V.A. Kuzmin,
Sov. Phys. JETP 22 (1966) 1051
[Zh. Eksp. Teor. Fiz. 49 (1965) 1532].

\bibitem{Bahcall:1997qw}
J.N. Bahcall et~al.,
Phys. Rev. Lett. 78 (1997) 171,
\href{http://arxiv.org/abs/astro-ph/9610250}{astro-ph/9610250}.

\bibitem{Fogli:2001vr}
G.L. Fogli et~al.,
Phys. Rev. D64 (2001) 093007,
\href{http://arxiv.org/abs/hep-ph/0106247}{hep-ph/0106247}.

\bibitem{Giunti:2001ws}
C. Giunti,
Phys. Rev. D65 (2002) 033006,
\href{http://arxiv.org/abs/hep-ph/0107310}{hep-ph/0107310}.

\bibitem{Apollonio:1999ae}
CHOOZ, M. Apollonio et~al.,
Phys. Lett. B466 (1999) 415,
\href{http://arxiv.org/abs/hep-ex/9907037}{hep-ex/9907037}.

\bibitem{Fogli:2002pt}
G.L. Fogli et~al.,
Phys. Rev. D66 (2002) 053010,
\href{http://arxiv.org/abs/hep-ph/0206162}{hep-ph/0206162}.

\bibitem{hep-ph/0309130}
M. Maltoni, T. Schwetz, M.A. Tortola and J.W.F. Valle,
\href{http://arxiv.org/abs/hep-ph/0309130}{hep-ph/0309130}.

\bibitem{hep-ph/0212147}
J.N. Bahcall, M.C. Gonzalez-Garcia and C. Pena-Garay,
JHEP 0302 (2003) 009,
\href{http://arxiv.org/abs/hep-ph/0212147}{hep-ph/0212147}.

\bibitem{hep-ph/0309174}
A. Bandyopadhyay et~al.,
\href{http://arxiv.org/abs/hep-ph/0309174}{hep-ph/0309174}.

\bibitem{Inoue:2003qs}
K. Inoue,
\href{http://arxiv.org/abs/hep-ex/0307030}{hep-ex/0307030}.

\bibitem{hep-ph/0302243}
A. Bandyopadhyay, S. Choubey and S. Goswami,
Phys. Rev. D67 (2003) 113011,
\href{http://arxiv.org/abs/hep-ph/0302243}{hep-ph/0302243}.

\bibitem{Bahcall:2003ce}
J.N. Bahcall and C. Pena-Garay,
\href{http://arxiv.org/abs/hep-ph/0305159}{hep-ph/0305159}.

\bibitem{hep-ph/0306017}
S. Choubey, S. Petcov and M. Piai,
\href{http://arxiv.org/abs/hep-ph/0306017}{hep-ph/0306017}.

\bibitem{Toshito:2001dk}
Super-Kamiokande, T. Toshito,
\href{http://arxiv.org/abs/hep-ex/0105023}{hep-ex/0105023}.

\bibitem{Scholberg:1999ar}
Super-Kamiokande, K. Scholberg,
\href{http://arxiv.org/abs/hep-ex/9905016}{hep-ex/9905016}.

\bibitem{hep-ex/0212035}
R.J. Wilkes,
eConf C020805 (2002) TTH02,
\href{http://arxiv.org/abs/hep-ex/0212035}{hep-ex/0212035}.

\bibitem{hep-ex/0210030}
Y. Oyama,
\href{http://arxiv.org/abs/hep-ex/0210030}{hep-ex/0210030}.

\bibitem{Kayser:2002ed}
B. Kayser,
Phys. Rev. D66 (2002) 010001,
2002 Review of Particle Physics.

\bibitem{hep-ph/0303064}
G. Fogli et~al.,
Phys. Rev. D67 (2003) 093006,
\href{http://arxiv.org/abs/hep-ph/0303064}{hep-ph/0303064}.

\bibitem{Nakaya:2002ki}
Super-Kamiokande, T. Nakaya,
eConf C020620 (2002) SAAT01,
\href{http://arxiv.org/abs/hep-ex/0209036}{hep-ex/0209036}.

\bibitem{Guler:2000bd}
OPERA, M.~Guler et~al.,
CERN-SPSC-2000-028.

\bibitem{Arneodo:2001tx}
ICARUS, F.~Arneodo et~al.,
\href{http://arxiv.org/abs/hep-ex/0103008}{hep-ex/0103008}.

\bibitem{Apollonio:1998xe}
CHOOZ, M. Apollonio et~al.,
Phys. Lett. B420 (1998) 397,
\href{http://arxiv.org/abs/hep-ex/9711002}{hep-ex/9711002}.

\bibitem{Reines:1953pu}
F. Reines and C.L. Cowan,
Phys. Rev. 92 (1953) 830.

\bibitem{Shi:1992zw}
X. Shi and D.N. Schramm,
Phys. Lett. B283 (1992) 305.

\bibitem{Bilenky:1998tw}
S.M. Bilenky and C. Giunti,
Phys. Lett. B444 (1998) 379,
\href{http://arxiv.org/abs/hep-ph/9802201}{hep-ph/9802201}.

\bibitem{Fogli:2002pb}
G.L. Fogli et~al.,
Phys. Rev. D66 (2002) 093008,
\href{http://arxiv.org/abs/hep-ph/0208026}{hep-ph/0208026}.

\bibitem{hep-ph/0212142}
W.L. Guo and Z.Z. Xing,
Phys. Rev. D67 (2003) 053002,
\href{http://arxiv.org/abs/hep-ph/0212142}{hep-ph/0212142}.

\bibitem{hep-ex/0210050}
C. Weinheimer,
\href{http://arxiv.org/abs/hep-ex/0210050}{hep-ex/0210050}.

\bibitem{Weinheimer:1999tn}
Mainz, C. Weinheimer et~al.,
Phys. Lett. B460 (1999) 219.

\bibitem{Lobashev:1999tp}
Troitsk, V.M. Lobashev et~al.,
Phys. Lett. B460 (1999) 227.

\bibitem{hep-ex/0109033}
KATRIN, A. Osipowicz et~al.,
\href{http://arxiv.org/abs/hep-ex/0109033}{hep-ex/0109033}.

\bibitem{Spergel:2003cb}
D.N. Spergel et~al.,
\href{http://arxiv.org/abs/astro-ph/0302209}{astro-ph/0302209}.

\bibitem{Gershtein:1966gg}
S.S. Gershtein and Y.B. Zeldovich,
JETP Lett. 4 (1966) 120,
[Pisma Zh. Eksp. Teor. Fiz. 4 (1966) 174].

\bibitem{Cowsik:1972gh}
R. Cowsik and J. McClelland,
Phys. Rev. Lett. 29 (1972) 669.

\bibitem{Hu:1998mj}
W. Hu, D.J. Eisenstein and M. Tegmark,
Phys. Rev. Lett. 80 (1998) 5255,
\href{http://arxiv.org/abs/astro-ph/9712057}{astro-ph/9712057}.

\bibitem{Wang:2001gy}
X. Wang, M. Tegmark and M. Zaldarriaga,
\href{http://arxiv.org/abs/astro-ph/0105091}{astro-ph/0105091}.

\bibitem{Hannestad:2002cn}
S. Hannestad,
Phys. Rev. D67 (2003) 085017,
\href{http://arxiv.org/abs/astro-ph/0211106}{astro-ph/0211106}.

\bibitem{Bennett:2003bz}
C.L. Bennett et~al.,
\href{http://arxiv.org/abs/astro-ph/0302207}{astro-ph/0302207}.

\bibitem{Colless:2003wz}
M. Colless et~al.,
\href{http://arxiv.org/abs/astro-ph/0306581}{astro-ph/0306581}.

\bibitem{astro-ph/0310133}
S. Hannestad,
\href{http://arxiv.org/abs/astro-ph/0310133}{astro-ph/0310133}.

\bibitem{astro-ph/0304237}
A. Blanchard et~al.,
\href{http://arxiv.org/abs/astro-ph/0304237}{astro-ph/0304237}.

\bibitem{astro-ph/0303076}
S. Hannestad,
JCAP 0305 (2003) 004,
\href{http://arxiv.org/abs/astro-ph/0303076}{astro-ph/0303076}.

\bibitem{astro-ph/0303089}
O. Elgaroy and O. Lahav,
JCAP 04 (2003) 004,
\href{http://arxiv.org/abs/astro-ph/0303089}{astro-ph/0303089}.

\bibitem{Klapdor-Kleingrothaus:2001yx}
H.V. Klapdor-Kleingrothaus et~al.,
Eur. Phys. J. A12 (2001) 147.

\bibitem{Aalseth:2002rf}
IGEX, C.E. Aalseth et~al.,
Phys. Rev. D65 (2002) 092007,
\href{http://arxiv.org/abs/hep-ex/0202026}{hep-ex/0202026}.

\bibitem{Pascoli:2002qm}
S. Pascoli, S.T. Petcov and W. Rodejohann,
Phys. Lett. B549 (2002) 177,
\href{http://arxiv.org/abs/hep-ph/0209059}{hep-ph/0209059}.

\bibitem{hep-ph/0310003}
S. Pascoli and S.T. Petcov,
\href{http://arxiv.org/abs/hep-ph/0310003}{hep-ph/0310003}.

\bibitem{hep-ph/0304276}
F.R. Joaquim,
Phys. Rev. D68 (2003) 033019,
\href{http://arxiv.org/abs/hep-ph/0304276}{hep-ph/0304276}.

\bibitem{hep-ph/0309114}
H. Murayama and C. Pena-Garay,
\href{http://arxiv.org/abs/hep-ph/0309114}{hep-ph/0309114}.

\bibitem{Cremonesi:2002is}
O. Cremonesi,
\href{http://arxiv.org/abs/hep-ex/0210007}{hep-ex/0210007}.

\bibitem{Zdesenko-RMP-02}
Y. Zdesenko,
Rev. Mod. Phys. 74 (2002) 663.

\bibitem{Gratta:LP03}
G. Gratta,
Talk presented at the
XXI International Symposium on Lepton Photon 2003, 11-16 August 2003,
Fermi National Accelerator Laboratory, Batavia, Illinois USA.

\bibitem{hep-ex/0210005}
D. Ayres et~al.,
\href{http://arxiv.org/abs/hep-ex/0210005}{hep-ex/0210005}.

\bibitem{Komatsu:2002sz}
M. Komatsu, P. Migliozzi and F. Terranova,
J. Phys. G29 (2003) 443,
\href{http://arxiv.org/abs/hep-ph/0210043}{hep-ph/0210043}.

\bibitem{Rubbia:2002rb}
A. Rubbia and P. Sala,
JHEP 09 (2002) 004,
\href{http://arxiv.org/abs/hep-ph/0207084}{hep-ph/0207084}.

\bibitem{Itow-ICHEP02}
Super-Kamiokande, Y. Itow,
Talk presented at the
31st International Conference on High Energy Physics, ICHEP 02,
Amsterdam, Holland, 24-31 July 2002.

\bibitem{hep-ph/0210192}
M. Apollonio et~al.,
\href{http://arxiv.org/abs/hep-ph/0210192}{hep-ph/0210192}.

\bibitem{Albright:2000xi}
C. Albright et~al.,
\href{http://arxiv.org/abs/hep-ex/0008064}{hep-ex/0008064}.

\bibitem{Zucchelli:2001gp}
P. Zucchelli,
\href{http://arxiv.org/abs/hep-ex/0107006}{hep-ex/0107006}.

\bibitem{hep-ex/0302007}
M. Mezzetto,
J. Phys. G29 (2003) 1771,
\href{http://arxiv.org/abs/hep-ex/0302007}{hep-ex/0302007}.

\bibitem{Blondel:2001jk}
A. Blondel et~al.,
Nucl. Instrum. Meth. A503 (2001) 173.

\bibitem{Harris:LP03}
D. Harris,
Talk presented at the
XXI International Symposium on Lepton Photon 2003, 11-16 August 2003,
Fermi National Accelerator Laboratory, Batavia, Illinois USA.

\bibitem{Minakata:2002jv}
H. Minakata et~al.,
\href{http://arxiv.org/abs/hep-ph/0211111}{hep-ph/0211111}.

\bibitem{hep-ph/0303232}
P. Huber, M. Lindner, T. Schwetz and W. Winter,
Nucl. Phys. B665 (2003) 487,
\href{http://arxiv.org/abs/hep-ph/0303232}{hep-ph/0303232}.

\bibitem{hep-ex/0306031}
M.H. Shaevitz and J.M. Link,
\href{http://arxiv.org/abs/hep-ex/0306031}{hep-ex/0306031}.

\bibitem{Schechter:1982bd}
J. Schechter and J.W.F. Valle,
Phys. Rev. D25 (1982) 2951.

\bibitem{Takasugi:1984xr}
E. Takasugi,
Phys. Lett. B149 (1984) 372.

\bibitem{Cisneros:1971nq}
A. Cisneros,
Astrophys. Space Sci. 10 (1971) 87.

\bibitem{Voloshin:1986ty}
M.B. Voloshin and M.I. Vysotsky,
Sov. J. Nucl. Phys. 44 (1986) 544.

\bibitem{Okun:1986na}
L.B. Okun, M.B. Voloshin and M.I. Vysotsky,
Sov. Phys. JETP 64 (1986) 446.

\bibitem{Okun:1986hi}
L.B. Okun, M.B. Voloshin and M.I. Vysotsky,
Sov. J. Nucl. Phys. 44 (1986) 440.

\bibitem{Akhmedov:1988nc}
E.K. Akhmedov,
Sov. J. Nucl. Phys. 48 (1988) 382.

\bibitem{Akhmedov:1988uk}
E.K. Akhmedov,
Phys. Lett. B213 (1988) 64.

\bibitem{Lim:1988tk}
C.S. Lim and W.J. Marciano,
Phys. Rev. D37 (1988) 1368.

\bibitem{Akhmedov:2002mf}
E.K. Akhmedov and J. Pulido,
Phys. Lett. B553 (2003) 7,
\href{http://arxiv.org/abs/hep-ph/0209192}{hep-ph/0209192}.

\bibitem{Yoo:2003rc}
Super-Kamiokande, J. Yoo et~al.,
\href{http://arxiv.org/abs/hep-ex/0307070}{hep-ex/0307070}.

\bibitem{hep-ph/0309191}
D.O. Caldwell and P.A. Sturrock,
\href{http://arxiv.org/abs/hep-ph/0309191}{hep-ph/0309191}.

\bibitem{Akhmedov:1989df}
E.K. Akhmedov,
Sov. Phys. JETP 68 (1989) 690.

\bibitem{Akhmedov:1991uk}
E.K. Akhmedov,
Phys. Lett. B255 (1991) 84.

\bibitem{Akhmedov:1993ta}
E.K. Akhmedov, S.T. Petcov and A.Y. Smirnov,
Phys. Lett. B309 (1993) 95,
\href{http://arxiv.org/abs/hep-ph/9301247}{hep-ph/9301247}.

\bibitem{Gando:2002ub}
Super-Kamiokande, Y. Gando et~al.,
Phys. Rev. Lett. 90 (2003) 171302,
\href{http://arxiv.org/abs/hep-ex/0212067}{hep-ex/0212067}.

\bibitem{Neganov:2001bn}
B.S. Neganov et~al.,
Phys. Atom. Nucl. 64 (2001) 1948,
\href{http://arxiv.org/abs/hep-ex/0105083}{hep-ex/0105083}.

\end{thebibliography}
\end{document}